**University of Nevada, Reno**

# Dynamic Response of Tunable Phononic Crystals and New Homogenization Approaches in Magnetoactive Composites

A dissertation submitted in partial fulfillment of the

requirements for the degree of Doctor of Philosophy in

Mechanical Engineering

By

Alireza Bayat

Dissertation Advisor:

Dr. Faramarz Gordaninejad

August, 2015

THE GRADUATE SCHOOL

We recommend that the dissertation
prepared under our supervision by

**ALIREZA BAYAT**

Entitled

**Dynamic Response Of Tunable Phononic Crystals**
**And**
**New Homogenization Approaches In Magnetoactive Composites**

be accepted in partial fulfillment of the
requirements for the degree of

DOCTOR OF PHILOSOPHY

Faramarz Gordaninejad, Ph.D., Advisor

Emil Geiger, Ph.D., Committee Member

Wanliang Shan, Ph.D., Committee Member

Mark A. Pinsky, Ph.D., Committee Member

Ahmad M. Itani, Ph.D., Graduate School Representative

David W. Zeh, Ph. D., Dean, Graduate School

August, 2015

# ABSTRACT

This research investigates dynamic response of tunable periodic structures and homogenization methods in magnetoelastic composites (MECs). The research on tunable periodic structures is focused on the design, modeling and understanding of wave propagation phenomena and the dynamic response of smart phononic crystals. High-amplitude wrinkle formation is employed to study a one-dimensional phononic crystal slab consists of a thin film bonded to a thick compliant substrate. Buckling induced surface instability generates a wrinkly structure triggered by a compressive strain. It is demonstrated that surface periodic pattern and the corresponding large deformation can control elastic wave propagation in the low thickness composite slab. Simulation results show that the periodic wrinkly structure can be used as a smart phononic crystal which can switch band diagrams of the structure in a transformative manner. A magnetoactive phononic crystal is proposed which its dynamic properties are controlled by combined effects of large deformations and an applied magnetic field. Finite deformations and magnetic induction influence phononic characteristics of the periodic structure through geometrical pattern transformation and material properties. A magnetoelastic energy function is proposed to develop constitutive laws considering large deformations and magnetic induction in the periodic structure. Analytical and finite element methods are utilized to compute dispersion relation and band structure of the phononic crystal for different cases of deformation and magnetic loadings. It is demonstrated that magnetic induction not only controls the band diagram of the structure but also has a strong effect on preferential directions of wave propagation. Moreover, a thermally controlled phononic crystal is designed using ligaments of bi-materials in the structure. Temperature difference is used to generate large deformations and affect the elastic moduli tensor of the structure. Phononic characteristics of the proposed structure are controlled by the applied temperature difference. The effect of temperature difference on the band diagrams of the structure is investigated.

Homogenization methods in periodic and random MECs are also investigated. A finite element method (FEM)-based homogenization approach is presented to simulate the nonlinear behavior of MECs under a macroscopic deformation and an external magnetic field. Micro-scale formulation is employed on a characteristic volume element, taking into account periodic boundary conditions. Periodic homogenization method is utilized to compute macroscopic properties of the MEC at different mechanical and magnetic loadings. A new efficient numerical scheme is used to develop the magnetoelastic tangent moduli tensors. In addition, the effective response of a random MEC under applied magnetic fields and large deformations is computed. The focus is on the spatially random distribution of identically circular inclusions inside a soft homogenous matrix. A FEM-based averaging process is combined with Monte-Carlo method to generate ensembles of randomly distributed MECs. The ensemble is utilized as a statistical volume element in a scale-dependent statistical algorithm to approach the desired characteristic volume element size.

# ACKNOWLEDGEMENTS

I would like to express my deepest gratitude to my academic advisor Professor Faramarz Gordaninejad, who supported me, advised me and provided me the opportunity to do the research I liked to do. I would like to thank his personal assistance during my stay in Reno and for helping me in adapting to the new environment. I look forward to many years of friendship and professional collaboration with him.

I was fortunate to enjoy the assistance and friendship of the members of Composite and Intelligent Materials Laboratory in the past few years. I am grateful to Xiaojie, Majid, Nima, Nich, Sevki, Ling and all those who have helped my professional as well as personal life.

I would like to thank my Ph.D. defense committee members for their time and consideration: Prof. Mark Pinsky, Prof. Ahmad Itani, Prof. Emil Geiger and Prof. Wanliang Shan.

Thank you to my lovely wife, Fatemeh, for being by my side throughout this process! Thank you to my family, who helped me, encouraged me, supported me and gave me cheer during all steps of my life!

# TABLE OF CONTENTS

# LIST OF FIGURES

# CHAPTER ONE

# Introduction

## 1.1 Overview

This dissertation consists of two main topics: 1) wave propagation in tunable phononic crystals, and 2) computational homogenization in periodic and random magnetoactive composites. New tunable periodic structures are proposed and analyzed to manipulate elastic wave propagation. Moreover, periodic and random homogenization approaches are addressed in magnetoelasticity framework. New numerical methods are presented to develop effective properties of magnetoelastic composites (MEC) utilizing homogenization approach. Based on the contents of the dissertation, the present Chapter provides brief introductions and backgrounds on different topics that are presented in following chapters.

## 1.2 Magnetoelastic composites

MECs are materials consisting of micro-size permeable particles randomly distributed in an elastomeric matrix. Upon the application of an external magnetic field, the MEC undergoes finite deformations due to the interaction of magnetizable particles and an elastomeric matrix [1-10]. In addition to the nonlinear geometrical deformation, the mechanical properties of the MEC can be controlled by the applied magnetic field. Extensive works have been devoted to develop mathematical formulation of MECs [3-28].

Recently, researches have been focused on media with electro-magneto-mechanical coupling interactions. Specifically, magnetorheological elastomers (MREs) have been developed as field-controllable composite materials capable of significant changes in their mechanical properties. MREs consist of micron size magnetic particles suspended in an elastic/hyperelastic elastomeric matrix that can endure finite deformations upon the application of an external magnetic field [15-23].

Curing the MRE by magnetic field causes the particles to form a chain-like column resulting in a transversely isotropic structure. Otherwise, a random distribution of magnetic particles can be considered as an isotropic MRE. The deformation in MREs is induced by the combined magnetic and mechanical interactions between iron particles as well as the hyperelastic matrix. The applied magnetic field is an added parameter to control the mechanical properties of the MRE [3, 14]. Figure 1 shows an SEM image of MREs with random distribution of magnetic particles, and particles aligned along the applied magnetic field used for curing, within the silicon matrix [11].

(a) (b)

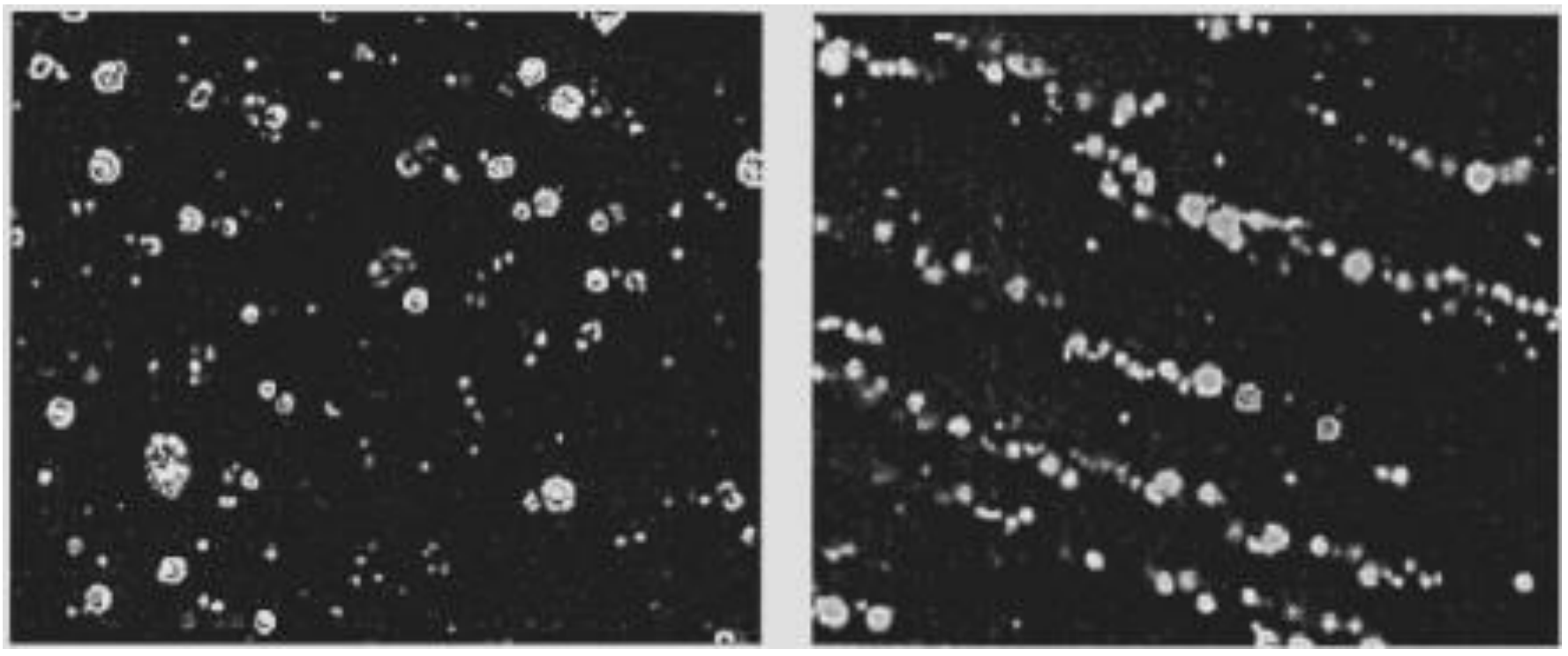

**Figure 1.** SEM images of MREs with **(a)** random distribution of magnetic particles, **(b)** particles aligned along the applied magnetic field used for curing, within the silicon matrix [11].

Theoretical modeling of MREs has been extensively studied based on the electrodynamics of continuous media [2, 15]. Several studies have been done to develop basic constitutive equations for nonlinear magnetoelastic relations [17-27]. Some efforts have been focused on the effect of magnetic fields and finite deformations on the wave propagation in MREs [20-28]. The effect of initial stress on the propagation of Rayleigh, Love and Stoneley waves in a magnetoelastic medium has also been studied in [24-26]. Recently, general theoretical framework for the analysis of incremental motions superimposed on a state of finite deformations subjected to the electromagnetic field has been studied [20-22].

## 1.3 Homogenization

Different homogenization approaches have been utilized to investigate the effective characteristics of MECs for random and periodic microstructures. Homogenization has been used as a tool to study the overall response of the composite and heterogeneous materials presumed to be statistically homogenous. Various techniques have been carried out to simulate the heterogeneous materials as an equivalent effective medium [29-36]. Homogenization offers an averaging process on the microstructure's characteristic volume element (CVE) to compute the effective behavior of the composite.

For periodic structures, CVE is known as the smallest spatial microstructural unit of the structure (i.e. a unit cell). For random heterogeneous materials, the challenge is to find an appropriate CVE to extract the overall properties of the composite material. For this purpose, a statistical analysis is required to find the CVE with randomly dispersed inclusions. The effective behavior of composite material depends on properties of microstructure's constituents. Exact mathematical procedures as well as numerical methods have been developed to study overall mechanical properties of adaptive composites such as magnetoelastic and electroelastic media [37-44].

Different types of micro-scale boundary conditions have been studied in the homogenization process [45-51]. Recently, a finite element method (FEM) based homogenization approach is used to study different mechanical and magnetic material parameters and boundary conditions on the overall response of the structure [46-47].

Homogenization has been widely used as a powerful tool in multi-scale modeling and $FE^2$ analysis of heterogeneous structures [50]. An algorithm for computation of consistent tangent moduli of electroelastic composites is presented in references [47-49, 51].

An example of multi-scale modeling in pure elasticity is shown in Figure 2. First, a CVE is chosen for the study. The CVE is loaded by the macroscopic deformation by incremental steps. Then, the effective constitutive laws and homogenized moduli tensors are driven to characterize the macroscopic properties of the continuum. The process is repeated in an iterative way until the convergence is achieved [52].

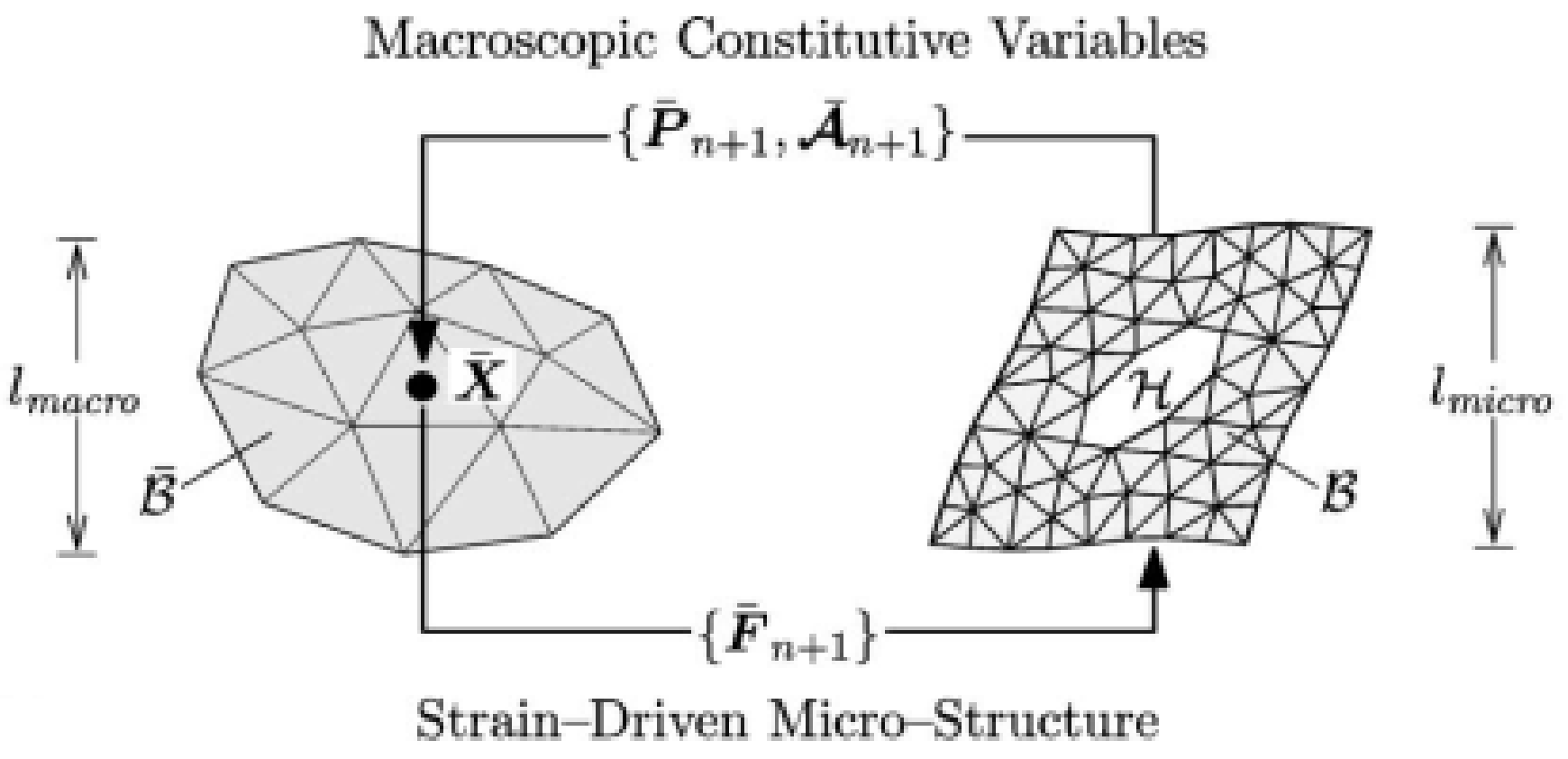

**Figure 2.** An example of multi-scale modeling for heterogeneous structures [52].

Many heterogeneous structures in engineering applications are identified by random distribution of their constituents within a matrix [52-61]. For brevity, “random composites” is referred to the heterogeneous structures with randomly dispersed inclusions. Likewise, “periodic composites” is used for the periodic distribution of the inclusions in a composite structure. Examples of random composites cover wide range of polymer, ceramic or metal matrix surrounding inclusions of different materials. The inclusions are mainly made of high material properties’ contrast with respect to matrix to enhance the effective properties of material. Recent studies on the heterogeneous structures focuses on evaluation of overall response of composites. Overall properties are also referred to effective or homogenized properties in the literature [52-61]. The challenge in study of mechanics of random composites is to identify a characteristic volume element (CVE) and appropriate boundary conditions. Since the classical CVE concept in the homogenization of periodic composites is not valid for random media. Several techniques have studied the homogenization of non-periodic media [52-69]. A statistical based scale-dependent homogenization approach is used to simulate mechanical behavior of a two phase particulate composite. A numerical convergence scheme is used to define an algorithm for determining the CVE size [53]. Monte-Carlo method is widely used to generate statistical ensembles of random composites [53-57]. Specific details are addressed in the determination of CVE size for linear elastic and nonlinear regimes under different boundary conditions [57]. A scale-dependent homogenization approach is presented for the study of linear and nonlinear thermoelastic random composites which uses variational methods to define hierarchy bounds on constitutive laws. This provides a statistical limit for determining CVE size [59-66]. Moreover, the mismatch between the material properties of composites constituents has shown to have strong effect on the CVE size [56-60]. In a recent study, a multi-scale strategy is proposed for nonlinear thermoelastic analysis of random composites with temperature dependent properties [66]. Random magnetoelastic composites have also been studied in [67-70] where different inclusion shapes, material properties and particle concentration are used in the calculations. A closed-form expression is derived for the effective constitutive laws through defining a homogenized energy function which takes into account the concentration, aspect ratio of fibers cross-section and their distribution [70].

### 1.4 Phononic crystals

Electromagnetic and elastic wave propagation through periodic structures has been extensively studied in recent years. A phononic crystal (PnC) is a periodic structure consists of different materials in an elastic medium designed to interact with elastic waves. PnCs are structures with one-dimensional (1D), two-dimensional (2D) or three-dimensional (3D) periodicity in their geometry and material properties. These structures have interesting applications, such as, frequency filters, beam splitters, sound or vibration isolators, acoustic mirrors and elastic waveguides [71-84].

Periodic structures can be designed to hinder wave propagation in some range of frequencies; namely, band-gaps. Existence of band-gaps is due to the periodic change of density and speed of wave in the structure [77-80]. Unit cell is the smallest structural unit of the periodic material which is spatially assembled together in 1D, 2D or 3D to build up the structure. Basically, the wave propagation characteristics in PnCs depend on unit cell's geometry and material properties (i.e. density, stiffness and bulk modulus).

While different approaches have been used to study wave propagation phenomena in periodic structures, the most common methods are plane wave expansion methods, finite element methods, and finite-difference time domain methods to explore band diagrams [72, 75, 77, 78]. Figure 3 shows examples of 1D, 2D and 3D PnCs and corresponding unit cells [85-87].

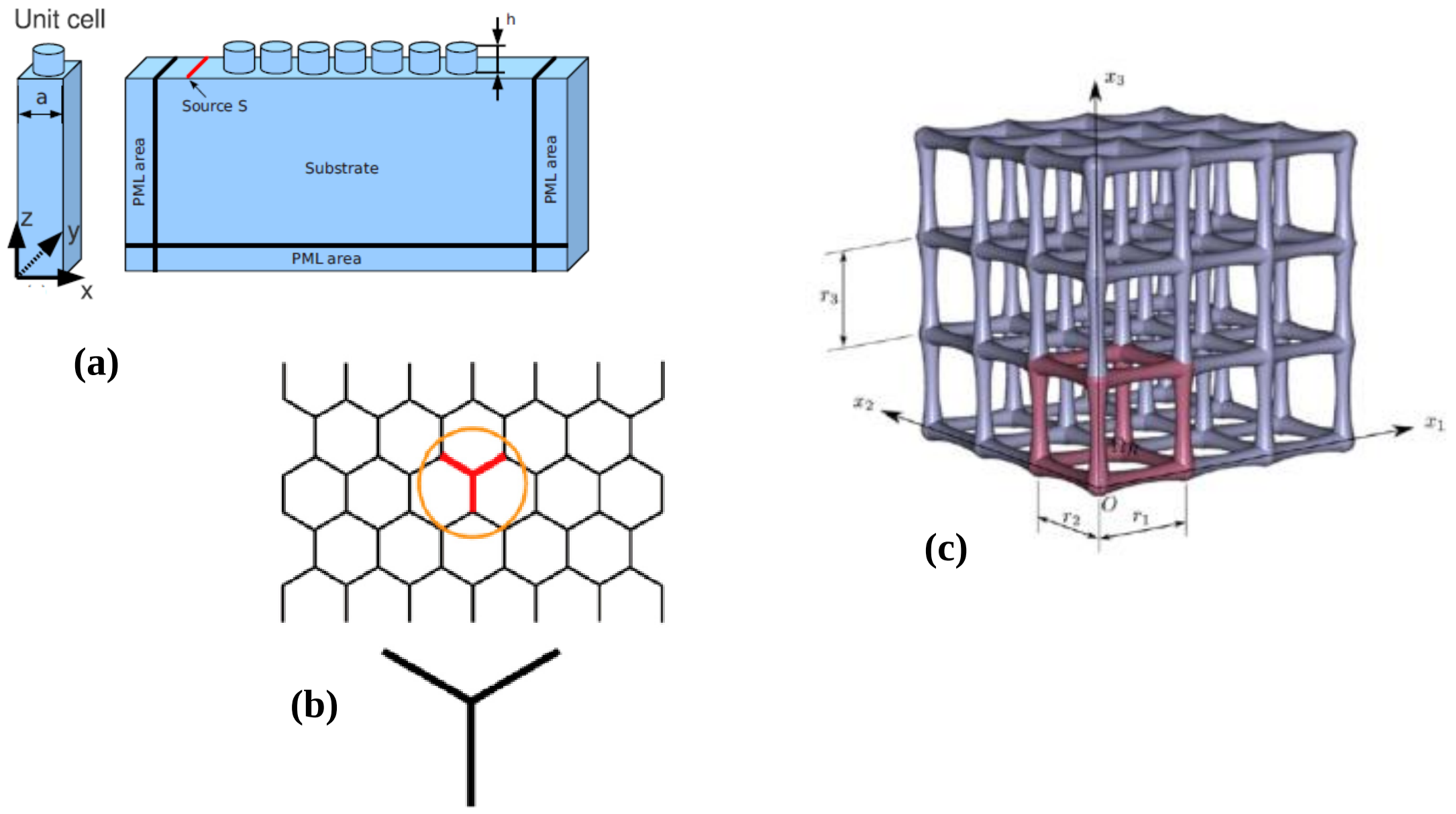

**Figure 3.** Examples of **(a)** a 1D [85], **(b)** a 2D [86] and **(c)** a 3D [87] PnC and corresponding unit cells chosen for wave propagation study.

Dynamic response of the structure can be controlled through the geometry and elastic properties of the unit cell. Adaptive periodic structures promise the ability to design tunable properties that can control band-gaps upon the application of an external stimulus. The elastic band structure of PnCs made of piezoelectric materials has been investigated through plane wave expansion method [83]. The phononic band-gap (PBG) and acoustic characteristics of the structure have been shown to significantly shift due to various electric and magnetic fields [84].

The band structure characteristics of lattices of three different geometries consisting of piezoelectric and piezomagnetic media were studied to understand the effects of different magnetic and electric fields through the plane wave expansion method [88-90]. It has been shown that tunable PnCs can be employed as a transmission switch for elastic waves when magnetic field passes a threshold [88]. Only limited work has been carried out on the band diagram calculations of MRE PnCs. In all previous work on MRE PnCs constant coefficients are presumed as magnetoelastic coefficients [91-94]. Parametric study of the effects of different geometry and configuration on the band structure of a set of parallel square-section columns regularly distributed in air has been investigated as a tunable periodic structure [90]. Periodic structures have also been studied theoretically and experimentally for super resolution at a narrow band of frequencies using flat lenses made of PnCs [95]. Figure 4 shows some applications of PnCs as waveguides and acoustic lenses [95-96].

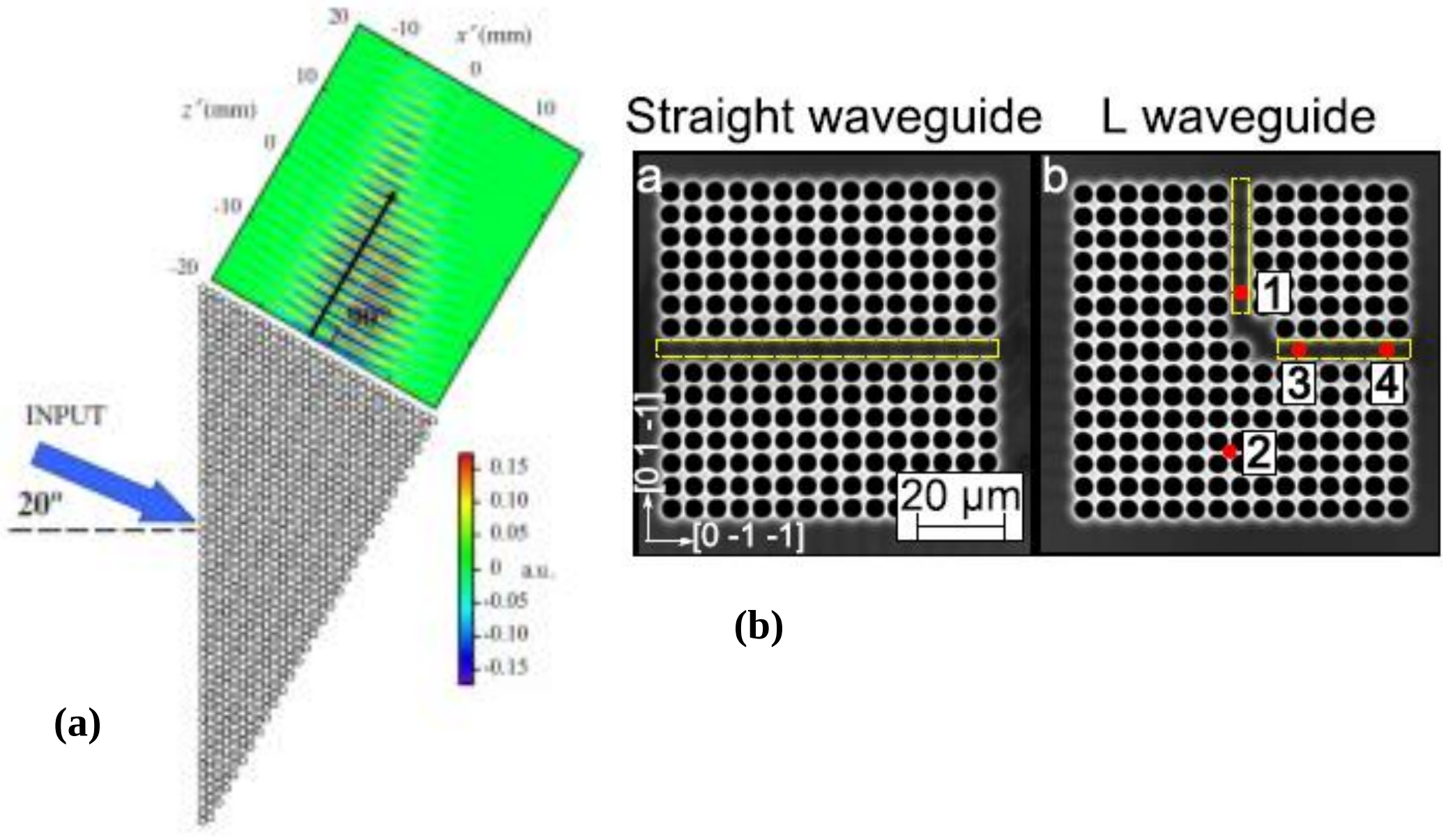

**Figure 4.** Application of PnCs as **(a)** acoustic lens [95] and **(b)** acoustic waveguides [96].

Deletion of existing band-gaps and creation of new band-gaps were reported due to the effects of large deformation and the microstructural elastic instability in periodic elastomers subjected to different types of mechanical loadings. When deformation reaches a critical value, strong revolution happens in the band structure due to transformation in structural pattern resulting in a tunable behavior of the structure in terms of the applied mechanical loading [97-98]. Figure 5 shows a tunable periodic structure which its phononic characteristics is tuned by the applied macroscopic deformation [99].

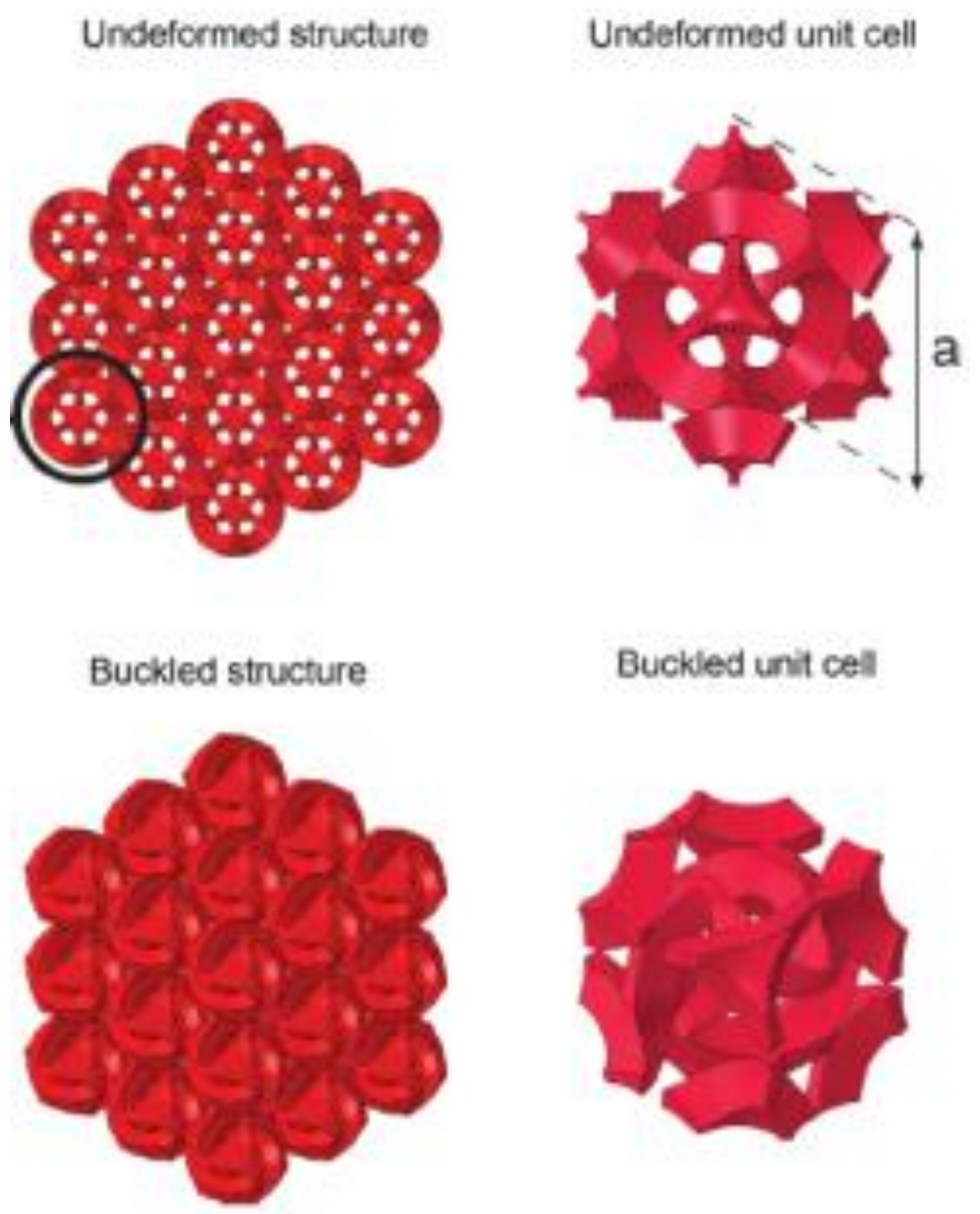

**Figure 5.** A tunable 3D PnC and corresponding unit cells in undeformed and deformed states, designed to manipulate elastic waves [99].

### 1.4.1 Wave propagation analysis using Bloch theorem

Elastic wave propagation in PnCs is studied based on the fundamentals of solid state physics in lattices utilizing Bloch theorem [74]. In PnCs, the desired quantities in wave propagation are extracted from displacement and stress fields. In this work, a 2D infinite size periodic lattice in $\boldsymbol{x_1}, \boldsymbol{x_2}$ plane, is considered. The periodic lattice is characterized by a unit cell and direct lattice vectors $a_1$ and $a_2$. Reciprocal lattice of any periodic structure is defined by reciprocal lattice vectors $b_1$ and $b_2$:

$$b_1 = \frac{2\pi(a_2 \times e)}{\|a_1 \times a_2\|} \quad , \quad b_2 = \frac{2\pi(e \times a_1)}{\|a_1 \times a_2\|} \ , \quad where \quad e = \frac{(a_1 \times a_2)}{\|a_1 \times a_2\|}$$

Hence, the position of each point, $Q$, of the lattice -with respect to the reference unit cell- can be expressed as

$$\boldsymbol{R} = q_1 a_1 + q_2 a_2$$

where $q_1, q_2$ are integers. The displacement of a point $Q$ in the lattice satisfies the periodic condition:

$$\boldsymbol{u}(\boldsymbol{X} + \boldsymbol{R}) = \boldsymbol{u}(\boldsymbol{R})$$

Bloch theorem states that the displacement of each point $Q$ follows the Bloch theorem:

$$\boldsymbol{u}(\boldsymbol{X} + \boldsymbol{R}) = \boldsymbol{u}(\boldsymbol{X}) e^{\boldsymbol{ik.R}}$$

where $\boldsymbol{k}$ is the Bloch wavenumber. Comparing the Bloch statement with the periodicity condition requires;

$$e^{\boldsymbol{ik.R}} = 1$$

Since $a_i . b_j = 2\pi\delta_{ij}$, (where $\delta_{ij}$ is the Kronecker delta) this statement is satisfied when $\boldsymbol{k}$ is represented in terms of reciprocal lattice vectors:

$$\boldsymbol{k} = k_1 b_1 + k_2 b_2$$

Clearly:

$$q_1 k_1 + q_2 k_2 = 1.$$

Figure 6 shows a typical periodic structure with corresponding unit cell, direct lattice vectors and reciprocal lattice vectors [74, 97-99].

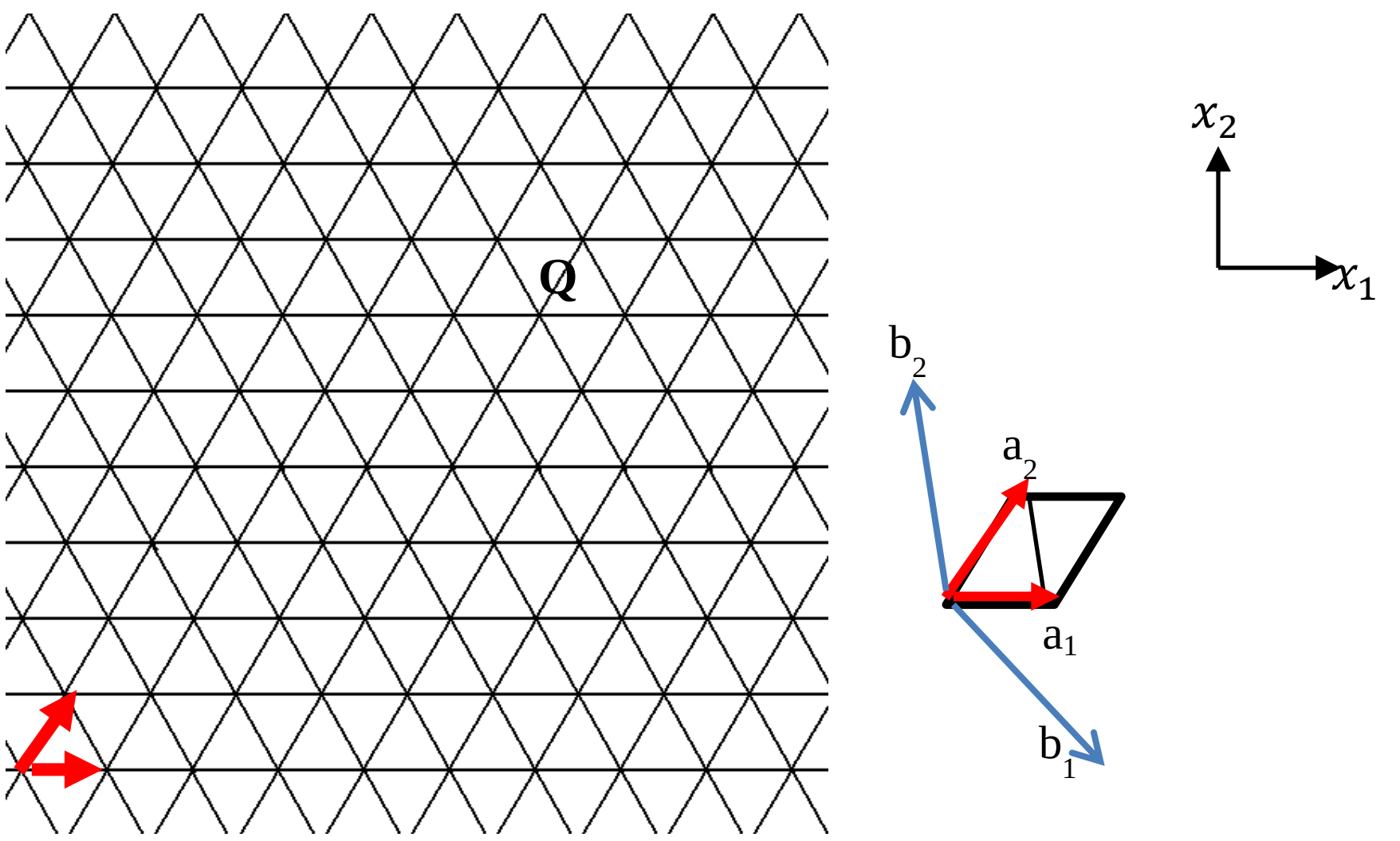

**Figure 6.** A typical periodic structure with corresponding unit cell, direct lattice vectors and reciprocal lattice vectors.

Bloch theorem states that the spatial field in each unit cell of the direct lattice presents the same distribution and does not depend on the cell's location. This fact reduces the problem of studying an infinite number of cells to one of considering only a single unit cell by applying appropriate boundary conditions. Several techniques are demonstrated on the wave propagation analysis in periodic structures. Mathematical methods such as plane wave expansion method have been used for simple geometry lattices while finite difference time domain method and FEM based solutions are usually used to study complex microstructures [71-99]. In FEM based approach, the unit cell is meshed and Bloch boundary conditions are applied on the opposite boundaries of the unit cell [77-78]. The wave propagation in the unit cell is described through the discretized equation of motion and by considering the cell interaction with the neighboring cells. Substituting Bloch solution in the wave equation results in an eigenvalue problem whose eigenvalues are the frequency of the solution. Wavenumber is a periodic function of the wavevector, ***k*** in the reciprocal lattice.

Hence, the dispersion relations are obtained by investigating the variation of the eigenfrequency, $\omega$, versus wavenumber, $k$, over a single period in reciprocal lattice. This requires identifying the single unit cell in the reciprocal lattice called first Brillouin zone. The dispersion relations are extracted from the eigenvalue problem where the wavenumber is swept on the boundaries of the first Brillouin zone. It has been shown that the calculation domain may further be reduced by taking the advantage of the symmetry of the first Brillouin zone. The reduced domain is referred as irreducible Brillouin zone (IBZ) [74-78]. Some examples of first Brillouin zone and IBZs are depicted in Figure 7 for some typical lattices [100].

Once the Brillouin zone is identified, the wavenumber, $k$, is swept on the boundaries of the IBZ and the corresponding eigenfrequencies are computed for each $k$. Band diagrams are then plotted as $\omega$ versus $k$ dispersion relations to analyze the band-gaps of the structure.

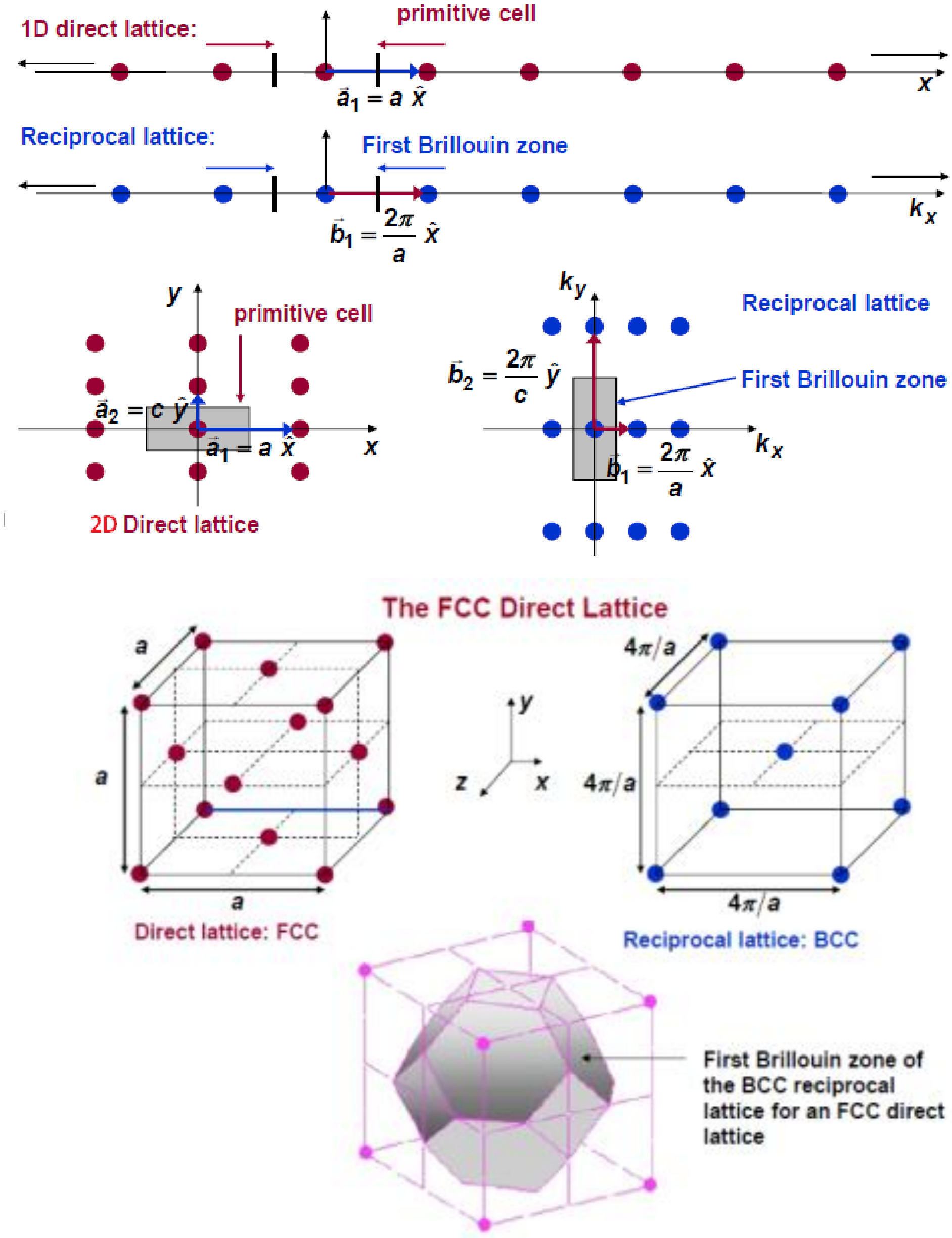

**Figure 7.** Examples of first Brillouin zone and irreducible Brillouin zone for 1D, 2D and 3D typical lattices [100].

## 1.5 Pattern change in soft structures

One of the concepts that have been utilized in the design of the PnCs in this dissertation is the pattern change in the soft structures. Recently, interests have been attracted to study the behavior of reconfigurable structures [101]. Once the loading condition is vanished the structure tends to return to the initial state in a reversible manner. Advances in reconfigurable materials have led to the next generation of adaptive and actuating materials. Various parameters can be employed to modulate the shape and geometry of the materials as a tunable material. The structures can be loaded by different external stimuli such as mechanical [102], electrical [103], magnetic [104] and thermal [105] excitations. The static and dynamic response of the structure then can be controlled through the applied stimuli. In this section, recent findings on pattern change in the reconfigurable structures are summarized.

Figure 8 shows a flower-shaped film that is capable of self-folding and resulting in a different enclosed shape. The hinged structure is combined with thermo-responsive self-folding capsules which are absorbed on the polymer bilayer at elevated temperatures. Cooling results in swelling of thermo-responsive polymer and folding of the capsules. Heating results in releasing the cells and unfolding the capsules provides a thermally modulated reconfigurable scheme in the structure [105].

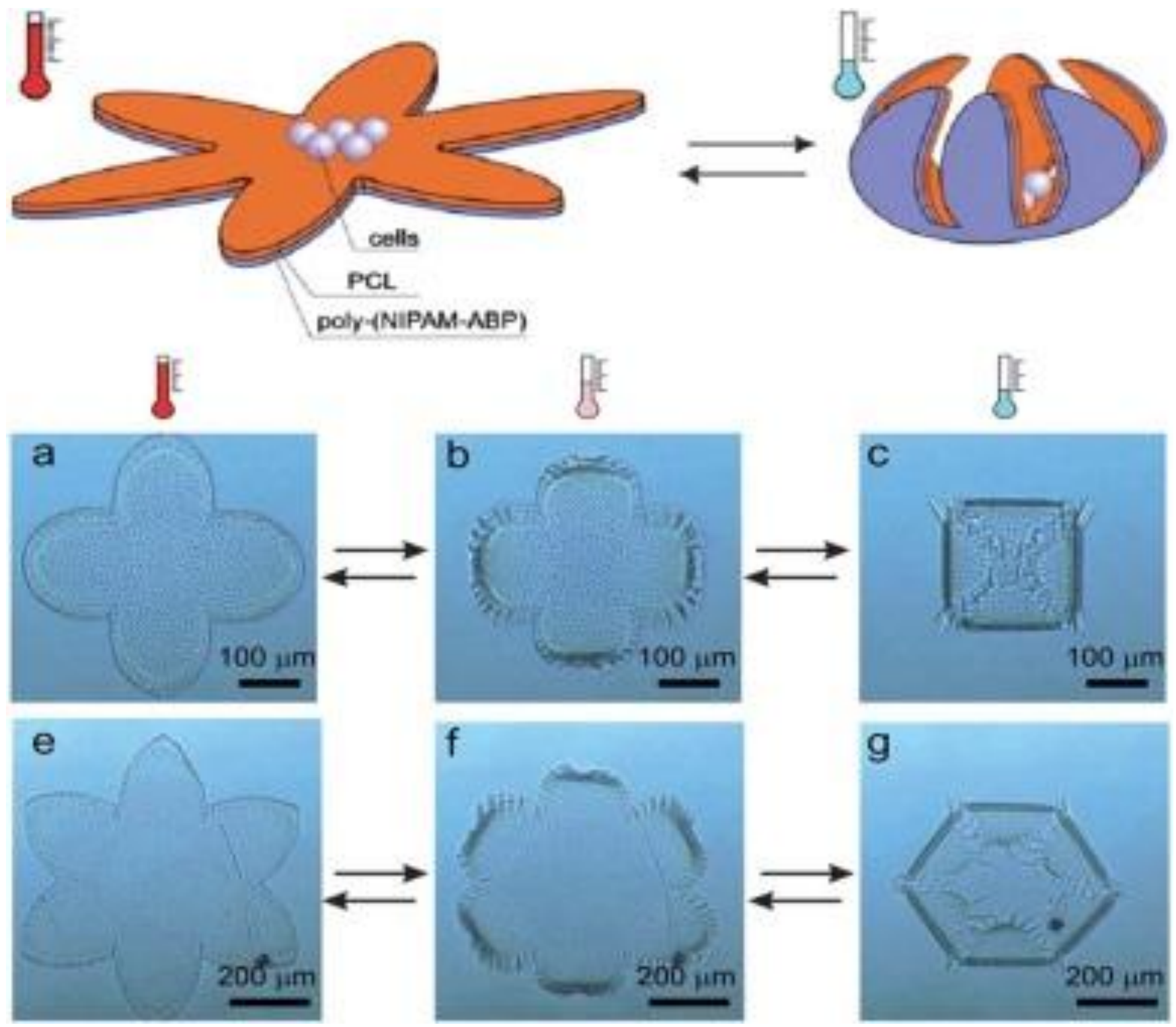

**Figure 8.** A thermally modulated reconfigurable structure [101, 105].

In a recent study, pattern change in a soft periodic lattice has shown to have negative Poisson's ratio effect. Buckling in the microstructure has been employed to contract the structure in the transverse direction when it is under compressive loading. The pattern change arises due to microstructural instability in the material. Figure 9 shows the 3D periodic lattice in undeformed and deformed state [106]. The structure is compressed by $\varepsilon = 0.3$ which demonstrates the contraction in the transverse direction.

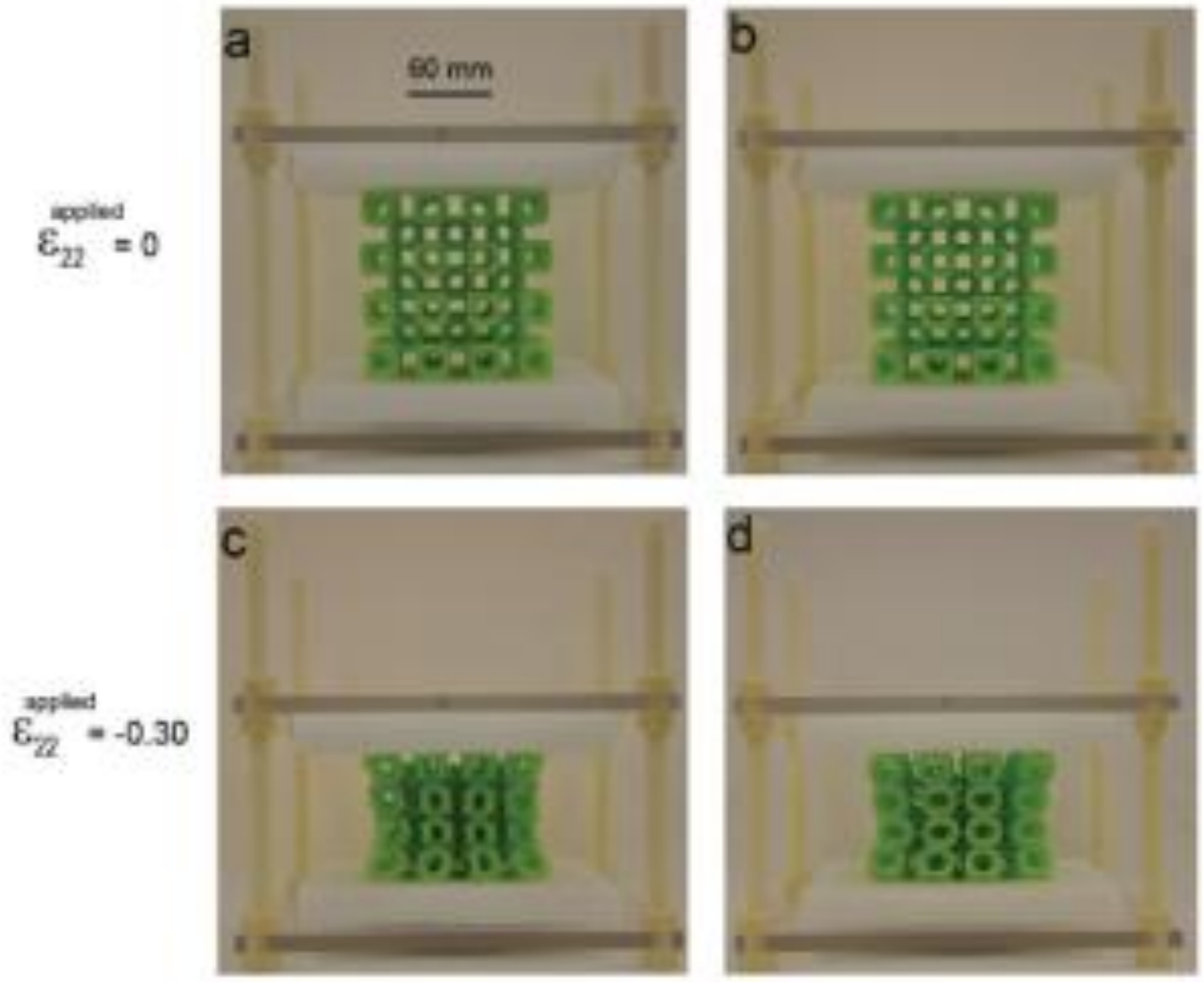

**Figure 9.** Transverse contraction of the periodic lattice under compressive strain due to microstructural buckling [106].

Magnetic field has been used as an external stimulus to change the pattern in magnetoactive polymers. The structure is made of 60 wt% vinyl cobalt nanoparticles within a PDMS matrix, under the excitation of a permanent magnet. The structure restores its initial shape once the magnetic field is vanished provides the opportunities to utilize as a reconfigurable structure [104].

Pattern change in elastic periodic structures has also been utilized in wave propagation systems [98, 107, 108]. The applied deformation transforms the structure to a new geometry which can switch the band diagrams in a reversible manner. Figure 10 shows the effect of compressive load on a granular structure. It has been shown that the phononic characteristics of the structure significantly changes under the influence of the large deformation and stress [107].

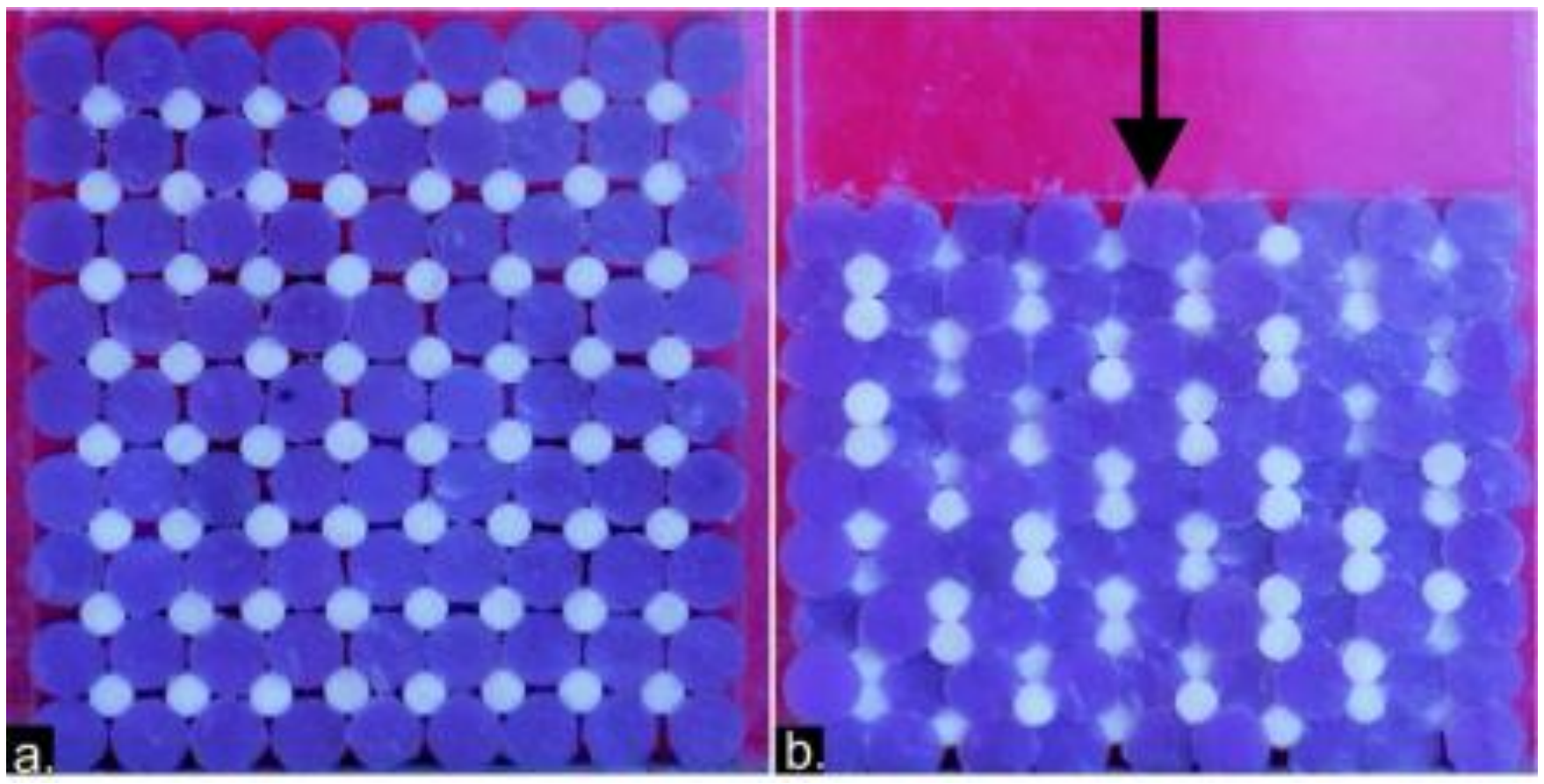

**Figure 10.** Pattern change in a granular structure under compressive loading [107].

One of the interesting pattern change phenomenon occurs in bilayer structures under compressive strain, where the contrast in the material properties of a thin layer with a thick substrate generates wrinkles at the surface of the structure. When a critical compressive strain is applied to an elastic thin film bonded to a compliant substrate, surface wrinkling occurs in the thin film due to surface buckling instability. Wrinkling can be employed as a novel methodology for creating 1D and 2D periodic micro and nanosurfaces [109-125]. Wrinkling has potential applications due to its highly ordered pattern, unique structure and convenient fabrication methods; from surface patterning [118-119] and smart adhesion [120-121] to optical surfaces [122-123] and flexible electronic devices [124]. It has been shown that the amplitude and periodicity of wrinkles depend on the material properties, geometry of layers and the applied compressive strain [110-111]. A recent study has developed a new method to fabricate high aspect ratio (amplitude over wavelength of wrinkles) wrinkles over flat surfaces which enhances the aspect ratio from 0.3 to 0.6 [125]. This has improved the surface contour area from 9% to 36%, doubled the local curvature, and increased the stretchability of wrinkly flexible electronic devices from 20% to 40%. Formation of high aspect ratio wrinkles results in a periodic pattern of scatterers at the surface of the structure. Figure 11 shows the surface wrinkling evolution of a bilayer composite of thin gold film on a PDMS substrate at different levels of macroscopic strain. Different wrinkle modes and patterns are generated by the increasing strain beyond $\varepsilon = 0.3$ [126].

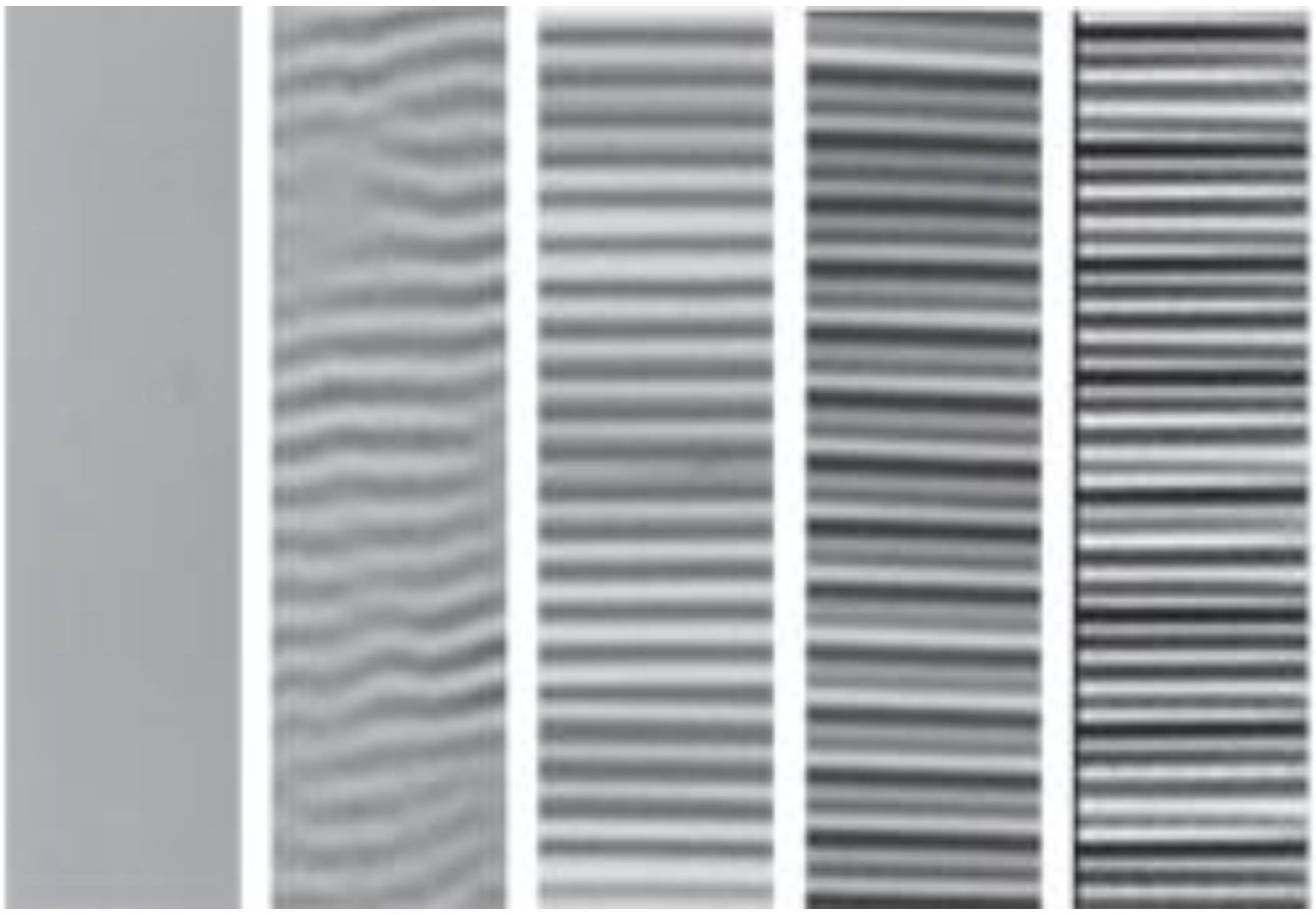

**Figure 11.** Surface wrinkling of a 66nm thick film of gold bonded on a PDMS substrate at different levels of applied compressive strain from $\boldsymbol{\varepsilon = 0}$ at left column to $\boldsymbol{\varepsilon = 0.5}$ at the right side [126].

In Chapter 2 the surface wrinkling pattern change is employed to design a smart PnC slab and analyze the wave propagation in the resultant periodic structure.

## 1.6 Notations

Here, the vector and tensor notations mainly used throughout this manuscript, are summarized. The summation notation for repeated indices is used. Scalars are in regular letters. Bold faces are used for vectors and tensors. The scalar product of two vectors $\boldsymbol{A}$ and $\boldsymbol{B}$ is denoted by: $\boldsymbol{A}.\boldsymbol{B} = A_i B_i$. The terms single contraction or dot product also refers to the scalar product. Cross product of two vectors is defined by: $\boldsymbol{A} \times \boldsymbol{B} = \epsilon_{ijk} A_i B_j$ where $\epsilon$ is the third-order permutation tensor.

The double contraction or scalar product of two second-order tensors $\boldsymbol{C}$ and $\boldsymbol{D}$ is denoted by: $\boldsymbol{C}:\boldsymbol{D} = C_{ij} D_{ij}$. Single contraction is also defined on two second-order tensors as; $\boldsymbol{C}.\boldsymbol{D} = C_{ij} D_{jk}$. Single contraction of a second order tensor $\boldsymbol{C}$ with a vector $\boldsymbol{A}$ is also denoted by; $\boldsymbol{C}.\boldsymbol{A} = C_{ij} A_j$. Double contraction and single contraction of two third-order tensors $\boldsymbol{E}$ and $\boldsymbol{F}$ are denoted by: $\boldsymbol{E}:\boldsymbol{F} = E_{ijk} C_{jkl}$ and $\boldsymbol{E}.\boldsymbol{F} = E_{ijk} C_{klm}$ , respectively. Double contraction of a third-order tensor $\boldsymbol{E}$ and a second-order tensor $\boldsymbol{C}$ is a vector denoted by; $\boldsymbol{E}:\boldsymbol{C} = E_{ijk} C_{jk}$. Scalar product of a third-order tensor $\boldsymbol{E}$ and a vector $\boldsymbol{A}$ is a second-order tensor represented by; $\boldsymbol{E}.\boldsymbol{C} = E_{ijk} A_k$. For a vector $\boldsymbol{A}$ and a second-order tensor $\boldsymbol{C}$ cross-product is a second-order tensor defined by; $\boldsymbol{A} \times \boldsymbol{C} = \epsilon_{ijk} A_i B_{jl}$ .

For three vectors $\boldsymbol{M}$, $\boldsymbol{N}$ and $\boldsymbol{P}$ tensor product is defined as; $(\boldsymbol{M}\otimes\boldsymbol{N})\boldsymbol{P} = (\boldsymbol{N}.\boldsymbol{P})\boldsymbol{M}$. Tensor product of two vectors $\boldsymbol{A}$ and $\boldsymbol{B}$ is a second-order tensor denoted by; $\boldsymbol{A}\otimes\boldsymbol{B} = A_i B_j$. Tensor product of two second-order tensors $\boldsymbol{C}$ and $\boldsymbol{D}$ is a fourth-order tensor denoted by; $\boldsymbol{C}\otimes\boldsymbol{D} = C_{ij}D_{kl}$. Tensor product is also defined for a vector $\boldsymbol{A}$ and a second-order tensors $\boldsymbol{C}$ by: $\boldsymbol{A}\otimes\boldsymbol{C} = A_i C_{jk}$ which represents a third-order tensor. Also $\boldsymbol{C}\otimes\boldsymbol{A} = C_{ij}A_k$. For a second order tensor $\boldsymbol{C}$ with components $C_{ij}$, the transpose is denoted by $\boldsymbol{C}^{\boldsymbol{T}}$ with the components $C_{ji}$. The trace of $\boldsymbol{C}$ is defined by; $tr\boldsymbol{C} \equiv C_{ii}$.

## 1.7 Research Goals

The goal for the tunable phononic crystals research is to explore:

- Novel tunable PnCs to employ in wave propagation systems.
- Combined effects of magnetic field and pattern change on the wave propagation in a tunable PnC.
- Surface acoustic band-gaps in a smart PnC.
- Thermally modulated pattern change effects on wave propagation in smart PnCs.

The goal for the homogenization methods in magnetoactive media is to:

- Identify the nonlinear magnetoelastic moduli tensors using a cost-effective computational method and available commercial FEM packages. Moduli tensors are of high importance in the study of instability, effective properties and multi-scale modeling of MECs. Only few existing work are reported on the calculation of the moduli tensors which requires high mathematical computations.

- Understand new computational methods for homogenized properties of MECs with randomly dispersed permeable inclusions using FE-based homogenization methods.

### 1.8 Uniqueness of this research

- Surface instability is employed to design a novel smart PnC slab. Band diagram of the resultant tunable structure is significantly transformed by the applied compressive strain.
- Combined effects of pattern change and magnetic excitation has shown significant effects on the band diagrams of the porous structure.
- Band diagrams of the magnetoelastic PnC is examined through defining magnetoelastic energy function. As a result, nonlinear moduli tensors are considered as functions of deformation gradient tensor and magnetic induction vector.
- The effect of magnetic field on the elastic wave directionality in the magnetoactive PnC is studied. Magnetic field has shown to strongly affect the preferential directions of wave propagation in magnetoactive PnCs.
- A thermally tunable PnC is designed and the effect of temperature difference on the wave propagation is investigated. Temperature difference has shown to transform the band diagram of the resulting tunable PnC.

- A new cost-effective algorithm for computation of nonlinear moduli tensors of MECs is presented.
- A FEM-based homogenization approach is combined with Monte-Carlo method to evaluate overall properties of MECs with random distribution of inclusions.

## 1.9 Structure of the dissertation

Chapter 2 introduces a new paradigm in designing tunable PnCs using surface instability. Buckling induced surface instability is used to generate wrinkly scatterers at the surface of a bilayer composite slab. The slab consists of a thin and stiff film bonded on a soft substrate. The effect of applied compressive strain on band diagrams of the PnC slab is investigated.

In Chapter 3 the effect of combined effects of mechanical deformation and magnetic field on dynamic response of a magnetoactive PnC is studied. First, the theoretical formulation of magnetoactive media is reviewed. A hyperelastic magnetic energy function is considered to develop the constitutive laws. Variational method is used to develop the coupled governing equations. Weak expressions of the coupled governing equations are derived to use in the FEM framework. Nonlinear moduli tensors are used as functions of deformation gradient tensor and magnetic field. The Chapter focuses on dynamic response of the tunable PnC under applied magneto-mechanical loadings.

Chapter 4 introduces a new thermally controlled PnC. Temperature difference is used to exploit large deformations and affect the phononic characteristics of the structure. Band diagrams are plotted to investigate the effect of temperature difference on band-gaps.

Theoretical modeling and governing equations of magnetoactive media is presented in Chapter 5. The Chapter employs the FEM-based homogenization approach to propose a novel and computationally cost-effective method for evaluation of homogenized tangent moduli tensors using sensitivity analysis.

Chapter 6 details the computational approach in the study of magnetoactive media with randomly dispersed particles. An FEM based homogenization method is combined with Monte-Carlo approach to evaluate the static response of the structure under macroscopic loadings. A statistical approach is presented to find the appropriate CVE size to compute the overall properties of the structure.

Finally, concluding remarks and some recommendations for future work are presented in Chapter 7.

# CHAPTER TWO

## Switching Band-Gaps of a Phononic Crystal Slab by Surface Instability[1]

High-amplitude wrinkle formation is employed to propose a one-dimensional phononic crystal slab consists of a thin film bonded to a thick compliant substrate. Buckling induced surface instability generates a wrinkly structure triggered by a compressive strain. It is demonstrated that surface periodic pattern and corresponding stress can control elastic wave propagation in the low thickness composite slab. Simulation results show that the periodic wrinkly structure can be used as a transformative phononic crystal which can switch band diagram of the structure in a reversible manner. Results of this study provide opportunities for smart design of tunable switch and frequency filters at ultrasonic and hypersonic frequency ranges.

In the present study, one of the unexplored features of wrinkle-based structure is studied; utilizing wrinkle formation at a surface of a slab as a tunable phononic crystal (PnC) in elastic wave propagation systems. The effect of generation and control of wrinkles on the propagation of elastic wave in thin film/substrate slabs is investigated. It is presumed that global buckling is prevented. Wrinkles are formed when the compressive strain

---

[1] Results of this chapter are published in:

- Bayat A and Gordaninejad F 2015 Switching band-gaps of a phononic crystal slab by surface instability *Journal of Smart Materials and Structures* **24** 075009.
- Bayat A and Gordaninejad F 2015 A Wrinkly Phononic Crystal slab *Proc. SPIE* San-Diego, USA, 9438, p. 943810.

reaches a critical strain. The compressive strain is used as a control parameter to tune the periodicity, amplitude and pattern of the wrinkly slab. Phononic characteristics of the periodic wrinkly slab will be controlled through both geometry and stress.

The wavelength $\gamma$ of uniform wrinkles is defined by $\gamma = 2\pi h_f \left(\frac{E_f(1-\nu_s^2)}{3E_s(1-\nu_f^2)}\right)^{1/3}$, where $E_f, \mu_f, \nu_f$ are the elastic modulus, shear modulus and Poisson's ratio of the film, respectively, and $E_s, \mu_s, \nu_s$ the elastic modulus, shear modulus and Poisson's ratio of the substrate, respectively. The wavelength is defined as the length of the structure divided by number of wrinkles in the undeformed state. The periodicity in a deformed configuration is defined by $\mathcal{P} = \gamma(1-\varepsilon)$ where $\varepsilon$ is the applied strain beyond the bifurcation point. The amplitude of wrinkles relates to the applied strain by $\Lambda = h_f\sqrt{\frac{\varepsilon}{\varepsilon_c} - 1}$ where $\varepsilon_c$ is the critical strain defined by $\varepsilon_c = \frac{1}{4}\left(\frac{3E_s(1-\nu_f^2)}{E_f(1-\nu_s^2)}\right)^{2/3}$.

### 2.1 Modeling

The post-bifurcation state of the structure as well as wave propagation in the deformed structure is investigated through numerical simulations. Simulations are performed using Finite element methods (FEM) through COMSOL Multiphysics. A 2D plane strain model is adopted to simulate the 1D wrinkling in a bilayer structure, beyond the bifurcation point. The schematic of the model, parameters and boundary conditions are shown in Figure 12(a). Loading of the structure will be applied through prescribed compressive strain, $\varepsilon$ in the $\boldsymbol{x_1}$ direction at the right edge. Free shear traction is imposed on all edges. Symmetry boundary condition is applied on the left and bottom edges to prevent normal displacements. The top surface has a free boundary condition. The thickness of the substrate is chosen to be sufficiently deep compared to the film thickness.

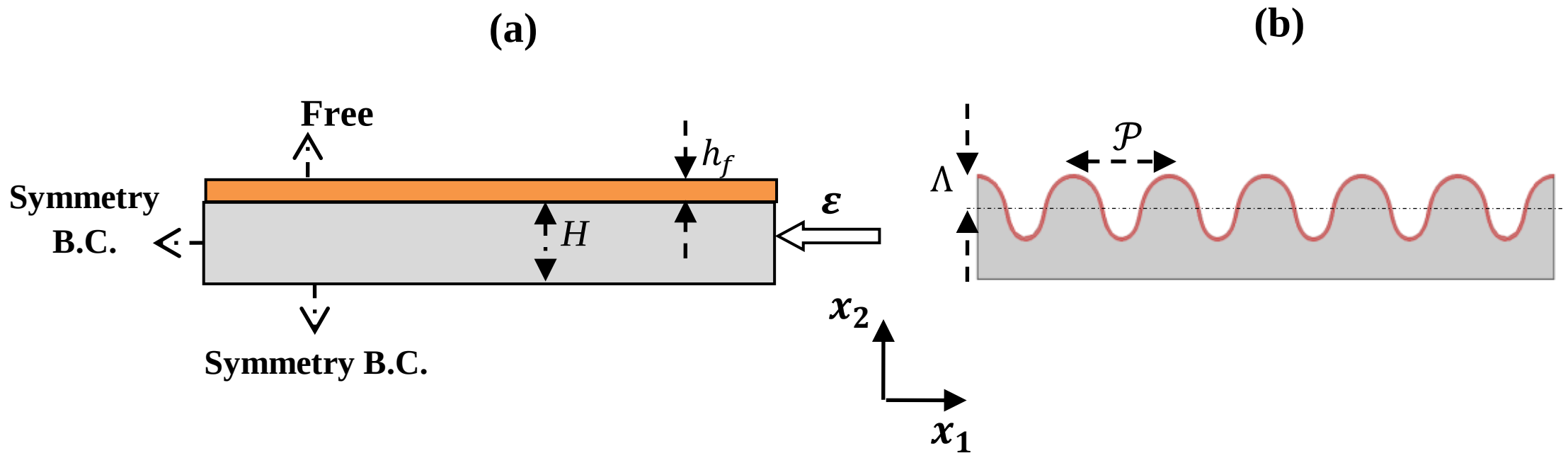

**Figure 12. (a)** Schematic of a bilayer slab and corresponding boundary conditions, and **(b)** the deformed wrinkly structure resulted from the compressive stretch, ε, due to surface instability (not to scale).

Lagrangian coordinates is chosen for the study. Two different cases of film-substrate bilayer composite slabs are considered here: a linear elastic film on a soft material (linear-hyperelastic) and a hyperelastic film on a soft substrate (hyperelastic-hyperelastic) hereafter will be referred as L-H and H-H bilayer slabs, respectively. A nearly incompressible neo-Hookean model is selected for hyperelastic materials. A linear buckling study is carried out on the bilayer structure to predict the mode shapes at compressive strains. To deal with small amplitude signals, a linear perturbation analysis is used for the solver to predict the mode shapes. Surface deformation modes will be a plane strain sinusoidal deflection in $\boldsymbol{x_1} - \boldsymbol{x_2}$ plane as schematically shown in Figure 12(b). Other mode shapes mainly differ in number of the periodicity. Two typical mode shapes are pictured in Figure 13. The first mode shape resulted from the buckling analysis for both L-H and H-H bilayer structures have similar patterns.

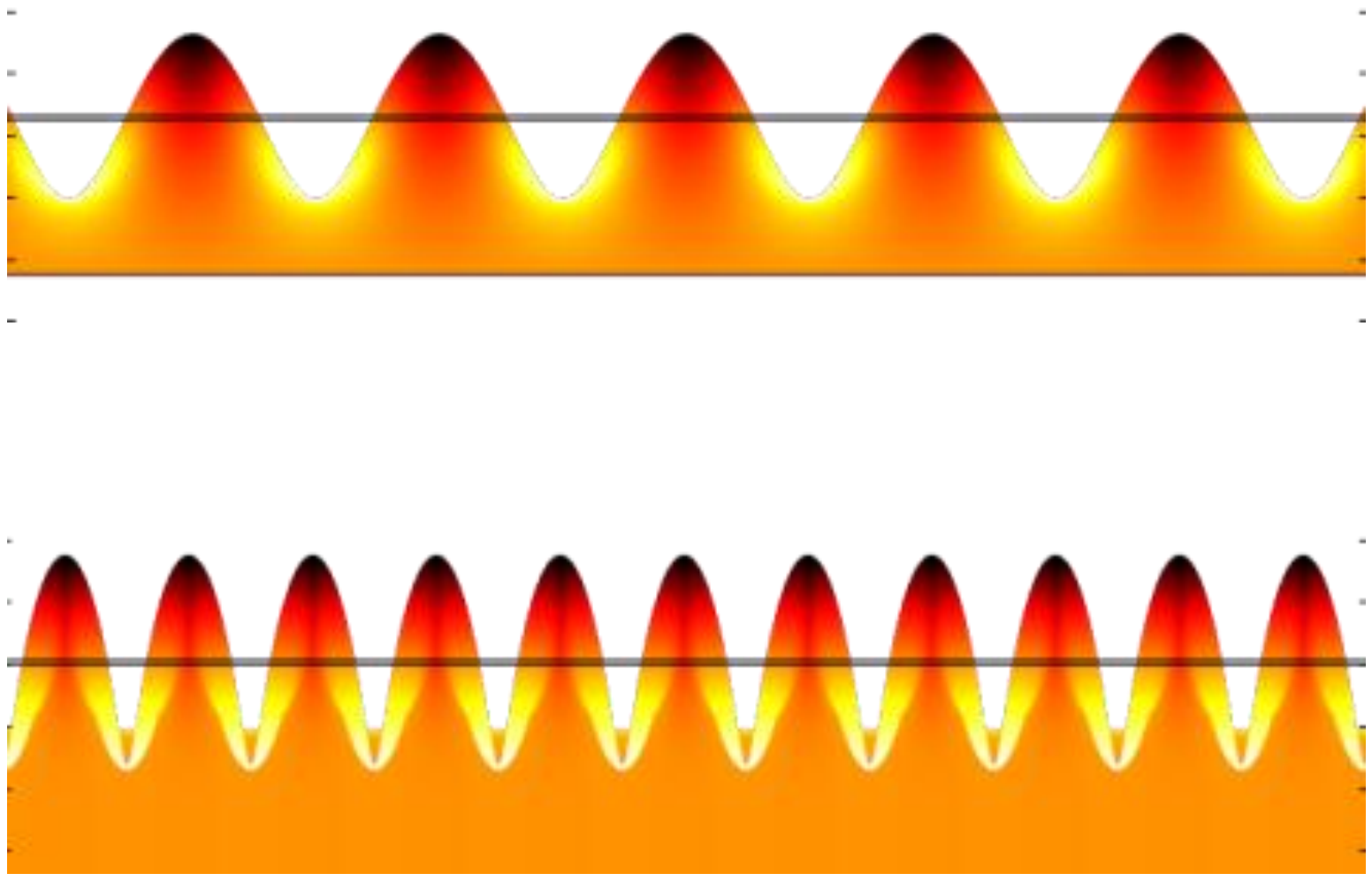

**Figure 13.** First and second mode shapes for a typical L-H structure

A nonlinear stationary study is carried out on the bilayer slab of finite thickness to predict post-buckled wrinkly shape under compressive strain. Slight sinusoidal imperfections are imposed on the geometry of the film layer to activate the buckling mode and capture the post-transformation shape. In buckling mechanisms, post-bifurcation states cannot be captured for an ideal system. External perturbations are required to excite the desired buckling modes. Imperfections can be applied through perturbations in: (1) mesh, (2) boundary conditions, and (3) geometry of the model. In this study, imperfections are applied through the geometry of the structure. Slight sinusoidal perturbation is imposed on the geometry of the film layer to activate the desired buckling mode and capture the post-transformation shape. Post-bifurcated wrinkles' shape is highly sensitive to the applied imperfection. Figure 14 shows the effect of imperfection amplitude on the bifurcation. A sensitivity analysis is conducted to find the appropriate imperfection weight. Similar post-bifurcation modes are resulted for the imperfection weight in the range of 0.02%-0.2%. No bifurcation occurs below 0.02%.

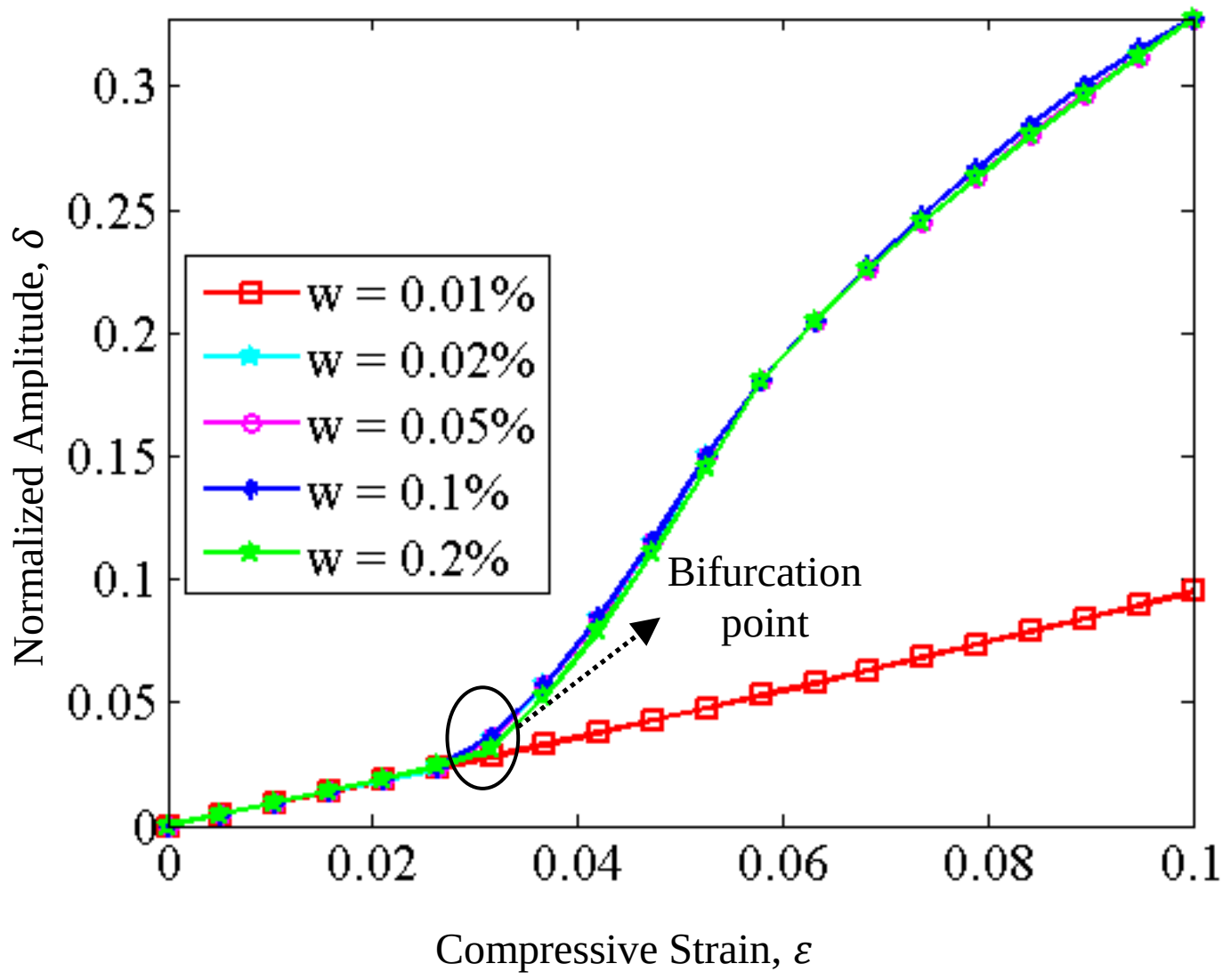

**Figure 14.** Effect of different imperfection weight (w) on the bifurcation. Post-buckling does not occur for w = 0.01% and less.

Results for wrinkling of the L-H and H-H bilayer structures at different levels of the applied strain are shown in Figure 15. For L-H structure, $E_f = 7\ GPa$ , $\nu_f = 0.2$, $\kappa_s = 1.16\ GPa$ , $\mu_s = 23.4\ MPa$, $\frac{h_f}{H} = 0.15$ and $\frac{H}{\gamma} = 0.25$ , are used as input parameters, where $\kappa_s$ and $\mu_s$ are bulk and shear moduli of hyperelastic substrate, respectively. For H-H structure, $\kappa_f = 3.53\ GPa$ , $\mu_f = 2.99\ GPa$, $\kappa_s = 1.16\ GPa$ , $\mu_s = 23.5\ MPa$ , $\frac{H}{\gamma} = 0.5$ and $\frac{h_f}{H} = 0.075$ are used in the model. Once the critical strain is reached, the surface initiates to buckle and evolves by the increasing applied strain. The amplitude of wrinkles continuously increases by the compressive strain. For the L-H composite slab, a sinusoidal mode appears in the structure for different levels of $\varepsilon$. In Figure 15(b), one observes that upon the increase of $\varepsilon$, different modes of post-bifurcation appears in the H-H slab. Once the $\varepsilon$ passes 0.33, the sinusoidal mode transfers to a second mode with a wavelength and a periodicity twice those of the first mode [113]. Further increase of the compressive strain beyond 0.37, results in a third mode with a wavelength and a periodicity three times as those of the first mode. It must be noted that all three post-bifurcation modes are triggered by the initial sinusoidal imperfection in both L-H and H-H models.

Figure 16 shows the variation of the dimensionless amplitude $\delta = \frac{A}{H+h_f}$ (left axis) and periodicity of wrinkles (right axis) versus the applied strain, $\varepsilon$, where $A$ is the maximum vertical displacement of the top surface boundary, obtained from the FEM modeling. The periodicity of wrinkles continuously decreases by the increasing strain for L-H (Figure 16(a)) and H-H (Figure 16(b)) structures and agrees with the theoretical estimation. Discontinuity of the periodicity graph in Figure 16(b), appears due to periodicity-doubling and tripling at $\varepsilon \approx 0.33$ and $\varepsilon \approx 0.37$, respectively.

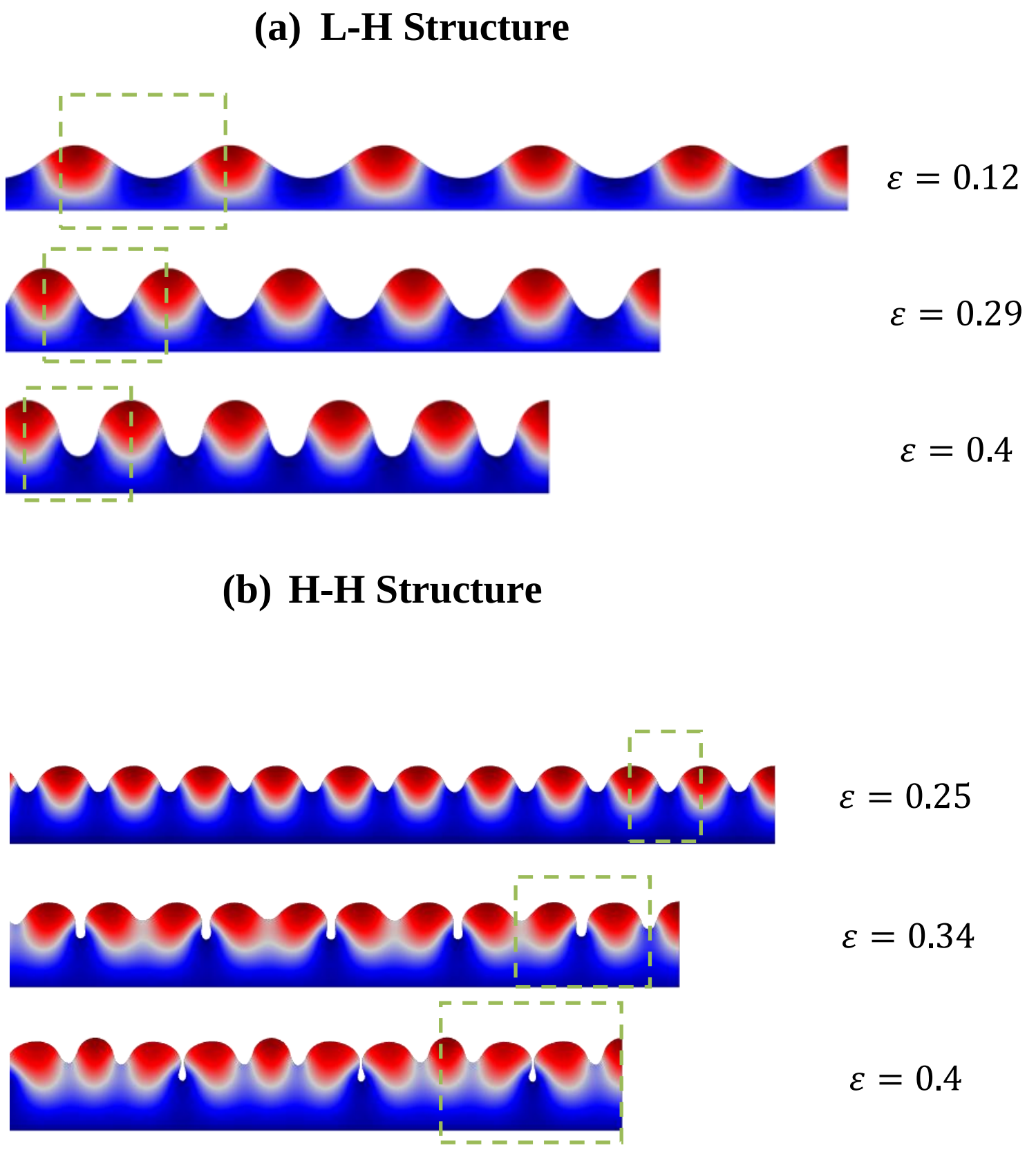

**Figure 15.** Simulation results for plane strain wrinkling of **(a)** L-H structure, with $\frac{\mu_f}{\mu_s} = \mathbf{125}$, $\boldsymbol{\nu_f} = \mathbf{0.2}, \boldsymbol{\nu_s} = \mathbf{0.49}$, $\frac{h_f}{H} = \mathbf{0.15}$, $\frac{H}{\gamma} = \mathbf{0.25}$, and **(b)** H-H structure, with $\frac{\mu_f}{\mu_s} = \mathbf{125}, \frac{\kappa_f}{\kappa_s} = \mathbf{3}$, $\frac{h_f}{H} = \mathbf{0.25}$, at different levels of compressive strain. Dashed rectangles show unit cells selected for wave propagation analysis.

The periodicity continues to decrease for second and third post-bifurcation modes. For L-H structure, the threshold strain to initiate the surface buckling is $\varepsilon_c = 0.0027$, which slightly differs from the theoretical prediction, $\varepsilon_c = 0.0026$. Likewise, for H-H structure, the critical strain is $\varepsilon_c = 0.0063$, while the theoretical critical strain predicts $\varepsilon_c = 0.0061$.

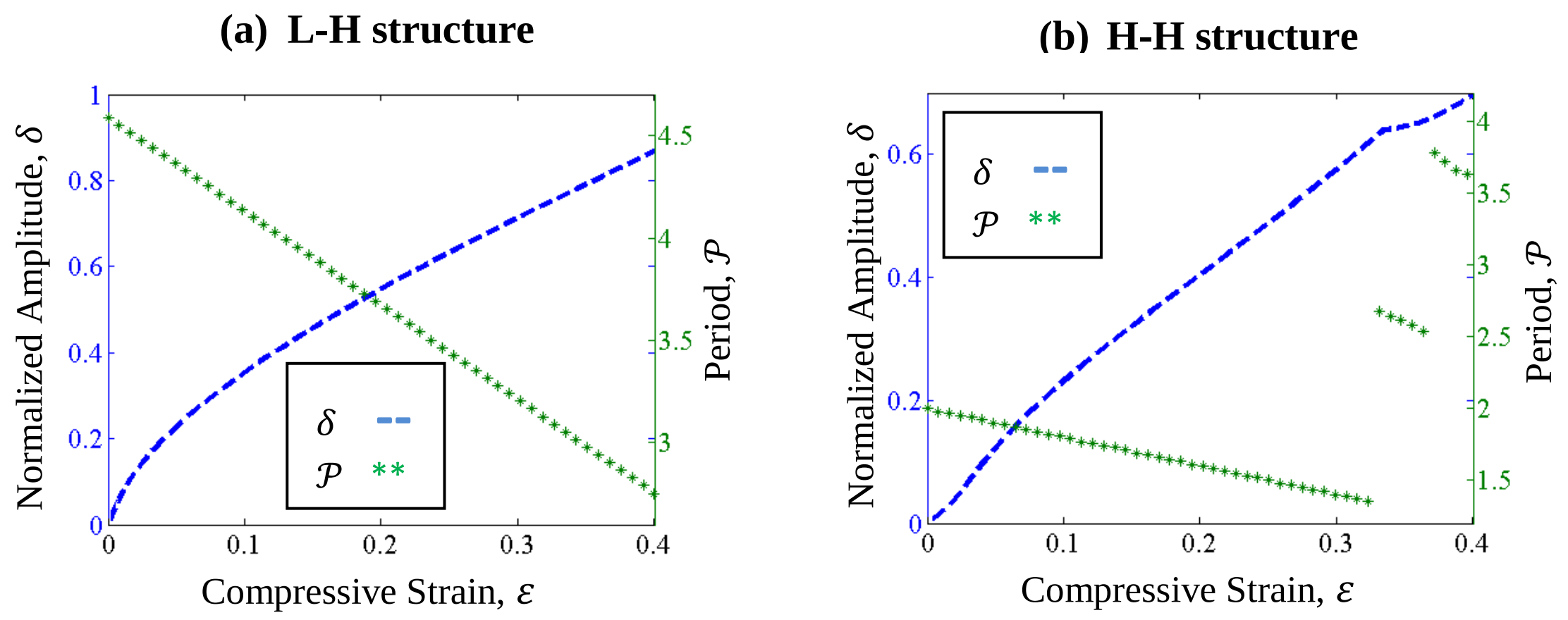

**Figure 16.** Variation of the dimensionless amplitude, $\boldsymbol{\delta}$ (left axis) and periodicity of the wrinkles, $\boldsymbol{\mathcal{P}}$ (right axis) vs. applied strain ε, for **(a)** L-H and **(b)** H-H composite slabs, computed form numerical simulations.

Consequently, it is demonstrated that the geometry, pattern and periodicity of wrinkles are significantly transformed by the applied compressive strain, creating opportunities to employ the wrinkly structure as a tunable 1D PnC. The elastic wave propagation in the periodic wrinkly structure is investigated through the FEM. Wave propagation in periodic structures depends on the unit cell's geometry, material properties and the lattice's periodicity. The incremental elastic wave propagation is influenced by the transformation of wrinkles' pattern due to large deformations and stress.

The deformation of the body is defined by the deformation gradient tensor, $\boldsymbol{F} = \frac{\partial \boldsymbol{x}}{\partial \boldsymbol{X}}$, which projects a point in the material coordinates, $\boldsymbol{X}$, to its Eulerian coordinates, $\boldsymbol{x}$. The constitutive law for linear elastic film at large deformation is defined by a stiffness matrix through $\boldsymbol{S} = \mathbb{D} : \varepsilon = \frac{1}{2}\mathbb{D}: (\boldsymbol{F}^T\boldsymbol{F} - \boldsymbol{I})$. The first Piola-Kirchoff stress tensor is defined by:

$$\boldsymbol{P} = \boldsymbol{FS} \tag{1}$$

The hyperelastic material is characterized by a neo-Hookean type strain energy function, as follows:

$$W = \frac{\mu}{2}(\boldsymbol{F} : \boldsymbol{F} - 3) - \mu \log J + \frac{\kappa}{2}(J-1)^2 \tag{2}$$

Thus, the first Piola-Kirchhoff stress tensor, $\boldsymbol{P} = \frac{\partial W}{\partial \boldsymbol{F}}$ is:

$$\boldsymbol{P} = \mu\boldsymbol{F} + (\kappa J(J-1) - \mu)\boldsymbol{F}^{-T} \tag{3}$$

where $\boldsymbol{F}^{-T}$ is the transpose matrix of the inverse of $\boldsymbol{F}$ and $J = det\boldsymbol{F}$. Equation governing the incremental motions superimposed on pre-deformed structures in Lagrangian coordinates is:

$$\nabla . \dot{\boldsymbol{P}} = \rho \frac{\partial^2 \dot{\boldsymbol{x}}}{\partial t^2} \tag{4}$$

where $\dot{\boldsymbol{x}}$ is the incremental displacemet, $\dot{\boldsymbol{P}}$ is the incremental first Piola-Kirchoff stress tensor, and $\rho$ is the density of each layer, hereafter is referred to as $\rho_f$ and $\rho_s$ for film and substrate layers, respectively. The increment of first Piola-Kirchoff stress tensor, $\dot{\boldsymbol{P}} = \mathbb{L} : \dot{\boldsymbol{F}}$ is a function of the incremental deformation gradient tensor, $\dot{\boldsymbol{F}} = \frac{\partial \dot{x}}{\partial \boldsymbol{X}}$ . The incremental moduli tensor is a fourth-order tensor defined by $\mathbb{L} = \frac{\partial^2 W}{\partial \boldsymbol{F} \partial \boldsymbol{F}}$ . For the material model defined in Equation (2) the incremental moduli tensor is:

$$\mathbb{L} = \mu\mathbb{I} + \kappa J(2J-1)\boldsymbol{F}^{-T} \otimes \boldsymbol{F}^{-T} + (\kappa J(J-1) - \mu)\mathbb{I}^T \tag{5}$$

Where $\mathbb{I}_{ijkl} = \frac{\partial \boldsymbol{F}_{ij}}{\partial \boldsymbol{F}_{kl}} = \delta_{ij}\delta_{kl}$ and $\otimes$ is the tensor product symbol. A solution of the wave equation in the form of plane wave $\dot{\boldsymbol{x}}(\boldsymbol{X},t) = \dot{\mathbb{x}}(\boldsymbol{X})e^{-i\omega t}$ is sought, where $\dot{\mathbb{x}}$ is the amplitude vector and $\omega$ is the angular frequency. Thus, the stress can be written in the following form:

$$\dot{\boldsymbol{P}} = \dot{\mathbb{P}}e^{-i\omega t} \tag{6}$$

Therefore, the equation of motion is an eigenvalue problem, as follows:

$$\nabla.\dot{\mathbb{P}} + \rho\omega^2\dot{\mathbb{x}} = 0 \tag{7}$$

where $\omega$ is the eigenfrequency of the system.

Wave propagation in PnCs is investigated through the application of Bloch type boundary conditions on parallel boundaries of the unit cell; the smallest repetitive structural element of the structure. The deformed unit cells are shown in Figure 15 for each compressive strain. To capture the first mode of the post-transformed unit cell, a rectangular geometry with a width of $\gamma = 2\pi h_f \left(\frac{E_f(1-\nu_s^2)}{3E_s(1-\nu_f^2)}\right)^{1/3}$, and a height of $H + h_f$ is chosen as the unit cell to be deformed through the compressive strain. For the second and the third post-bifurcation modes, unit cells with widths of $2\gamma$ and $3\gamma$ are modeled, respectively. For each unit cell, appropriate perturbations are imposed on the geometry of top surface to excite the desired mode. The periodicity, $\mathcal{P}$ of the lattice equals the wavelength of the wrinkle in the deformed state. For 1D wave propagation, Bloch type displacement boundary conditions $\dot{\mathbb{x}}^{+}(\boldsymbol{X}+\boldsymbol{r}) = \dot{\mathbb{x}}^{-}(\boldsymbol{X})e^{i(\boldsymbol{k}.\boldsymbol{r})}$, are applied on the left and the right boundaries of the unit cell, where $\boldsymbol{k}$ is the Bloch wave vector and $\boldsymbol{r} = \mathcal{P}\boldsymbol{i}$ is the distance vector between parallel boundaries. Superscripts + and – denote corresponding right and left opposite boundaries, respectively. It follows that the traction on the right and the left boundaries follows the Bloch relation. The only difference appears due to the fact that tractions on parallel boundaries are in opposite directions. The wave vector takes values of $\boldsymbol{k} = k\boldsymbol{b_1}$ along the irreducible Brillouin zone (IBZ), where $0 \leq k \leq 0.5$ is a real number and $\boldsymbol{b_1} = \frac{2\pi}{\mathcal{P}}$ is the lattice vector of the 1D reciprocal space [74].

It is noted that in FEM solvers, once the Bloch displacement boundary conditions are applied, the traction boundary conditions are automatically satisfied. Implementation of the plane wave form solution in the equation of motion and FEM discretization of the problem results in an eigenvalue problem whose stiffness matrix is a function of deformation gradient tensor and wavenumber, $\boldsymbol{k}$. Prestressed eigenfrequency study of the software is used to take into account the effect of deformation and prestress on the wave propagation analysis. This analysis is provided by the software to compute eigenfrequencies influenced by a prior static loading. By sweeping $k$ on the boundaries of IBZ, dispersion relations are investigated and eigenfrequencies are extracted as a function of wavenumber, $\boldsymbol{k}$ to plot band diagrams. Dispersion curves are plotted as normalized frequency, $\Omega = \frac{\omega\mathcal{P}}{2\pi c_0}$ versus reduced wavenumber, $k$, where $c_0 = \frac{h_f c_f + H c_s}{H + h_f}$, $c_f = \sqrt{\frac{\kappa_f + 2\mu_f}{\rho_f}}$ and $c_s = \sqrt{\frac{\kappa_s + 2\mu_s}{\rho_s}}$. All dispersion curves and band diagrams are computed for the eigenfrequencies larger than 1*MHz*.

## 2.2 Results and discussion

Band diagrams at different levels of the applied strain are shown in Figure 17, for both L-H and H-H structures. Band-gaps are shown in shaded regions. $\frac{\rho_f}{\rho_s} = 2.2$, $\frac{\mu_f}{\mu_s} = 2750$, $\frac{h_f}{H} = 0.12$, $\frac{H}{\gamma} = 0.25$, $\kappa_s = 1.0\ GPa$, and $\nu_f = 0.2$ are inputs of the L-H model. Geometry and material parameters used in the H-H model are $\frac{\rho_f}{\rho_s} = 2.2$, $\frac{\mu_f}{\mu_s} = 125$, $\frac{h_f}{H} = 0.21$, $\frac{H}{\gamma} = 0.25$, and $\frac{\kappa_f}{\kappa_s} = 3.0$, where $\gamma$ is the wavelength of wrinkles at first bifurcation mode.

Figure 17(a-d) demonstrates that band-gaps are generated and shifted by increasing compressive strain for L-H periodic slab. Two gaps are observed at $\varepsilon = 0.1$ and $\varepsilon = 0.2$. Five band-gaps are created in frequency spectrum of the structure when the strain is increased to $\varepsilon = 0.3$ and $\varepsilon = 0.4$. Analogously, for H-H composite, number and width of the band-gaps are switched by increasing strain. Figure 17(g-h) shows that when the periodicity doubling and tripling take place at $\varepsilon = 0.35$ and $\varepsilon = 0.40$, respectively, significant transformation of the wave propagation characteristics is observed. Results show that the dispersion curves tend to be flat bands at higher strains, providing more band-gap properties at second and third post-bifurcation modes. All results presented in Figure 17 are computed for a constant height of the slab, $\frac{H}{\gamma} = 0.25$.

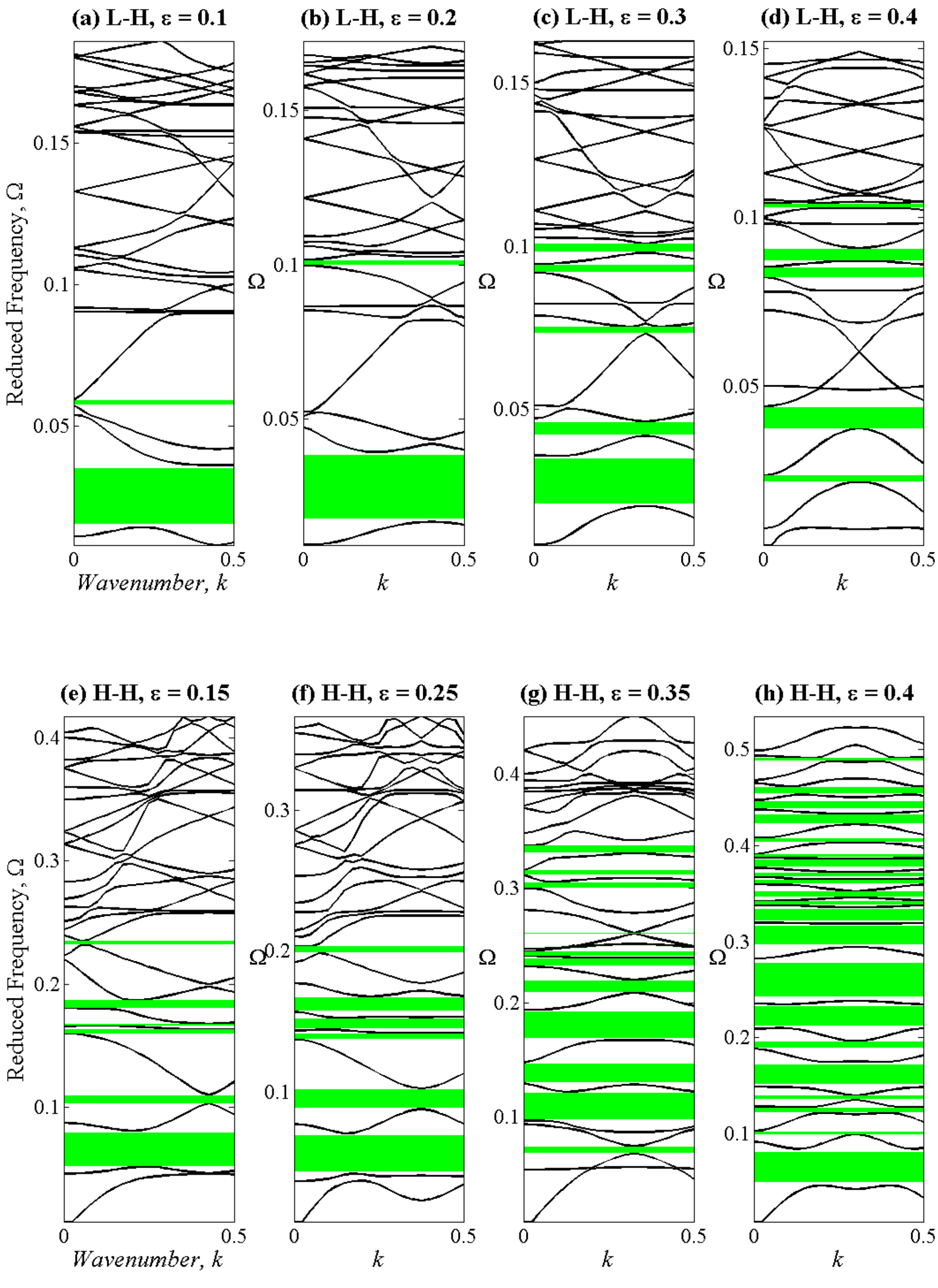

**Figure 17.** Numerical results for band diagram plots for **(a-d)** L-H and **(e-h)** H-H composite structure at different levels of the compressive strain, $\boldsymbol{\varepsilon}$. Band-gaps are shown in shaded regions. Dispersion curves are computed for eigenfrequencies larger than $\mathbf{1}\boldsymbol{MHz}$.

Figure 18 illustrates the effect of height of the slab on band diagrams of the structure. Slabs with different heights of $\frac{H}{\gamma} = 0.25,\ 0.5,\ 1$ and 3 are modeled for both L-H and H-H composites. Input parameters for the L-H composite are $\frac{\rho_f}{\rho_s} = 2.2,\ \frac{\mu_f}{\mu_s} = 2750,$ $\nu_f = 0.2,\ \kappa_s = 1.0\ GPa,\ h_f = 0.075\ \mu m,\ \gamma = 3.9\ \mu m$ and $\varepsilon = 0.3$. For the H-H model, $\frac{\rho_f}{\rho_s} = 2.2,\ \frac{\mu_f}{\mu_s} = 125,\ \frac{\kappa_f}{\kappa_s} = 3.0,\ h_f = 0.075\ \mu m,\ \gamma = 1.4\ \mu m$ and $\varepsilon = 0.3$ are inputs of the model.

Figure 18(a-d) shows how dynamic response of the L-H composite is influenced by height of the substrate. Band-gaps at higher frequencies vanish when $H = \gamma$. Similar trend is observed for H-H model, as is shown in Figure 18(e-h). For both L-H and H-H composites, band-gaps moves to lower frequencies and the high frequency gaps vanish for $H = \gamma$. For $H = 3\gamma$ no band-gap appears in band diagrams. Increasing the height of the slab results in a high contrast between the wrinkles' amplitude and thickness of the structure, so as wrinkles barely affect the wave propagation. One observes from Figures 18(d) and 18(h) that band diagrams are most likely a bulk material rather than a periodic structure. Red dashed line in Figures 18(d) and 18(h) is the sound line limiting the sound cone given by the transverse phase velocity of the substrate. Two narrow surface band-gaps are observed for both L-H and H-H structure as shown in the Figures 18(d) and 18(h), respectively.

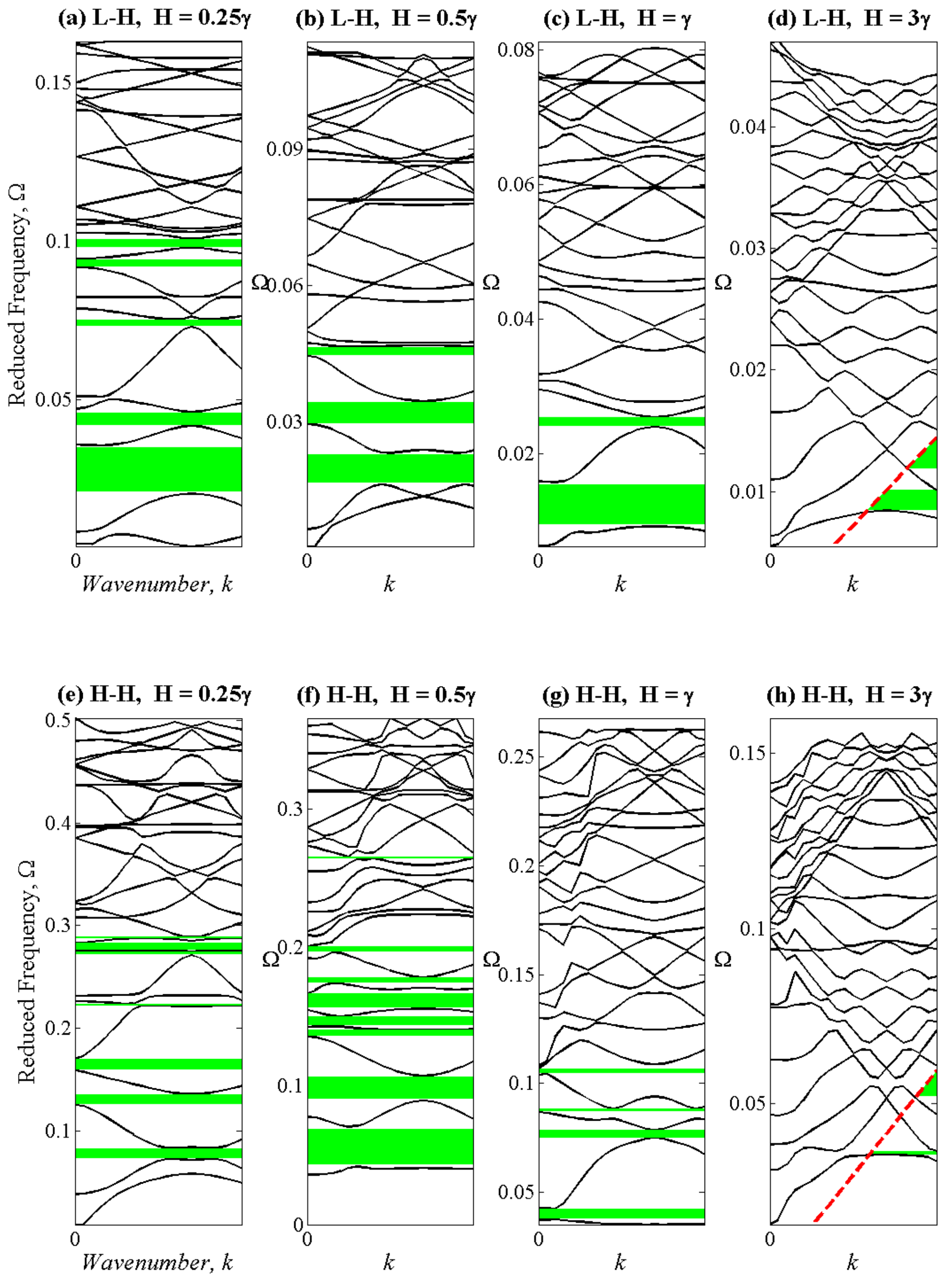

**Figure 18.** Numerical results for band diagram plots for **(a-d)** L-H and **(e-h)** H-H periodic structure for different heights of composite slabs. Band-gaps are shown in shaded regions. Dispersion curves are computed for eigenfrequencies larger than $\mathbf{1}\boldsymbol{MHz}$.

Figure 19(a-b) shows the evolution of band-gaps by the increasing density contrast ratio, $\frac{\rho_f}{\rho_s}$. Inputs of the L-H model are $\frac{\mu_f}{\mu_s} = 2750,\ \nu_f = 0.2,\ \kappa_s = 1.0\ GPa,\ \frac{h_f}{H} = 0.12,\ \frac{H}{\gamma} = 0.25,$ and $\varepsilon = 0.3$. Inputs of H-H model are, $\frac{\mu_f}{\mu_s} = 125,\ \frac{\kappa_f}{\kappa_s} = 3.0,\ \frac{h_f}{H} = 0.21,\ \frac{H}{\gamma} = 0.25,$ and $\varepsilon = 0.3$. One observes that for both L-H and H-H structures, density contrast ratio has slight effect on band-gap properties of the structure.

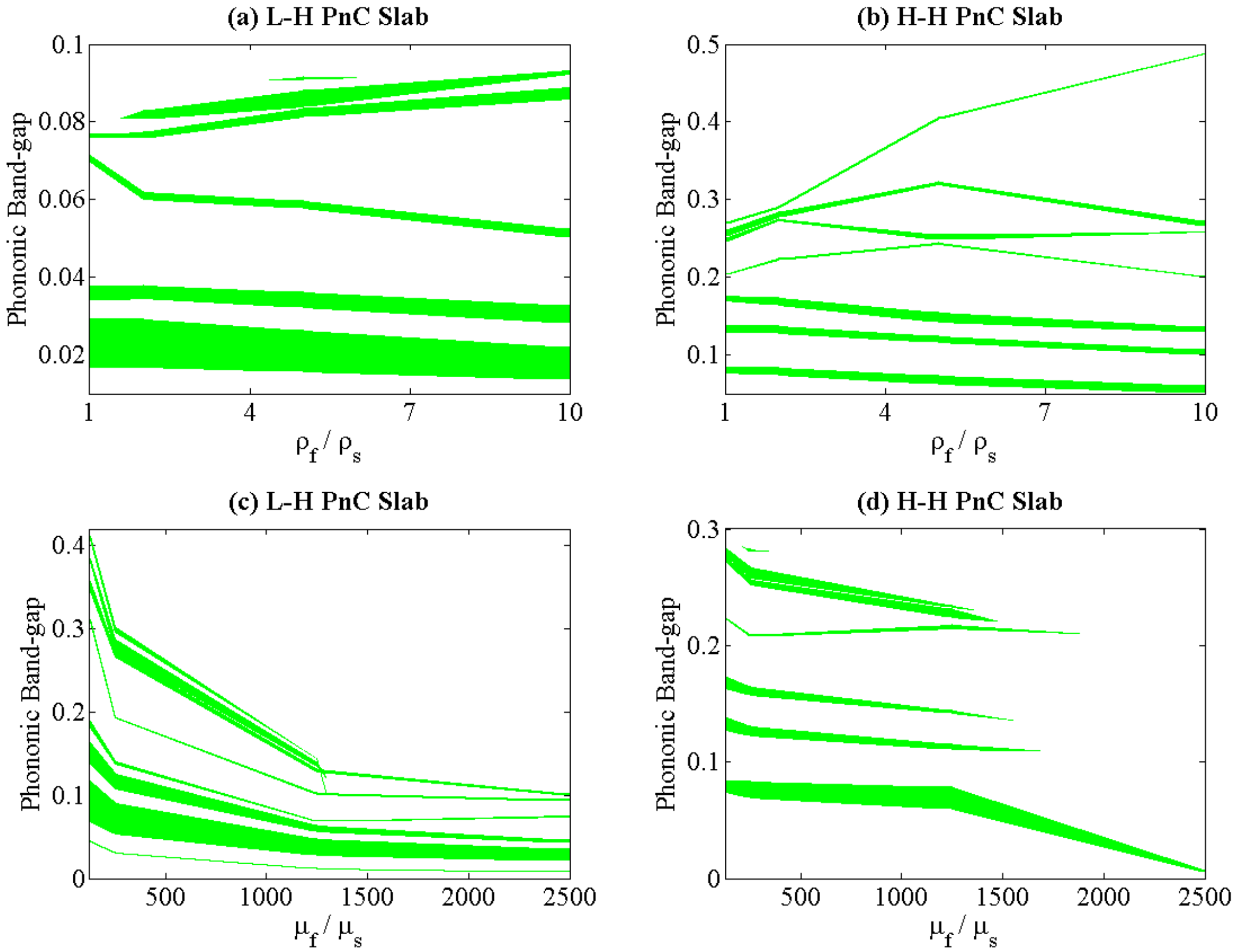

**Figure 19.** Evolution of Phononic band-gap vs. density contrast ratio, $\frac{\rho_f}{\rho_s}$ for (a) L-H and (b) H-H composites. Phononic band-gap vs. shear modulus contrast ratio, $\frac{\mu_f}{\mu_s}$ for (c) L-H and (d) H-H composites.

Effects of shear modulus contrast ratio, $\frac{\mu_f}{\mu_s}$ on band diagrams of L-H and H-H structures are shown in Figure 19(c-d). $\frac{\rho_f}{\rho_s} = 2.2,\ \nu_f = 0.2,\ \kappa_s = 1.0\ GPa,\ \frac{h_f}{H} = 0.12,\ \frac{H}{\gamma} = 0.25$ and $\varepsilon = 0.3$ are inputs of the L-H model and $\frac{\rho_f}{\rho_s} = 2.2,\ \frac{\kappa_f}{\kappa_s} = 3.0,\ \frac{h_f}{H} = 0.21,\ \frac{H}{\gamma} = 0.25$ and $\varepsilon = 0.3$ are input parameters of the H-H model. It is important to note that the width of the unit cell, $\gamma$, depends on the input parameters $\mu_f$ and $\mu_s$. For low shear modulus contrast ratio, more band-gap properties are observed at high frequencies. By increasing $\frac{\mu_f}{\mu_s}$ ratio, band-gaps tend to shift to low frequencies. At $\frac{\mu_f}{\mu_s} > 1500$, high frequency band-gaps vanish and only a single band-gap remains for H-H composite structure.

Figure 20 represents examples of harmonic response of an L-H slab compressed by $\varepsilon = 0.4$ at different frequencies. Prestressed frequency domain study is utilized to capture the dynamic response of the structure on a priori deformed structure. This study takes into account the effect of large deformations and stress on the frequency response of the structure. Harmonic perturbation is applied on the right edge of the model. The amplitude of the perturbation is chosen as 10% of the applied displacement. Line graphs show the strain energy density along middle surface of the structure. Contour plots represent in-plane vertical component of displacement resulted from a harmonic perturbation superimposed on the deformed structure. Figures 20(a-d) illustrates the harmonic perturbation response at $\Omega = 0.030, 0.040, 0.060$ and $0.085$, respectively. While Figures 20(a) and 20(c) show the wave propagation in the deformed structure, Figures 20(b) and 20(d) demonstrate that the wave is attenuated along the wrinkly structure. This is due to the fact that $\Omega = 0.040$ and $\Omega = 0.085$ correspond to the center of band-gaps of the composite slab, as is shown in Figure 20(d). Accordingly, frequency domain analysis verifies the results obtained from the Bloch wave analysis on the deformed unit cell of the composite slab.

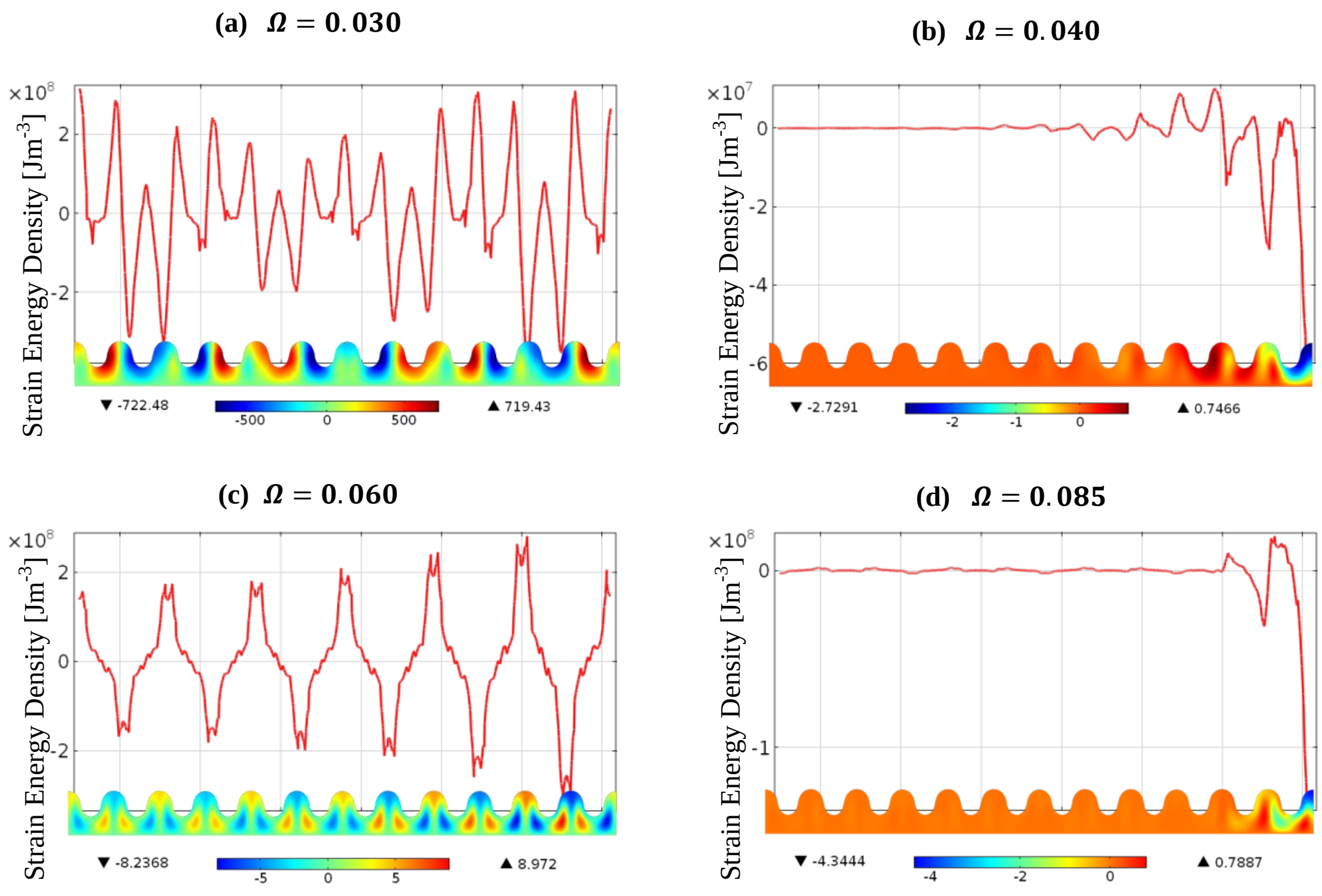

**Figure 20.** Line graphs showing the strain energy density along the middle surface of the L-H composite slab compressed by ε=0.4 at different normalized frequencies. Contour plots represent the vertical component of harmonic perturbation. Modeling parameters are identical to the ones used for band diagram investigation in Figure 17(d).

## 2.3 Summary and conclusion

The ability of high aspect ratio wrinkle formation to be utilized as a highly tunable PnC is designed and demonstrated through FEM simulations. Two material models are considered for the analysis; L-H and H-H composite slabs. Amplitude, period and geometry of wrinkles are controlled by the applied compressive strain. Wave propagation is controlled through the combined effect of large deformations and stress in the moderately thick periodic wrinkly slab. For H-H structure, three different mode shapes are triggered at different levels of the applied compressive strain. Band diagram analysis demonstrates that the applied compressive strain can control and transform band-gap properties of the designed PnC.

Conversion of post-bifurcation modes of the H-H structure has shown to have significant effects on phononic characteristics of the wrinkly slab. Moreover, the dynamic response of the wrinkly structure for different thickness of the slab is presented and possibility of surface elastic band-gaps is discussed. A frequency sweep analysis is performed on a typical L-H structure to examine the harmonic response at different frequencies. Results of the harmonic perturbation analysis verified the band diagram analysis.

The proposed structure demonstrates the possibility of creating large elastic band-gaps by the applied compressive stretch. The designed PnC can function as a 1D pillar based PnC [127] with capability of tuning the phononic band-gaps. Due to wide range of fabrication methods and growing interest in micro and nano-wrinkle structures [109-112], results of this study can be employed in developing micro and nano-scale tunable acoustic switches [128], acoustic filters [128], elastic isolators and acoustic sensors [128-129] to be utilized in ultrasonic and hypersonic applications. Moreover, the wrinkly tunable PnC can be employed in developing PnC-based high quality factor micro-mechanical resonators [131] which has high applications in optomechanical systems [132], sensing and communication devices [130-132].

# CHAPTER THREE

## Dynamic Response of a Tunable Phononic Crystal under Applied Mechanical and Magnetic Loadings[1]

Dynamic response of a tunable PnC consisting of a porous hyperelastic magnetoelastic elastomer subjected to a macroscopic deformation and an external magnetic field is theoretically investigated. Finite deformations and magnetic induction influence phononic characteristics of the periodic structure through geometrical pattern transformation and material properties. A magnetoelastic energy function is proposed to develop constitutive laws considering large deformations and magnetic induction in the periodic structure. Analytical and finite element methods are utilized to compute dispersion relation and band structure of the PnC for different cases of deformation and magnetic loadings. It is demonstrated that magnetic induction not only controls the band diagram of the structure but also has strong effect on preferential directions of wave propagation.

In this study, a porous periodic structure of a soft magnetoelastic medium is designed to control wave propagation characteristics via the combined effects of finite deformations

---

[1] Results of this chapter are published in:

- Bayat A and Gordaninejad F 2015 Band-gap of a Soft Magnetorheological Phononic Crystal *ASME Journal of Vibrations and Acoustics* **137** 011013.
- Bayat A and Gordaninejad F 2015 Dynamic Response of a Tunable Phononic Crystal under Applied Mechanical and Magnetic Loadings *Journal of Smart Materials and Structures* **24,** 065027.
- Bayat A and Gordaninejad F 2014 A Magnetically Field-Controllable Phononic Crystal Proc. SPIE, San-Diego, USA, 9057, p. 905713.

and magnetic fields. Figure 21 illustrates the solution procedure. A soft matrix provides the reversible pattern transformation of the structure due to large deformations. The magnetic field contributes to the change in mechanical properties of the structure. In addition, constitutive relations for incremental motions superimposed on a predeformed structure is developed following the approaches used in Ref. [19-22]. Moreover, wave propagation analysis is performed and the variational formulation for magnetoelastic wave equations considering the Bloch boundary conditions is developed. Finite element (FE) methods are used to solve the coupling magnetoelastic equations and results are presented as band diagrams as well as isofrequency contours at different cases of loadings. The nonlinear FE solver, COMSOL Multiphysics 4.4 is utilized for numerical simulations.

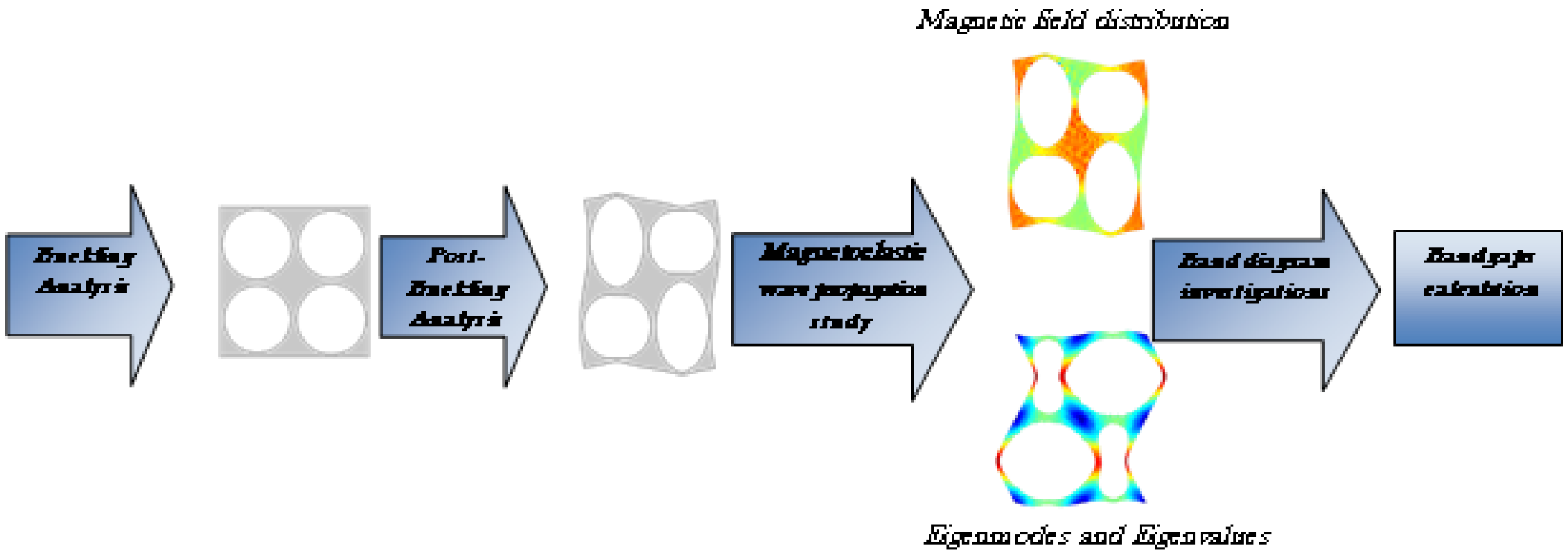

**Figure 21**. Schematic of the computational approach proceeded to study the dynamic response of the structure.

### 3.1 Governing equations

The continuum is considered as a magnetoelastic hyperelastic body that is initially in an undeformed state, denoted by $\mathbb{C}_\mathrm{r}$with boundary $\partial\mathbb{C}_\mathrm{r}$ as the reference configuration. When the body is subjected to time-dependent magnetic and mechanical loadings, it deforms. Thus, the region occupied by the continuum $\mathbb{C}_\mathrm{t}$, with boundary $\partial\mathbb{C}_\mathrm{t}$, at a given time *t* is the deformed configuration. Let X and x be the position vectors of the material particle at reference and deformed configuration respectively, where $\mathrm{x} = \chi(\mathrm{X}, \mathrm{t})$ and $\chi: \mathbb{C}_\mathrm{r} \to \mathbb{C}_\mathrm{t}$ is called the deformation mapping. In this study, a multi-scale approach is followed to consider the geometrical nonlinearity in the deformed structure as well as a linear wave propagation analysis in the deformed structure. Figure 22 illustrates the reference, deformed and incrementally deformed configurations.

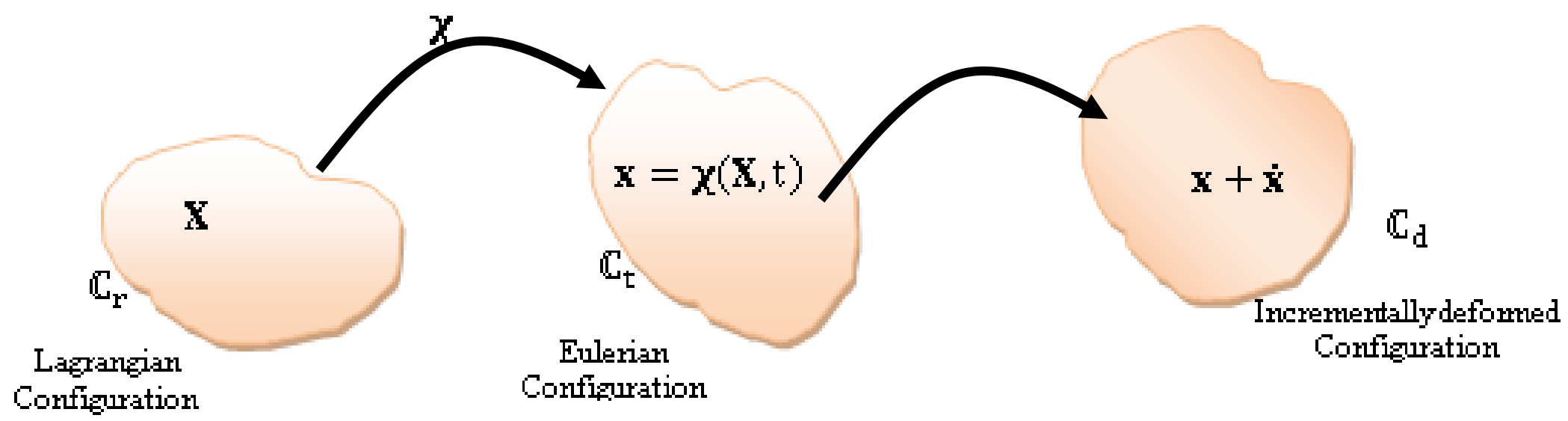

**Figure 22.** Reference (undeformed), deformed and superimposed incrementally deformed configurations.

The continuum first undergoes large deformations due to the macroscopic deformation gradient tensor and magnetic induction from $\mathbb{C}_r$ to $\mathbb{C}_\mathrm{t}$ to take into account the pattern change in the structure. Next, the incremental wave motion superimposed on the deformed body is studied at $\mathbb{C}_{\mathrm{d}}$ configuration.

The deformation gradient tensor is defined as $\boldsymbol{F} = Grad\, \boldsymbol{x} = \partial \boldsymbol{\chi} / \partial \boldsymbol{X}$, where $Grad$ is gradient operator with respect to material coordinates $\boldsymbol{X}$. The notations $Grad, Div\ and\ Curl$ are used for differential operators in Lagrangian coordinate $\boldsymbol{X},$ and $grad, div\ and\ curl$ are used as the corresponding operators in Eulerian coordinates with respect to $\boldsymbol{x}$. Also, $J = det\, \boldsymbol{F}$ with $J > 0$. In Eulerian form, the magnetic field and magnetic induction vectors are denoted by $\boldsymbol{H} = \boldsymbol{H}(\boldsymbol{x})$ and $\boldsymbol{B} = \boldsymbol{B}(\boldsymbol{x})$, respectively.

It is assumed that the magnetic field is stationary and the non-conducting magnetoelastic material is initially at the static configuration and subjected to only magnetic and mechanical interactions. Thus, $\boldsymbol{H}\ and\ \boldsymbol{B}$ are independent of time and equations of magnetostatics can be written, as follows:

$$curl\, \boldsymbol{H} = \boldsymbol{0}, \qquad div\, \boldsymbol{B} = 0, \tag{1}$$

In the absence of distributed charges or current density, the electric field and the displacement vector can be neglected. Let us assign $\boldsymbol{H}_l = \boldsymbol{F}^T \boldsymbol{H}$ and $\boldsymbol{B}_l = J\boldsymbol{F}^{-1}\boldsymbol{B}$ for the Lagrangian counterpart of the magnetic field and the magnetic induction vectors, (defined in the reference configuration $\mathbb{C}_{\mathrm{r}}$) respectively, where $\boldsymbol{F}^T$ is the transpose of the deformation tensor. Using the standard kinematic identities: $Div(J\boldsymbol{F}^{-1}\boldsymbol{B}) = Jdiv\boldsymbol{B}$ and $\boldsymbol{F}Curl(\boldsymbol{F}^T\boldsymbol{H}) = Jcurl\boldsymbol{H},$ Equation (1) can be written in the Lagrangian form, as follows:

$$Curl\, \boldsymbol{H}_l = \boldsymbol{0}, \qquad Div\, \boldsymbol{B}_l = 0, \tag{2}$$

The equation of motion governing a continuum in the presence of electromagnetic and body forces are,

$$div\, \boldsymbol{\tau} + \rho \boldsymbol{f} = \rho \frac{\partial^2 \boldsymbol{x}}{\partial t^2}, \qquad \text{in } \mathbb{C}_t \tag{3}$$

where $\boldsymbol{\tau}$ is the total Cauchy stress tensor which incorporates electromagnetic body forces, $\rho$ is the density, and $\boldsymbol{f}$ is the mechanical body force density per unit mass. In Lagrangian form, the equation of motion is,

$$Div\, \boldsymbol{T} + \rho_r \boldsymbol{f} = \rho_r \frac{\partial^2 \boldsymbol{X}}{\partial t^2}, \qquad \text{in } \mathbb{C}_r \tag{4}$$

where $\boldsymbol{T}$ is the total nominal stress tensor and $\rho_r$ is the reference mass density, which are related to their Eulerian counterparts by:

$$\boldsymbol{\tau} = J^{-1} \boldsymbol{F} \boldsymbol{T}, \qquad \rho_r = \rho J \tag{5}$$

Here, a nonlinear magnetoelastic hyperelastic material is considered with total energy density $\boldsymbol{\Psi} = \boldsymbol{\Psi}(\boldsymbol{F}, \boldsymbol{B}_l)$ as a function of deformation tensor and magnetic induction vector, defined per unit volume in $\mathbb{C}_r$. For a compressible material, the constitutive relations for total nominal stress and magnetic field in the Lagrangian coordinates are:

$$\boldsymbol{T} = \frac{\partial \boldsymbol{\Psi}}{\partial \boldsymbol{F}}, \qquad \boldsymbol{H}_l = \frac{\partial \boldsymbol{\Psi}}{\partial \boldsymbol{B}_l}, \qquad \text{in } \mathbb{C}_r \tag{6}$$

The Eulerian counterparts of Equation (6) are:

$$\boldsymbol{\tau} = J^{-1} \boldsymbol{F} \frac{\partial \boldsymbol{\Psi}}{\partial \boldsymbol{F}}, \qquad \boldsymbol{H} = \boldsymbol{F}^{-T} \frac{\partial \boldsymbol{\Psi}}{\partial \boldsymbol{B}_l}, \qquad \text{in } \mathbb{C}_t \tag{7}$$

For an isotropic magnetoelastic material, $\boldsymbol{\Psi}$ is an isotropic function of two tensors $\boldsymbol{c} = \boldsymbol{F}^{\boldsymbol{T}}\boldsymbol{F}$ (right Cauchy–Green deformation tensor) and $\boldsymbol{B}_l \otimes \boldsymbol{B}_l$ which can be expressed in terms of six independent invariants, as follows:

$$\boldsymbol{I}_1 = tr\boldsymbol{c}, \quad \boldsymbol{I}_2 = \frac{1}{2}[(tr\boldsymbol{c})^2 - tr(\boldsymbol{c})^2], \quad \boldsymbol{I}_3 = det\boldsymbol{c} = J^2,$$
$$\boldsymbol{I}_4 = \boldsymbol{B}_l \cdot \boldsymbol{B}_l, \quad \boldsymbol{I}_5 = (\boldsymbol{c}\boldsymbol{B}_l) \cdot \boldsymbol{B}_l, \quad \boldsymbol{I}_6 = (\boldsymbol{c}^2\boldsymbol{B}_l) \cdot \boldsymbol{B}_l, \tag{8}$$

$\boldsymbol{I}_1, \boldsymbol{I}_2$ and $\boldsymbol{I}_3$ are principal invariants of $\boldsymbol{c}$ and $\boldsymbol{I}_4, \boldsymbol{I}_5$ and $\boldsymbol{I}_6$ are invariants which are function of $\boldsymbol{B}_l$. Thus, the total nominal stress and the Lagrangian magnetic field can be expressed, as follows:

$$\boldsymbol{T} = \sum_{i=1}^{6} \boldsymbol{\Psi}_{\boldsymbol{i}} \frac{\partial \boldsymbol{I}_i}{\partial \boldsymbol{F}}, \quad \boldsymbol{H}_l = \sum_{i=3}^{6} \boldsymbol{\Psi}_{\boldsymbol{i}} \frac{\partial \boldsymbol{I}_i}{\partial \boldsymbol{B}_l}, \tag{9}$$

where $\boldsymbol{\Psi}_{\boldsymbol{i}} = \frac{\partial \boldsymbol{\Psi}}{\partial \boldsymbol{I}_{\boldsymbol{i}}}$ are the derivatives of the energy function with respect to independent invariants.

Assuming that the deformation $\boldsymbol{x} = \boldsymbol{\chi}(\boldsymbol{X}, \boldsymbol{t})$ and the applied magnetic field in $\mathbb{C}_t$ are priori known, an incremental displacement is superimposed on the deformed body in the form of $\boldsymbol{u}(\boldsymbol{x}, t) = \dot{\boldsymbol{x}} = \dot{\boldsymbol{\chi}}(\boldsymbol{X}, t)$, which causes both the magnetic field and deformation undergo incremental changes within the material. In the following, a superimposed dot $\dot{}$, represents an infinitesimal change in the quantity concerned. The new configuration is denoted by $\mathbb{C}_{\mathrm{d}}$. Considering $\dot{\boldsymbol{F}}$ and $\dot{\boldsymbol{B}}_l$ as increments in independent variables $\boldsymbol{F}$ and $\boldsymbol{B}_l$, by considering increments of constitutive Equations (8), one can express the linearized form of these equations, as follows:

$$\dot{\boldsymbol{T}} = \mathbb{A}\dot{\boldsymbol{F}} + \mathbb{K}\dot{\boldsymbol{B}}_l, \quad \dot{\boldsymbol{H}}_l = \mathbb{K}^{\boldsymbol{T}}\dot{\boldsymbol{F}} + \mathbb{L}\dot{\boldsymbol{B}}_l, \tag{10}$$

where $\dot{\boldsymbol{F}} = Grad\boldsymbol{u} = (grad\boldsymbol{u})\boldsymbol{F}$ and $\mathbb{A}, \mathbb{K}$ and $\mathbb{L}$ are incremental magnetoelastic moduli tensors defined by:

$$\mathbb{A} = \frac{\partial^2 \boldsymbol{\Psi}}{\partial \boldsymbol{F} \partial \boldsymbol{F}}\,, \qquad \mathbb{K} = \frac{\partial^2 \boldsymbol{\Psi}}{\partial \boldsymbol{F} \partial \boldsymbol{B}_l}\,, \qquad \mathbb{K}^{\boldsymbol{T}} = \frac{\partial^2 \boldsymbol{\Psi}}{\partial \boldsymbol{B}_l \partial \boldsymbol{F}}\,, \qquad \mathbb{L} = \frac{\partial^2 \boldsymbol{\Psi}}{\partial \boldsymbol{B}_l \partial \boldsymbol{B}_l}, \tag{11}$$

$\mathbb{A}, \mathbb{K}$ and $\mathbb{L}$ are fourth-, third- and second-order tensors, with corresponding symmetries, respectively. The expressions for incremental moduli tensors in terms of the six invariants of the total energy function are well documented in the literature [19-21]. The products in Equation (10) are defined as:

$$\begin{gathered}(\mathbb{A}\dot{\boldsymbol{F}})_{\boldsymbol{\alpha i}} = \mathbb{A}_{\alpha i \beta j}\dot{\boldsymbol{F}}_{j\beta}\,, \qquad (\mathbb{K}\dot{\boldsymbol{B}}_l)_{\alpha i} = \mathbb{K}_{\alpha i|\beta}\dot{\boldsymbol{B}}_{l\beta}\,, \\ (\mathbb{K}^T\dot{\boldsymbol{F}})_{\beta} = \mathbb{K}_{\beta|\boldsymbol{\alpha i}}\dot{\boldsymbol{F}}_{\boldsymbol{i\alpha}}\,, \qquad (\mathbb{L}\dot{\boldsymbol{B}}_l)_{\boldsymbol{\alpha}} = \mathbb{L}_{\alpha\beta}\dot{\boldsymbol{B}}_{l\beta},\end{gathered} \tag{12}$$

By considering the increment of the Lagrangian form of governing Equations (2) and (3), one has:

$$\begin{gathered}Div\,\dot{\boldsymbol{T}} + \rho_r\dot{\boldsymbol{f}} = \rho_r\frac{\partial^2\boldsymbol{u}}{\partial t^2} \\ Div\,\dot{\boldsymbol{B}}_l = 0 \qquad\qquad Curl\,\dot{\boldsymbol{H}}_l = \boldsymbol{0}\end{gathered} \tag{13}$$

The Eulerian counterparts of the Equation (13) can be obtained through the following transformations:

$$\dot{\boldsymbol{T}}^{\boldsymbol{e}} = J^{-1}\boldsymbol{F}\dot{\boldsymbol{T}}\,, \qquad \dot{\boldsymbol{H}}^{\boldsymbol{e}} = \boldsymbol{F}^{-T}\dot{\boldsymbol{H}}_l, \qquad \dot{\boldsymbol{B}}^{\boldsymbol{e}} = J^{-1}\boldsymbol{F}\dot{\boldsymbol{B}}_l \tag{14}$$

where the superscript $\boldsymbol{e}$ indicates the Eulerian form of the quantity concerned, when the reference configuration is updated from $\mathbb{C}_r$ to $\mathbb{C}_t$, after the increments are formed. Hence Equation (13) in the Eulerian form can be written as:

$$div\,\dot{\boldsymbol{T}}^{\boldsymbol{e}} + \rho\dot{\boldsymbol{f}} = \rho\frac{\partial^2\boldsymbol{u}}{\partial t^2},$$

$$div\,\dot{\boldsymbol{B}}^{\boldsymbol{e}} = 0 \qquad\qquad curl\,\dot{\boldsymbol{H}}^{\boldsymbol{e}} = \boldsymbol{0} \tag{15}$$

Combining Equations (10), (12) and (14), one has:

$$\dot{\boldsymbol{T}}^{\boldsymbol{e}} = \mathbb{A}^{\boldsymbol{e}}\boldsymbol{d} + \mathbb{K}^{\boldsymbol{e}}\dot{\boldsymbol{B}}^{\boldsymbol{e}} \qquad\qquad \dot{\boldsymbol{H}}^{\boldsymbol{e}} = \mathbb{K}^{\boldsymbol{e}^{\boldsymbol{T}}}\boldsymbol{d} + \mathbb{L}^{\boldsymbol{e}}\dot{\boldsymbol{B}}^{\boldsymbol{e}} \tag{16}$$

where $\boldsymbol{d} = grad\boldsymbol{u}$ is the displacement gradient and the updated magnetoelastic moduli tensors are defined by:

$$\mathbb{A}^{\boldsymbol{e}}_{jqip} = J^{-1}\boldsymbol{F}_{\boldsymbol{i\alpha}}\boldsymbol{F}_{\boldsymbol{j\beta}}\mathbb{A}_{\beta q\alpha p}\,, \quad \mathbb{K}^{\boldsymbol{e}}_{ij|k} = \boldsymbol{F}_{i\alpha}\boldsymbol{F}^{-1}_{k\beta}\mathbb{K}_{\alpha j|\beta}, \quad \mathbb{L}^{\boldsymbol{e}}_{ij} = J\boldsymbol{F}^{-1}_{\alpha i}\boldsymbol{F}^{-1}_{\beta j}\mathbb{L}_{\alpha\beta}\,, \tag{17}$$

It is noted that the updated moduli tensors possess the same symmetry as the ones in Equation (11).

### 3.2 Buckling Analysis

In periodic structures, shape transformation due to buckling instabilities has been utilized to explore tunable PnCs [97, 98]. Under a deformation, the periodicity of an infinite size periodic porous structure breaks down, due to the first occurrence of bifurcation in the microstructure. Pattern transformation of PnCs occurs due to either: i) local buckling modes, or ii) global modes of instability, namely microscopic and macroscopic instability. Microscopic instability is based on the investigation of all possible bounded modes of instability within the range of wavelengths comparable to the unit cell's size.

The macroscopic instability is a measure of instability modes with wavelengths much longer than the unit cell's size [133].

In this study, a buckling analysis is performed on finite size PnCs to find eigenmodes and post-buckling periodicity of the deformed structure, following the approach documented in References [133]. The structure consists of a two-dimensional (2D) square array of circular holes of radius $r_c = 2.2\,mm$. The initial cell of the undeformed structure is a square with a side of $a = 5\,mm$, that is used as the direct lattice vectors in $\boldsymbol{x}$ and $\boldsymbol{y}$ directions (Figure 23**(a)**). A nearly incompressible Neo-Hookean hyperelastic model is assumed for numerical simulations. The density, $\rho = 1{,}000\,kg\,m^{-3}$, shear modulus, $\mu = 0.25\,MPa$, and initial bulk modulus of $\kappa = 10^4\mu$ are assumed for FE simulations [30].

The deformation boundary conditions are applied on all boundaries of the models through; $\boldsymbol{\mathcal{U}}^+ - \boldsymbol{\mathcal{U}}^- = \boldsymbol{F}(\boldsymbol{X}^+ - \boldsymbol{X}^-)$, where $\boldsymbol{\mathcal{U}}$, $\boldsymbol{F}$ and $\boldsymbol{X}$ are the displacement vector, the deformation gradient tensor and the position vector in Lagrangian coordinates, respectively. The superscripts (+) and (–) are associated with the nodes on the opposite boundaries (right (top) and left (bottom) in Figure 23**(a)**) of the rectangular model. The linear buckling study is employed to study the buckling eigenmodes of the structure. The buckling analysis on several PnCs of $2 \times 2$, $4 \times 4$, $8 \times 8$, and $20 \times 20$ cells, subjected to deformation boundary conditions, agreed with the previous findings on the instability modes [97, 98, 133]. Figure 23 shows the first mode pattern transformation results from the buckling analysis of a finite size $8 \times 8$ periodic structure under uniaxially compressive stretch.

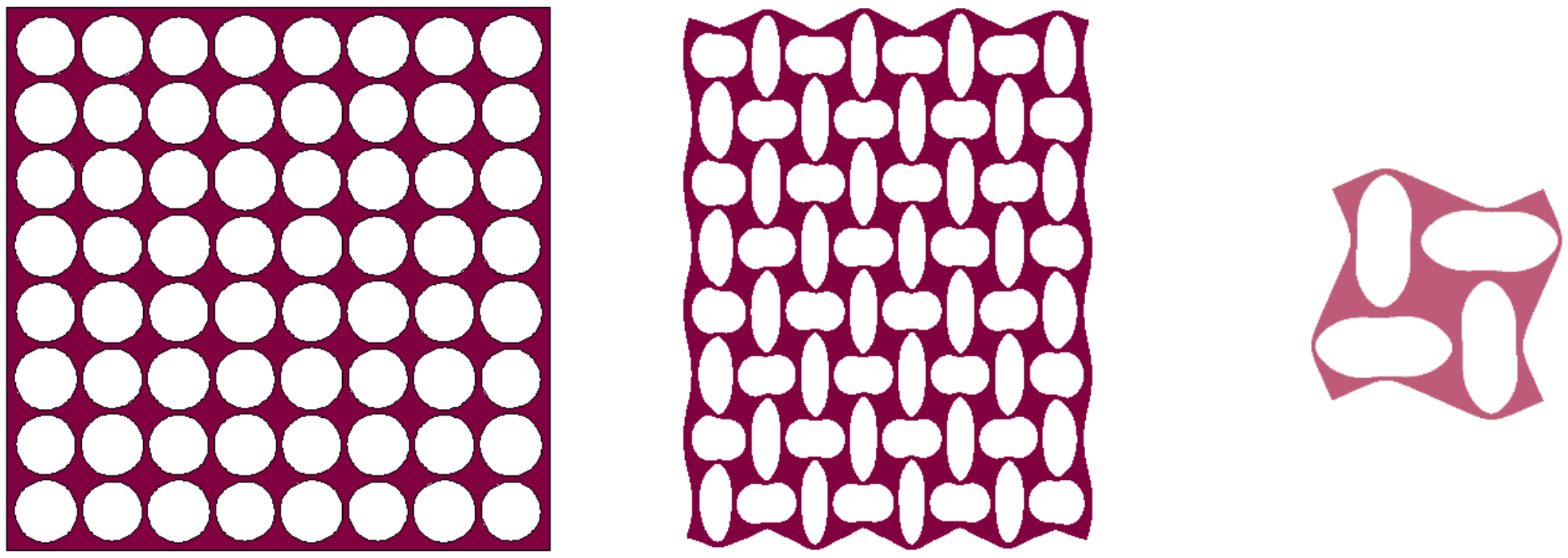

(a) (b) (c)

**Figure 23.** The FEM results for pattern transformation in the phononic crystal due to buckling instability under the applied deformation. **(a)** An undeformed 8×8 periodic structure, **(b)** the first buckling mode of the structure subjected to uniaxially compressive stretch in horizontal direction, and **(c)** the first buckling eigenmode of the enlarged unit cell.

Hence, buckling instabilities changes the configuration to a new enlarged unit cell with periodicity of $2 \times 2$ times the initial unit cell (Figure 23**(c)**). The new unit cell will be referred to as the representative unit cell (RUC). A post-transformation analysis is performed on the RUC to obtain the deformed geometry for the wave propagation study.

### 3.3 Wave propagation analysis

In this study, the propagation of infinitesimal harmonic plane waves is considered in a medium subjected to a pre-existing homogeneous deformation gradient tensor, $\boldsymbol{F}$, and a uniform magnetic induction vector, $\boldsymbol{B}_l$. In this work, the focus is on the two-dimensional elastic wave propagation in periodic structure of infinite size. PnCs are characterized by a unit cell that is defined through direct lattice vectors, $\boldsymbol{a_1}$ and $\boldsymbol{a_2}$. These vectors are the periodicity of the lattice in $\boldsymbol{x}$ and $\boldsymbol{y}$ directions. The RUC and the corresponding deformed geometry for a horizontal uniaxial stretch $\lambda_x = 0.9$, considered for wave propagation study, are shown in Figure 24.

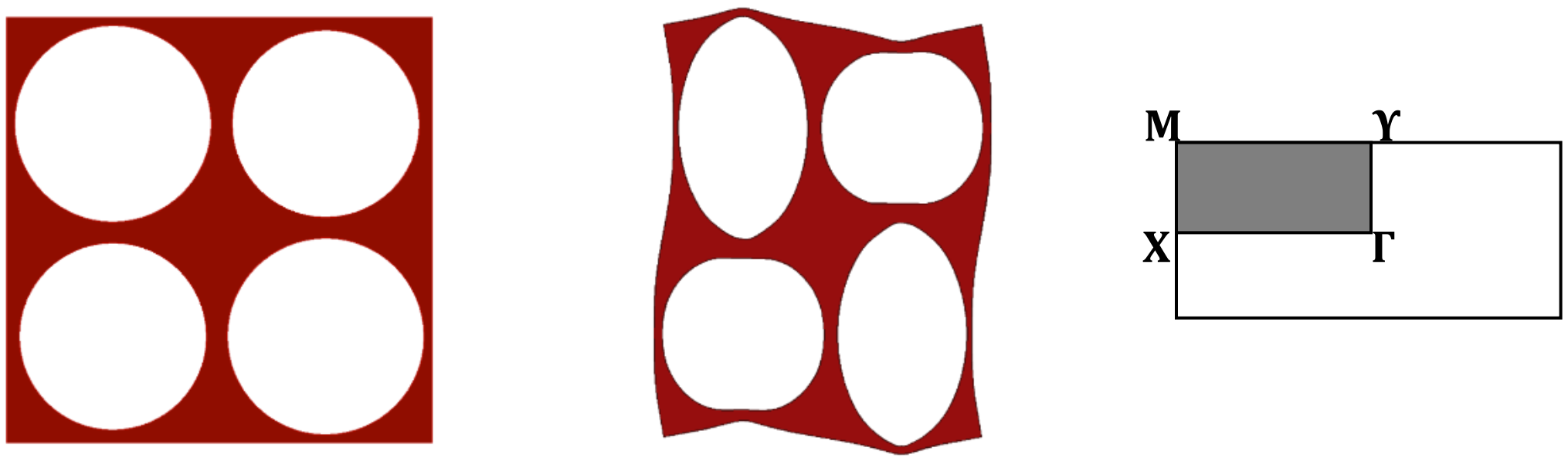

(a) (b) (c)

**Figure 24**. **(a)** The RUC selected for the uniaxial compression, **(b)** the corresponding deformed geometry for uniaxial stretch $\boldsymbol{\lambda_x = 0.9}$, and **(c)** reciprocal lattice's unit cell selected for the wave propagation study. Irreducible Brilluine zone is shown in the region bounded by $\boldsymbol{\Gamma - \Upsilon - M - X - \Gamma}$.

A solution of field Equation (15) in the form of:

$$\boldsymbol{u} = \boldsymbol{u}(\boldsymbol{x})\boldsymbol{e}^{-i\omega t}, \qquad \dot{\boldsymbol{B}}^{\boldsymbol{e}} = \boldsymbol{q}(\boldsymbol{x})\boldsymbol{e}^{-i\omega t}, \tag{18}$$

is considered, where $\boldsymbol{u}$ and $\boldsymbol{q}$ are amplitudes and $\omega$ is the angular frequency. Substituting Equation (18) into Equation (15) in the absence of body forces yields the updated Equations as a function of $\boldsymbol{u}, \boldsymbol{q}$ and $\omega$. To develop a FE model, the variational formulation of the coupled Equations (15) is derived. $div\, \dot{\boldsymbol{B}}^{\boldsymbol{e}} = div\, \boldsymbol{q} = \boldsymbol{0}$ is used to define a vector potential $\boldsymbol{A}$ such that $\boldsymbol{q} = curl\, \boldsymbol{A}$. The weak form of governing Equations is derived by taking the inner product of Equations $(15)_1$ and $(15)_3$ with an arbitrary test function. Let us consider $\delta\boldsymbol{u}$ and $\delta\boldsymbol{A}$ be arbitrary variations of $\boldsymbol{u}$ and $\boldsymbol{A}$, respectively, that satisfies the boundary conditions on $\partial V$. Taking the variational form of the $(15)_1$ and $(15)_3$, integrating by parts and using the divergence theorem yields:

$$\int_V \dot{\boldsymbol{T}}^e : grad\ \delta\boldsymbol{u}\ dV - \int_{\partial V} \dot{\boldsymbol{\tau}} \cdot \delta\boldsymbol{u}\ dS = -\rho\omega^2 \int_V \boldsymbol{u}\delta\boldsymbol{u}\ dV \tag{19}$$

$$\int_V \dot{\boldsymbol{H}}^{\boldsymbol{e}} : curl\ \delta\boldsymbol{A}\ dV = \int_{\partial V} (\dot{\boldsymbol{H}}^{\boldsymbol{e}} \times \boldsymbol{n}) \cdot \delta\boldsymbol{A}\ dS \tag{20}$$

Equations (19) and (20) are defined on $V$, the RUC's domain, and $\partial V$ represents the perimeter of the RUC. It is noted that the weak form in Equation (20) results from a stationary magnetic field condition and the non-conducting magnetoelastic medium in the absence of the surface current density on $\partial V$. Moreover, $\dot{\boldsymbol{t}} = \dot{\boldsymbol{T}}^e\boldsymbol{n} = \dot{\boldsymbol{\tau}}(\boldsymbol{x})\boldsymbol{e}^{-i\omega t}$ is the incremental surface traction normal to the boundaries of the deformed RUC, where $\boldsymbol{n}$ is the unit vector normal to boundaries in the outward direction. Additionally, $(\dot{\boldsymbol{H}}^{\boldsymbol{e}} \times \boldsymbol{n})$ represents the tangent magnetic field at boundaries. Natural boundary conditions in Equations (19) and (20) appear as work terms applied to the boundaries $\partial V$, which arise from the tangent magnetic field and the normal traction force. It is evident from Figure 24 that the boundary integrals $\int_{\partial V} \dot{\boldsymbol{\tau}} \cdot \delta\boldsymbol{u}\ dS$ in Equation (19) and $\int_{\partial V} (\dot{\boldsymbol{H}}^{\boldsymbol{e}} \times \boldsymbol{n}) \cdot \delta\boldsymbol{A}\ dS$ in Equation (20) vanish, because the normal unit vector $\boldsymbol{n}$ acts in opposite directions on the parallel boundaries of the deformed RUC. The final weak forms of the constitutive equations are:

$$\int_V \dot{\boldsymbol{T}}^e : grad\delta\boldsymbol{u}\ dV = -\rho\omega^2 \int_V \boldsymbol{u}\delta\boldsymbol{u}\ dV \tag{21}$$

$$\int_V \dot{\boldsymbol{H}}^{\boldsymbol{e}} : curl\delta\boldsymbol{A}\ dV = 0 \tag{22}$$

Weak forms in Equations (21) and (22) define the coupling constitutive behavior of MREs. The nodal values of the vector potential $\boldsymbol{A} = [A_1, A_2, A_3]$ and displacement $\boldsymbol{u} = [u_1, u_2, u_3]$ are the unknowns.

Bloch displacement boundary conditions are applied on the opposite boundaries of the deformed RUC so that $\boldsymbol{u}^{+}(\boldsymbol{x}+\boldsymbol{r}) = \boldsymbol{u}^{-}(\boldsymbol{x})\boldsymbol{e}^{\boldsymbol{i}(\boldsymbol{k}.\boldsymbol{r})}$, where $\boldsymbol{k}$ is the Bloch wave vector and $\boldsymbol{r}$ denotes the distance vector between parallel boundaries. The superscripts (+) and (–) denote the corresponding opposite boundaries; right (top) and left (bottom) in Figure 24, respectively. It should be noted that in Equation (16) the tractions on boundaries follow the Bloch relation. The only difference appears due to the fact that the tractions on the parallel boundaries are in the opposite directions. Hence, the Bloch type boundary conditions are $\dot{\boldsymbol{t}}^{+}(\boldsymbol{x}+\boldsymbol{r}) = -\dot{\boldsymbol{t}}^{-}(\boldsymbol{x})\boldsymbol{e}^{\boldsymbol{i}(\boldsymbol{k}.\boldsymbol{r})}$. The deformed RUC and the corresponding unit cell in reciprocal lattice including irreducible Brillouin zone (IBZ) are shown in Figure 24. The wave vector takes the values of $\boldsymbol{k} = k_1\boldsymbol{b_1} + k_2\boldsymbol{b_2}$ along the edges $\Gamma - \Upsilon - M - X - \Gamma$, where $0 \le k_1 \le 0.5\ and\ 0 \le k_2 \le 0.5$ are real numbers $\boldsymbol{b_1}$ and $\boldsymbol{b_2}$ are the lattice vectors of the reciprocal space [74]:

$$\boldsymbol{b_1} = \frac{2\pi(\boldsymbol{a_2}\times\boldsymbol{e})}{\|\boldsymbol{a_1}\times\boldsymbol{a_2}\|} \quad , \quad \boldsymbol{b_2} = \frac{2\pi(\boldsymbol{e}\times\boldsymbol{a_1})}{\|\boldsymbol{a_1}\times\boldsymbol{a_2}\|} \ , \qquad \text{where} \quad \boldsymbol{e} = \frac{(\boldsymbol{a_1}\times\boldsymbol{a_2})}{\|\boldsymbol{a_1}\times\boldsymbol{a_2}\|},$$

Since the boundary conditions are in complex space, thus $\boldsymbol{A}$ and $\dot{\boldsymbol{t}}$ are generally complex. It is noted that in FE solvers, once the Bloch displacement boundary conditions are applied, the traction boundary conditions are automatically satisfied. The Bloch type boundary condition is applied through the essential or Dirichlet boundary condition node in the partial differential equations (PDE) interface of the FE solver. A computational code is implemented in the FE model to investigate the incremental moduli tensors defined by Equation (17) in terms of deformation gradient and the applied magnetic induction. The magnetoelastic interaction is characterized by a stiffness matrix, which is a function of magnetic induction, stretch and wave vector.

The field equations for 2D motions are considered; $u_3 = \dot{B}_3^e = 0$. Hence, solutions which only depend on the in-plane variables $\boldsymbol{x}$ and $\boldsymbol{y}$ are sought. The FE discretization of the coupling problem results in an eigenvalue problem. Eigenfrequency study is used for investigating the eigenfrequencies, $\omega$, of the coupled equations. Eigenfrequencies of the wave equation are obtained to examine band diagrams and plot the iso-frequency contours at different levels of the applied magnetic induction. Dispersion relations are calculated sweeping the boundaries of IBZ. The reduced frequency, $\Omega = \frac{\omega\ell}{2\pi c_t}$, is plotted as a function of the wave vector where $\ell = \frac{a_1+a_2}{2}$ and $c_t = \sqrt{\mu/\rho}$.

### 3.4 Results and discussions

A compressible Neo-Hookean type energy function is proposed to extract constitutive relations and magnetoelastic moduli tensors, as follows:

$$\boldsymbol{\Psi} = \frac{\mu}{2}(\boldsymbol{I_1} - 3) - \mu\, lnJ + \frac{\lambda}{2}(lnJ)^{\boldsymbol{2}} + p\boldsymbol{I_4} + q\boldsymbol{I_5} \tag{23}$$

where $\mu = 0.25\ MPa$ and $\lambda = 10^4\mu$ are the two Lame constants of the hyperelastic material in the absence of the magnetic induction and is assumed to be constant. $\boldsymbol{I_1}$ is principal invariant of right Cauchy–Green deformation tensor and $\boldsymbol{I_4}$ and $\boldsymbol{I_5}$ are invariants dependent on the magnetic induction vector which exhibit the nonlinear coupling behavior of the material. One observes that the parameter $p$ does not affect the stress tensor, while $q$ if positive, enhances the material stiffness in the direction of the applied magnetic induction. In the contrary, $p$ provides a measure of how the magnetic field is affected by the macroscopic deformation. Magnetic permeability in vacuum, $\mu_0 = 1.256 \times 10^{-6} NA^{-2}$, $\nu = 0.499$ and $p = 2q = 2/\mu_0$ are assumed for numerical simulations which accounts for 10% by volume of iron particles in the soft elastomer [17-23]. Moreover, the dissipation is neglected.

The constitutive laws in Equations (16) are expanded as:

$$\begin{aligned}
\dot{T}^e_{11} &= \mathbb{A}^e_{1111}u_{1,1} + \mathbb{A}^e_{1112}u_{2,1} + \mathbb{A}^e_{1121}u_{1,2} + \mathbb{A}^e_{1122}u_{2,2} + \mathbb{K}^e_{111}\dot{B}^e_1 + \mathbb{K}^e_{112}\dot{B}^e_2 \\
\dot{T}^e_{12} &= \mathbb{A}^e_{1211}u_{1,1} + \mathbb{A}^e_{1212}u_{2,1} + \mathbb{A}^e_{1221}u_{1,2} + \mathbb{A}^e_{1222}u_{2,2} + \mathbb{K}^e_{121}\dot{B}^e_1 + \mathbb{K}^e_{122}\dot{B}^e_2 \\
\dot{T}^e_{21} &= \mathbb{A}^e_{2111}u_{1,1} + \mathbb{A}^e_{2112}u_{2,1} + \mathbb{A}^e_{2121}u_{1,2} + \mathbb{A}^e_{2122}u_{2,2} + \mathbb{K}^e_{211}\dot{B}^e_1 + \mathbb{K}^e_{212}\dot{B}^e_2 \\
\dot{T}^e_{22} &= \mathbb{A}^e_{2211}u_{1,1} + \mathbb{A}^e_{2212}u_{2,1} + \mathbb{A}^e_{2221}u_{1,2} + \mathbb{A}^e_{2222}u_{2,2} + \mathbb{K}^e_{221}\dot{B}^e_1 + \mathbb{K}^e_{222}\dot{B}^e_2 \\
\dot{H}^e_1 &= \mathbb{K}^e_{111}u_{1,1} + \mathbb{K}^e_{121}\left(u_{2,1} + u_{2,1}\right) + \mathbb{K}^e_{221}u_{2,2} + \mathbb{L}^e_{11}\dot{B}^e_1 + \mathbb{L}^e_{12}\dot{B}^e_2 \\
\dot{H}^e_2 &= \mathbb{K}^e_{112}u_{1,1} + \mathbb{K}^e_{122}\left(u_{2,1} + u_{2,1}\right) + \mathbb{K}^e_{222}u_{2,2} + \mathbb{L}^e_{12}\dot{B}^e_1 + \mathbb{L}^e_{22}\dot{B}^e_2
\end{aligned} \tag{24}$$

The energy function in Equation (23) is used to calculate the incremental magnetoelastic moduli tensors (Equation (17)) in order to apply in corresponding constitutive laws in Equation (24). For a uniaxial deformation gradient tensor $\boldsymbol{F} = \begin{bmatrix} \lambda_x & 0 \\ 0 & \lambda_y \end{bmatrix}$ and uniaxial magnetic induction vector $\boldsymbol{B}_l = \begin{bmatrix} 0 \\ B_y \end{bmatrix}$, substituting Equation (11) in Equation (17) and vanishing the zero terms, the corresponding moduli tensors will be:

$$
\begin{aligned}
&\mathbb{A}^e_{1112} = \mathbb{A}^e_{1211} = \mathbb{A}^e_{1121} = \mathbb{A}^e_{1222} = \mathbb{A}^e_{2111} = \mathbb{A}^e_{2122} = \mathbb{A}^e_{2212} = \mathbb{A}^e_{2221} = 0\,, \\
&\mathbb{A}^{\boldsymbol{e}}_{1111} = \frac{2\boldsymbol{\Psi}_1 b_{11}}{J} = \frac{0.25\times10^6 {\lambda_x}^2}{J}, \\
&\mathbb{A}^e_{1122} = \mathbb{A}^e_{2211} = \frac{2\boldsymbol{\Psi}_3 J^2}{J} = \frac{2.5e9}{J}, \\
&\mathbb{A}^e_{1212} = \frac{2\boldsymbol{\Psi}_1 b_{11}}{J} = \frac{0.25\times10^6 {\lambda_x}^2}{J}, \\
&\mathbb{A}^e_{1221} = \mathbb{A}^e_{2112} = -\frac{2\boldsymbol{I}_3\boldsymbol{\Psi}_3}{J} = -\frac{2.5e9}{J}, \\
&\mathbb{A}^e_{2121} = \frac{2\boldsymbol{\Psi}_1 b_{22} + \boldsymbol{I}_3\boldsymbol{\Psi}_5 B_2 B_2}{J} = \frac{0.25\times10^6 {\lambda_y}^2 + 0.8\times10^6 (B_y)^2 J^2}{J}, \\
&\mathbb{A}^e_{2222} = \frac{2\Psi_1 b_{22} + I_3\Psi_5 B_2 B_2}{J} = \frac{0.25\times10^6 {\lambda_y}^2 + 0.8\times10^6 (B_y)^2 J^2}{J}, \\
&\mathbb{K}^e_{111} = \mathbb{K}^e_{112} = \mathbb{K}^e_{221} = \mathbb{K}^e_{122} = \mathbb{K}^e_{212} = 0\,, \\
&\mathbb{K}^e_{121} = \mathbb{K}^e_{211} = 1.6\times10^6 J B_y, \\
&\mathbb{K}^e_{222} = 3.2\times10^6 J B_y, \\
&\mathbb{L}^e_{12} = \mathbb{L}^e_{21} = 0 \\
&\mathbb{L}^e_{11} = J(3.2\times10^6 {\lambda_y}^2 + 1.6\times10^6), \\
&\mathbb{L}^e_{22} = J(3.2\times10^6 {\lambda_x}^2 + 1.6\times10^6)
\end{aligned}
\tag{25}
$$

where $\boldsymbol{b} = \boldsymbol{F}\boldsymbol{F}^{\boldsymbol{T}} = \begin{bmatrix} {\lambda_x}^2 & 0 \\ 0 & {\lambda_y}^2 \end{bmatrix}$ is the left Cauchy-green deformation tensor.

To illustrate the effect of the applied magneto-mechanical loadings, the PnC is first deformed through the macroscopic deformation tensor, $\boldsymbol{F}$, and then is subjected to an external magnetic field. A plane strain uniaxial compression is applied to deform the RUC. The deformation gradient tensor is $\boldsymbol{F} = [\lambda_x \;\; 0; 0 \;\; \lambda_y]$, where $\lambda_x$ and $\lambda_y$ are the applied stretches in $\boldsymbol{x}$ and $\boldsymbol{y}$ directions, respectively. Since the deformation occurs before the magnetic field is applied, $\lambda_y$ is determined from the equilibrium condition, such that the pure mechanical stress component in $\boldsymbol{y}$ direction vanishes. $\lambda_x$ is assumed to be 0.9 for the numerical simulations, hence $\lambda_y = 1.11$.

A nonlinear stationary study is performed on the RUC to capture the large deformation effect. The deformation boundary conditions are applied on all boundaries of the RUC. Small distortions are implemented to perturb the initial geometry and obtain the post-bifurcation state of the RUC. The deformed RUC is extracted and imported in a new model for the coupled wave propagation analysis. No pre-stress exists in the new model. The RUC and the deformed RUC for uniaxially compressive stretch $\lambda_x = 0.9$ are shown in Figures 24**(a)** and 24**(b)**, respectively.

Once the deformation is applied, the RUC is subjected to different levels of magnetic inductions. The corresponding band diagrams are shown in Figure 25. To illustrate the effect of the applied magnetic induction, the first 30 in-plane modes are computed and plotted in band diagrams. As can be seen in Figure 24, the range of the frequency spectrum of the band structure expands as the magnitude of the applied magnetic induction increases.

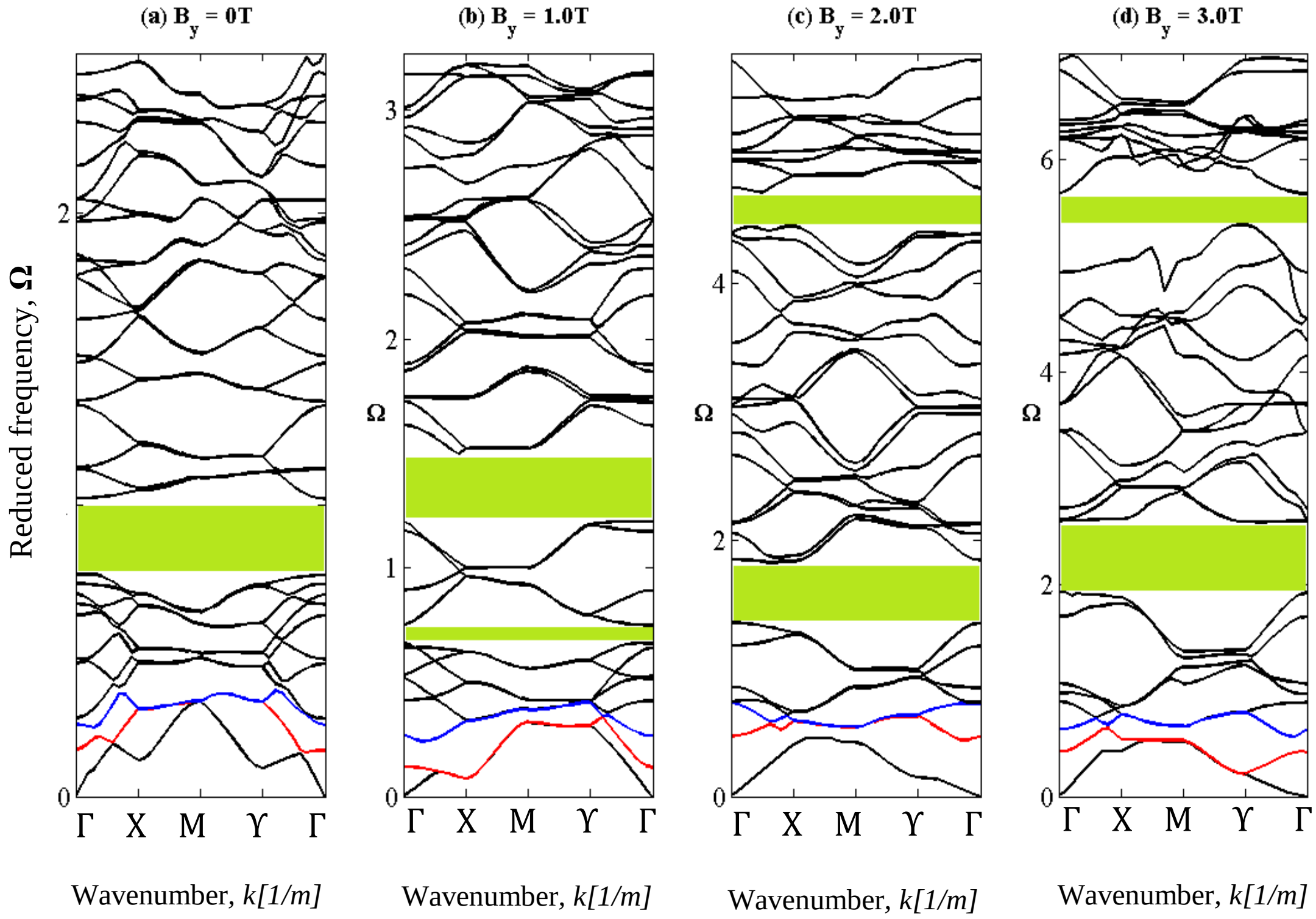

**Figure 25**. In-plane band diagrams for uniaxially compressive stretch $\lambda_x = 0.9$ at **(a)** No magnetic induction, **(b)** 1.0T, **(c)** 2.0T, and **(d)** 3.0T unidirectionally applied magnetic induction. PBGs are shown in shaded regions. The second and third modes represented by red and blue bands, respectively, are selected for directionality analysis.

The range of the reduced frequency spectrum increases from $\Omega = 0 - 2.55$ in the absence of magnetic stimulus, to $\Omega = 0 - 7.00$ for the 3.0T applied magnetic induction. Figure 25**(a)** shows the band diagram in the absence of magnetic induction. A Phononic bandgap (PBG) is captured at the range of $\Omega = 0.76 - 1.05$. No considerable change in PBG is observed at 0.25T applied magnetic induction. As the magnetic induction increases to 1.0T, 2.0T and 3.0T, the first PBG transforms to $\Omega = 1.22 - 1.58$, $\Omega =$ $1.35 - 1.85$ and $\Omega = 1.90 - 2.60$, respectively (Figure 25**(b-d)**). Upon the application of the 1.0T magnetic induction a new PBG is created at $\Omega = 0.65 - 0.75$, but vanishes at 2.0T and 3.0T applied magnetic inductions. Also, a new PBG is generated at $\Omega = 4.44 - 4.72$ at 2.0T and widens and shifted to $\Omega = 5.35 - 5.74$ as the magnetic induction increases to 3.0T (Figure 25**(c)** and **(d)**). During the application of the magnetic induction, the first and second PBGs orientation is widened and considerably shifted after about $B_y = 2.0T$. Additionally, band diagram investigation on the effect of magnetic induction in the undeformed structure demonstrates that the first PBGs in the band diagram appear due to the applied deformation while the second PBGs appear due to the applied magnetic induction. In Figure 25, second and third modes, depicted by red and blue lines, respectively, are selected to study the directional behavior of the structure.

While the band diagram exhibits the compact representation of the consecutive eigenmodes, it cannot illustrate the directionality of the wave propagating in the structure. Dispersion relations are also presented in the form of iso-frequency contours to fully illustrate the wave propagation characteristics of the PnCs. The iso-frequency contours identify the frequency of free wave motion of the PnC in the $(k_x, k_y)$ plane. The solution $\omega = \omega(k_x, k_y)$ of the magnetoelastic eigenvalue problem for all combinations of $k_x = k_1\boldsymbol{b_1}$ and $k_y = k_2\boldsymbol{b_2}$ defines the dispersion surfaces of the structure. The symmetry of the phase constant surfaces allows to limit the iso-frequency plots in the first Brillouin zone.

Preferential directions of the free wave propagation, phase velocity and group velocity can be derived from iso-frequency contours of the phase constant surfaces. By definition, phase velocity is related to the speed of the wave crests while the group velocity represents the speed of the wave envelope and lies along the normal to the corresponding iso-frequency contour in the $(k_x, k_y)$ plane. Phase velocity and group velocity are defined by $\boldsymbol{c_p} = \frac{\omega}{k^2}\boldsymbol{k}$ and $\boldsymbol{c_g} = \frac{\partial\omega}{\partial\boldsymbol{k}}$, respectively, where $k$ is the magnitude of the wave vector [134-135]. The group velocity determines the direction of the wave propagation, while directionality represents the flow of energy in the structure. Directionality of PnCs also provides information about the directions in the structure that wave does not propagate. These unique features of PnCs depend on the unit cell's geometry, material properties and frequency and provide useful information on the wave propagation in the periodic structures. Directionality can be considered as a measure of stiffness of a PnC structure. For instance, in the vicinity of a long wavelength limit, the anisotropy of a PnC can be expressed in terms of the phase and group velocity diagrams [135-136].

Figure 26 represents the effect of uniaxial compression, $\lambda_x = 0.9$ and different magnitudes of magnetic induction on dispersion relations in terms of normalized iso-frequency contours of the second eigenmode shown by red line in the band diagrams. It is necessary to mention that isofrequency contours are represented for 0.25T to demonstrate the transition in the contours for the loading from 0 to 1.0T magnetic induction. Since the isofrequency plots are similar for the applied magnetic inductions beyond 2.0T, the results for 3.0T are not presented in the paper. The dispersion relations in case of no magnetic field are plotted in Figure 26**(a)** and shows that wave propagates at different directions with different speeds. In this case, qualitative behavior of the contour lines verifies the results documented in the literature [97].

By increasing the magnetic loading, the dispersive behavior of the contour lines changes and tends to become parallel to the $k_x$ axis, particularly at higher frequencies. The pattern of contour lines does not change qualitatively when the magnitude of the magnetic induction vector increases beyond 2.0T. It is interesting to note that frequency spectrum covers a higher range by increasing the magnetic excitation. In Figure 26, arrows represent the direction and magnitude of the group velocity vectors. Group velocity vector fields confirm that the structure shows a directional behavior at higher levels of magnetic loading.

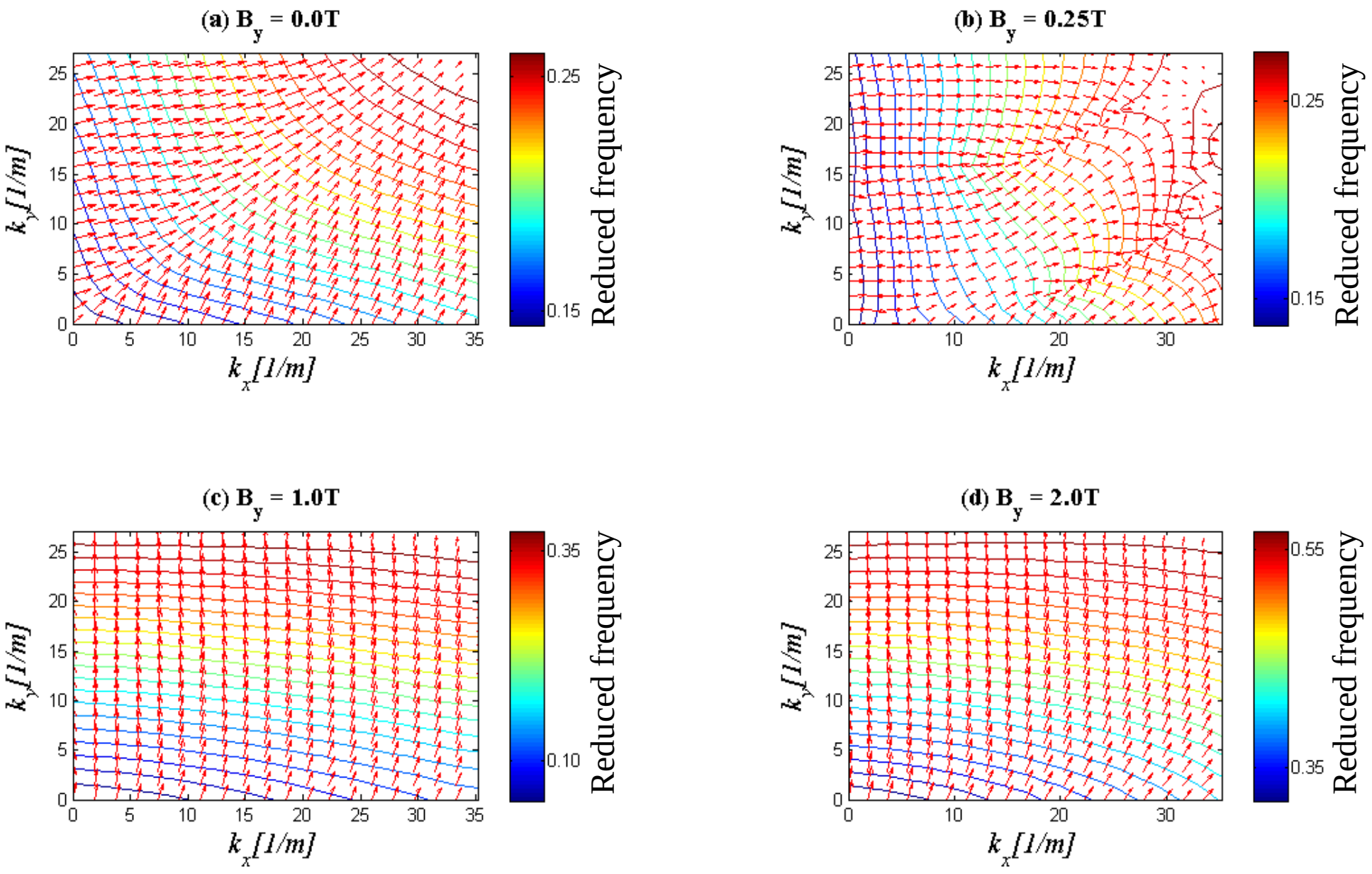

**Figure 26**. Normalized iso-frequency contours associated with the second modes for uniaxially compressive stretch $\boldsymbol{\lambda_x = 0.9}$ at **(a)** No magnetic induction, **(b)** 0.25T, **(c)** 1.0T, and **(d)** 2.0T unidirectionally applied magnetic induction. The contours are associated with the red band in Figure 25.

The iso-frequency contour plots of the third modes (represented by blue line in the band diagrams) are shown in Figure 27 for uniaxial compression, $\lambda_x = 0.9$ and different levels of magnetic induction. A totally different pattern is observed for the third modes. At 1.0T and 2.0T wave propagation has a directional behavior only at low frequencies.

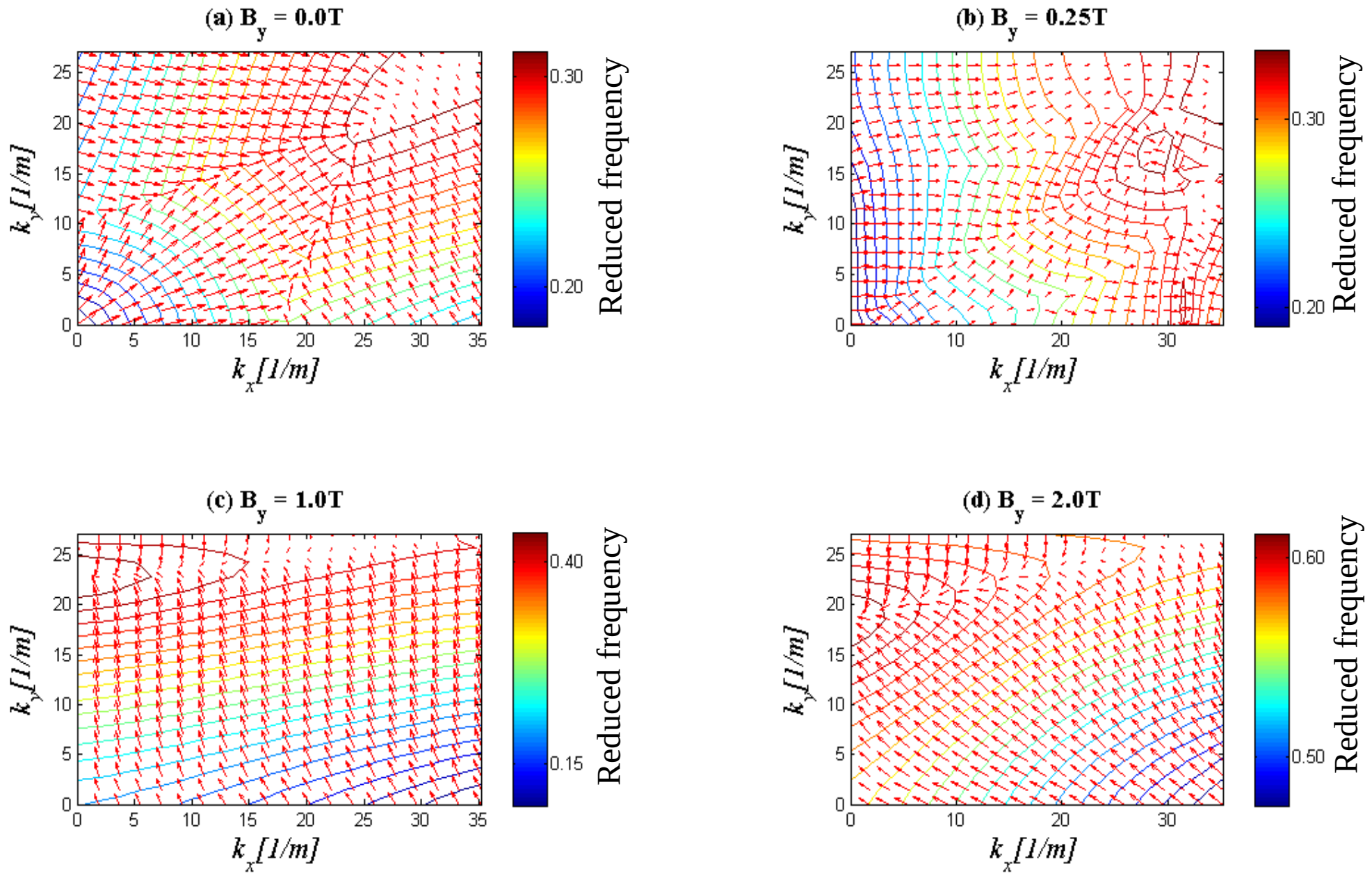

**Figure 27**. Normalized iso-frequency contours associated with the third modes for uniaxially compressive stretch $\boldsymbol{\lambda_x = 0.9}$ at **(a)** No magnetic induction, **(b)** 0.25T, **(c)** 1.0T, and **(d)** 2.0T unidirectionally applied magnetic induction. The contours are associated with the blue band in Figure 25.

Directionality, quantified as $\boldsymbol{D} = \frac{\pi}{2} + tan^{-1}\frac{\partial k_y}{\partial k_x}$ at each frequency, is represented by polar plots in Figures 28 and 29 associated with results presented in Figures 26 and 27 respectively. Frequency and directionality is plotted in radial and tangential directions, respectively.

Figure 28**(a-b)** illustrates that the frequency-dependent directionality of the structure covers a large angular range. Upon the application of the 0.25T magnetic induction, the directional behavior of the structure is initiated. At high frequencies an isotropic pattern is observed. Figure 28**(c-d)** exhibits directional behavior of the structure due to the increasing magnetic excitation. The directionality spans a narrow angular range demonstrating a higher speed of wave in the direction of the applied magnetic field.

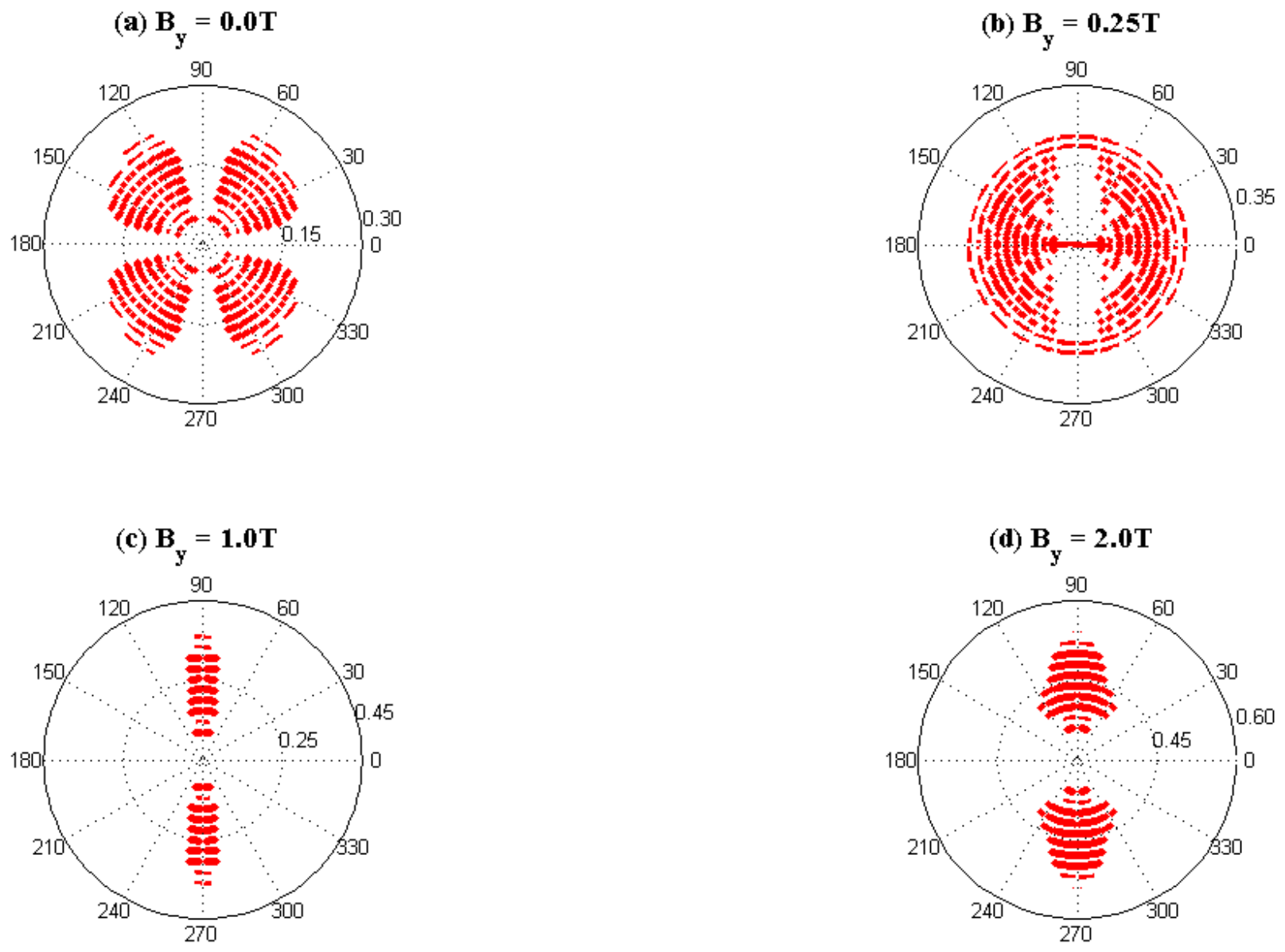

**Figure 28**. Directionality plots associated with the second modes for uniaxially compressive stretch $\boldsymbol{\lambda_x = 0.9}$ at **(a)** No magnetic induction, **(b)** 0.25T, **(c)** 1.0T, and **(d)** 2.0T unidirectionally applied magnetic induction. Frequency and direction is plotted in radial and angular directions, respectively.

Figure 29 shows the directionality plot associated with the third modes at different levels of the magnetic induction. At 1.0T, structure shows directional behavior only for low frequencies. No strong preference in directions is captured for third modes by increasing magnetic induction beyond 1.0T.

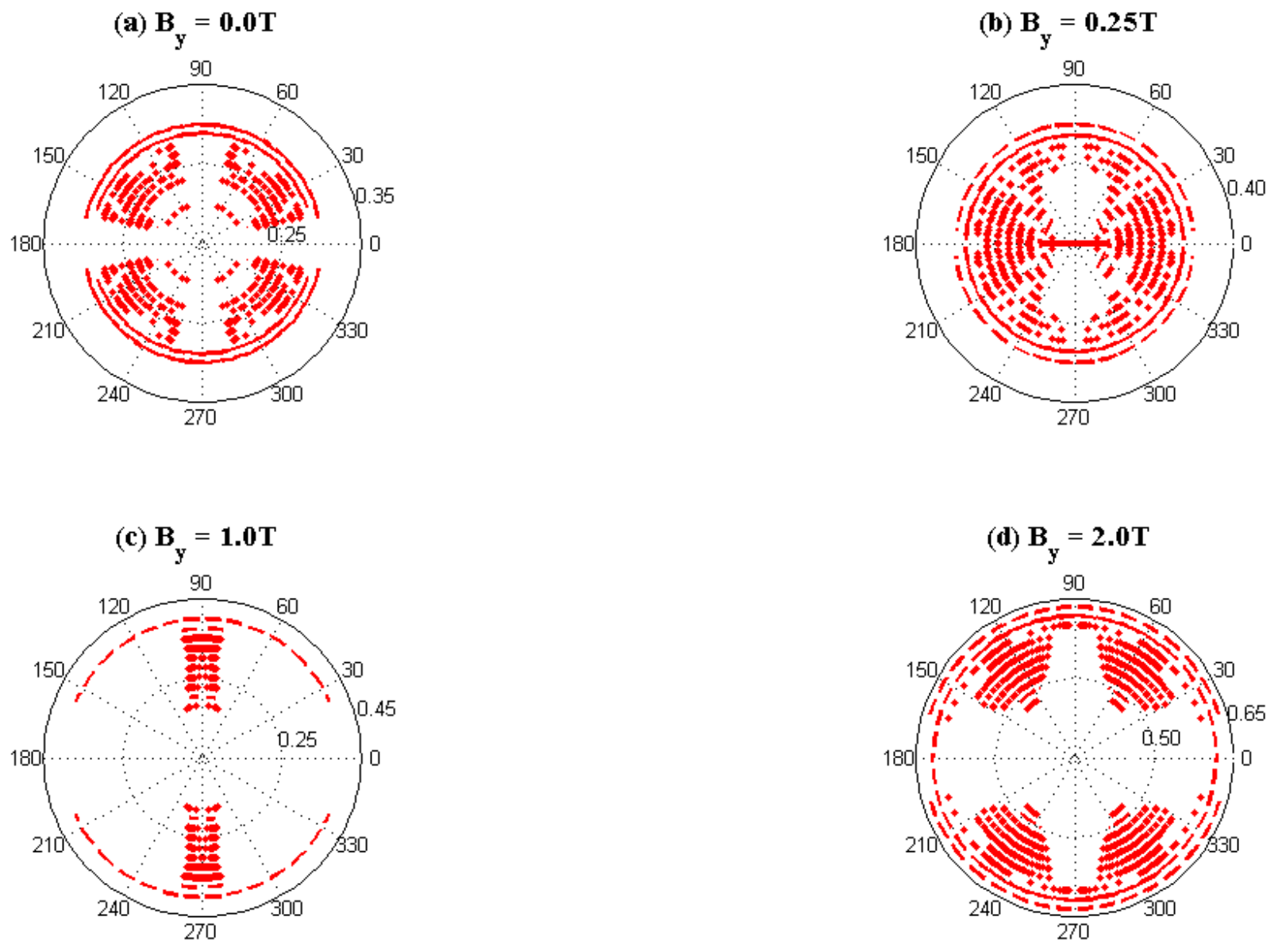

**Figure 29.** Directionality plots associated with the third modes for uniaxially compressive stretch $\boldsymbol{\lambda_x = 0.9}$ at **(a)** No magnetic induction, **(b)** 0.25T, **(c)** 1.0T, and **(d)** 2.0T unidirectionally applied magnetic induction Frequency and direction is plotted in radial and angular directions, respectively.

Next, we study the effect of the macroscopic deformation loading case on the band diagrams. Different macroscopic deformation tensors generate different patterns in the porous structure. Here, a Mooney-Rivlin type energy function is used to extract constitutive relations and magnetoelastic moduli tensors, as follows:

$$\Psi = \frac{\mu}{4}[(1+\nu)(I_1-3)+(1-\nu)(I_2-3)]+pI_4+qI_5, \tag{26}$$

where $\mu = 25e4\ \mathrm{N/m^2}$ is the shear modulus in the absence of the magnetic induction, and assumed to be constant. $I_1$ and $I_2$ are principal invariants of right Cauchy–Green deformation tensor and $I_4$ and $I_5$ are invariants dependent on the magnetic induction vector. Here, it is assumed that $\mu_0 = 1.256e-6\mathrm{NA^{-2}}$ for magnetic permeability in vacuum, $\nu = 0.499$ and $p = 2q = 2/\mu_0$ for numerical simulations [20-22].

To illustrate the effect of the applied magneto-mechanical loadings, the PnC is first deformed through the macroscopic deformation tensor $F$, and then is subjected to an external magnetic field. Two different cases of deformation are considered:

- A plane strain uniaxial compression: In this case, the RUC is compressed in $x_1$ direction. The deformation gradient tensor is defined as: $F = [\lambda_1\ \ 0; 0\ \ \lambda_2]$. Where $\lambda_1$ and $\lambda_2$ are the applied stretches in $x_1$ and $x_2$ directions, respectively. Since the deformation is occurred before the magnetic field applies, the $\lambda_2$ is determined from the equilibrium condition so that the pure mechanical stress component in $x_2$ direction vanishes. $\lambda_1 = 0.9$ is assumed for the numerical simulations, hence $\lambda_2 = 1.11$.
- A plane strain equally biaxial compression: In this case, the RUC is equally compressed in $x_1$ and $x_2$ directions. The deformation gradient tensor is defined as: $F = [\lambda\ \ 0; 0\ \ \lambda]$, where $\lambda$ is the applied stretch, and considered to be 0.9 for numerical simulations.

The RUC and the deformed RUC for uniaxially compressive stretch $\lambda_1 = 0.9$ are shown in Figures 30(a) and 30(b), respectively. For the biaxial compression, different RUC is selected, due to convergence issue. The corresponding RUC and deformed RUC are shown in Figures 30(c) and 30(d), respectively.

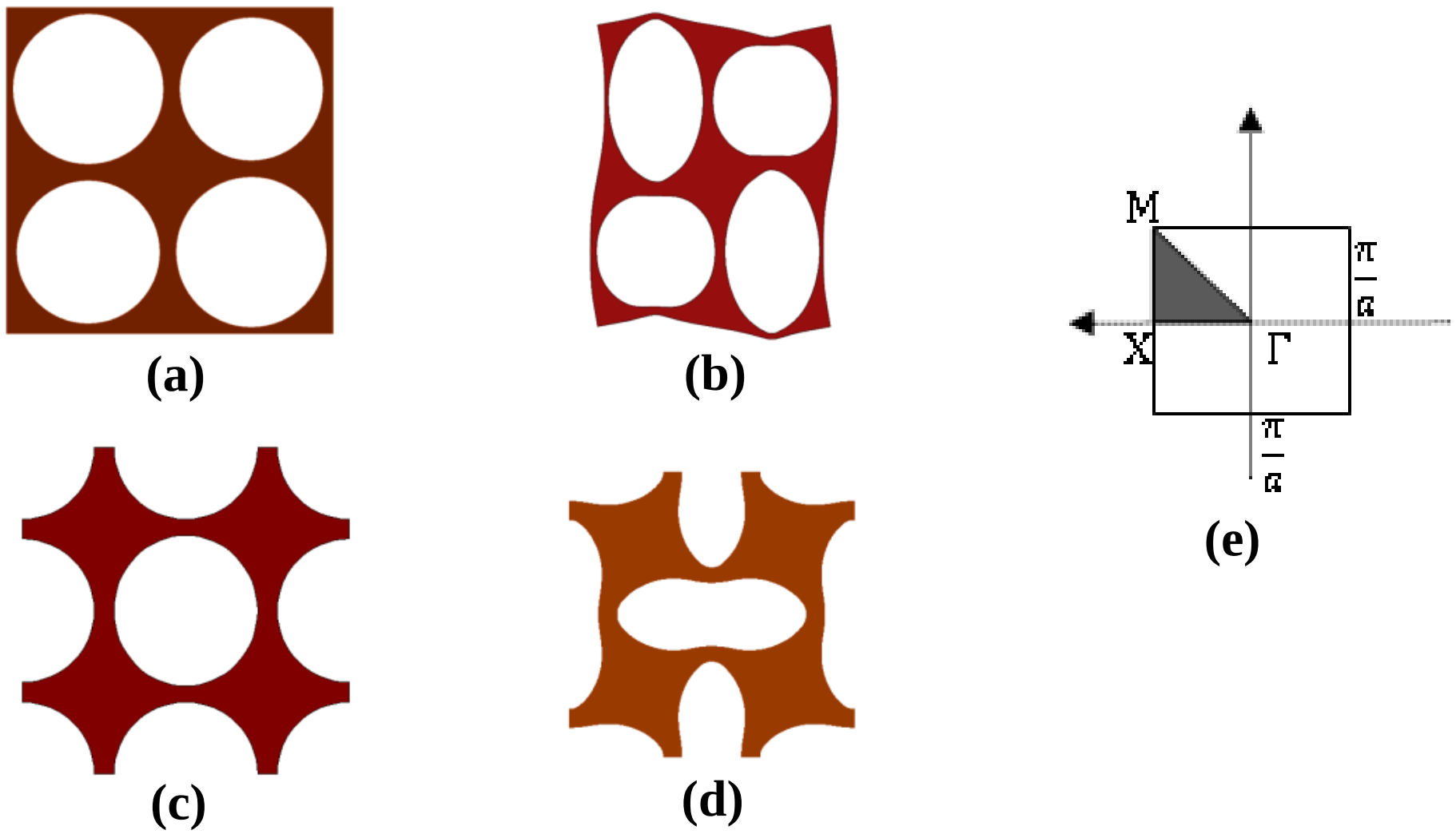

**Figure 30.** **(a)** The RUC selected for the uniaxial compression, **(b)** the corresponding deformed geometry for $\lambda_1 = 0.9$ , **(c)** The RUC selected for equally biaxial compression **(d)** the corresponding deformed geometry for $\lambda = 0.9$, **(e)** reciprocal lattice's unit cell selected for the wave propagation study. Irreducible Brilluine zone is shown in the region bounded by $\Gamma - X - M - \Gamma$.

Once the deformation is applied, the RUC is subjected to different levels of magnetic inductions. Figure 31 reports the evolution of PBGs with the applied magnetic induction for different types of macroscopic deformation.

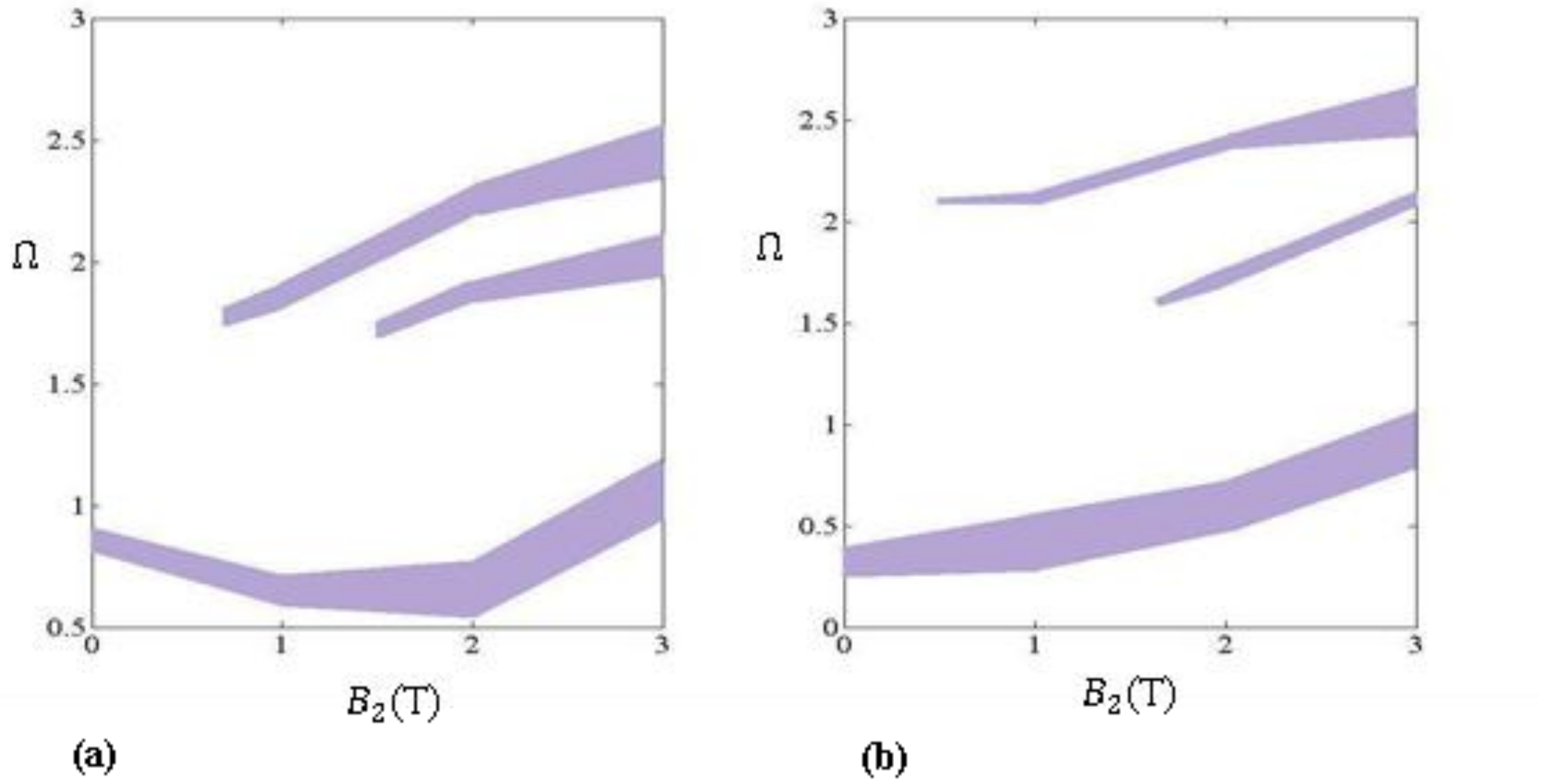

**Figure 31.** PBG vs. applied magnetic induction in y direction, for the square array PnC in: **(a)** uniaxially compressive stretch $\boldsymbol{\lambda_1 = 0.9}$, and **(b)** in biaxially compressive stretch $\boldsymbol{\lambda = 0.9}$.

This plot is resulted from the band diagram simulation at different levels of the magnetic induction. During the application of the magnetic induction on the uniaxially compressed structure (Figure 31(a)), the first PBG orientation is widened and considerably shifted after about $B_2 = 2.0T$. The width of the second and third PBGs remains fairly constant with increasing the magnetic induction up to 2.0T. The three PBGs gradually widens when the magnetic field reaches to $B_2 = 2.0T$. The PBG diagram obeys a fairly similar trend for the case of biaxial compression (Figure 31(b)). PBGs gradually widen and shift to higher frequencies during the increase of the magnetic induction.

### 3.5 Summary and conclusion

In pursuing of the smart design for isolators and frequency filters, a tunable PnC design is illustrated through a porous pattern of the soft magnetoelastic PnC. In summary, the ability of a periodic magnetoelastic structure to be used as a tunable PnC is demonstrated. A buckling analysis is carried out to demonstrate the change of periodicity of the structure due to the applied deformation. When the magneto-mechanical loading is applied, the transformation of eigenmodes occurs due to both the geometric pattern transformation and the material property changes. Results reveal that the designed PnC has the potential to control and switch the PBGs with the applied macroscopic deformation and magnetic stimulus. Upon the application of the magnetic field, new PBGs is created and the position and width of the PBGs is altered. Interestingly, the band diagram investigations reveal that by increasing the magnetic induction the frequency spectrum of the bands widens and PBGs shift to higher frequencies. Strong preferential directions in wave propagation are observed by increasing the magnetic induction, particularly for the second modes. Results confirm that the considered PnC possesses highly directional dispersive phononic characteristics compare to the previously studied mechanically tunable PnC [97]. The benefit of this approach is that it combines the effect of magnetic induction and deformation based pattern change in the microstructure and provides an added degree of freedom in the control of the dynamic response of the structure. However, 1.0T magnetic induction is required to capture the desired band diagram and directionality in the deformed structure, which is a high value, but achievable.  Results demonstrate the potential of the proposed PnC to be utilized in wave propagation systems, such as wave filters, beam steering and waveguides.

# CHAPTER FOUR

# A Thermally Tunable Phononic Crystal

A thermally tunable PnC is designed and analyzed through analytical and FEM simulations. Bimaterial ligaments composed of two strips with contrast in their thermal expansion coefficient are employed to design local resonators inside a periodic structure. The thermally induced large deformations are utilized to exploit pattern change in the structure to control elastic wave propagation. Once the temperature difference is removed the structure tends to return to the initial state providing opportunities to be used as thermally tunable acoustic switches and filters.

## 4.1 Modeling

Figure 32(a) shows the geometry of the periodic structure considered in this study. The unit cell is a resonator consisting of a square core with four identical bimaterial ligaments. The frame and core are modeled as elastic materials with negligible thermal expansion coefficients. Once the temperature difference is applied on the structure the bimaterial ligaments are deformed and generate pattern change in the periodic structure, as is shown in Figure 32(b).

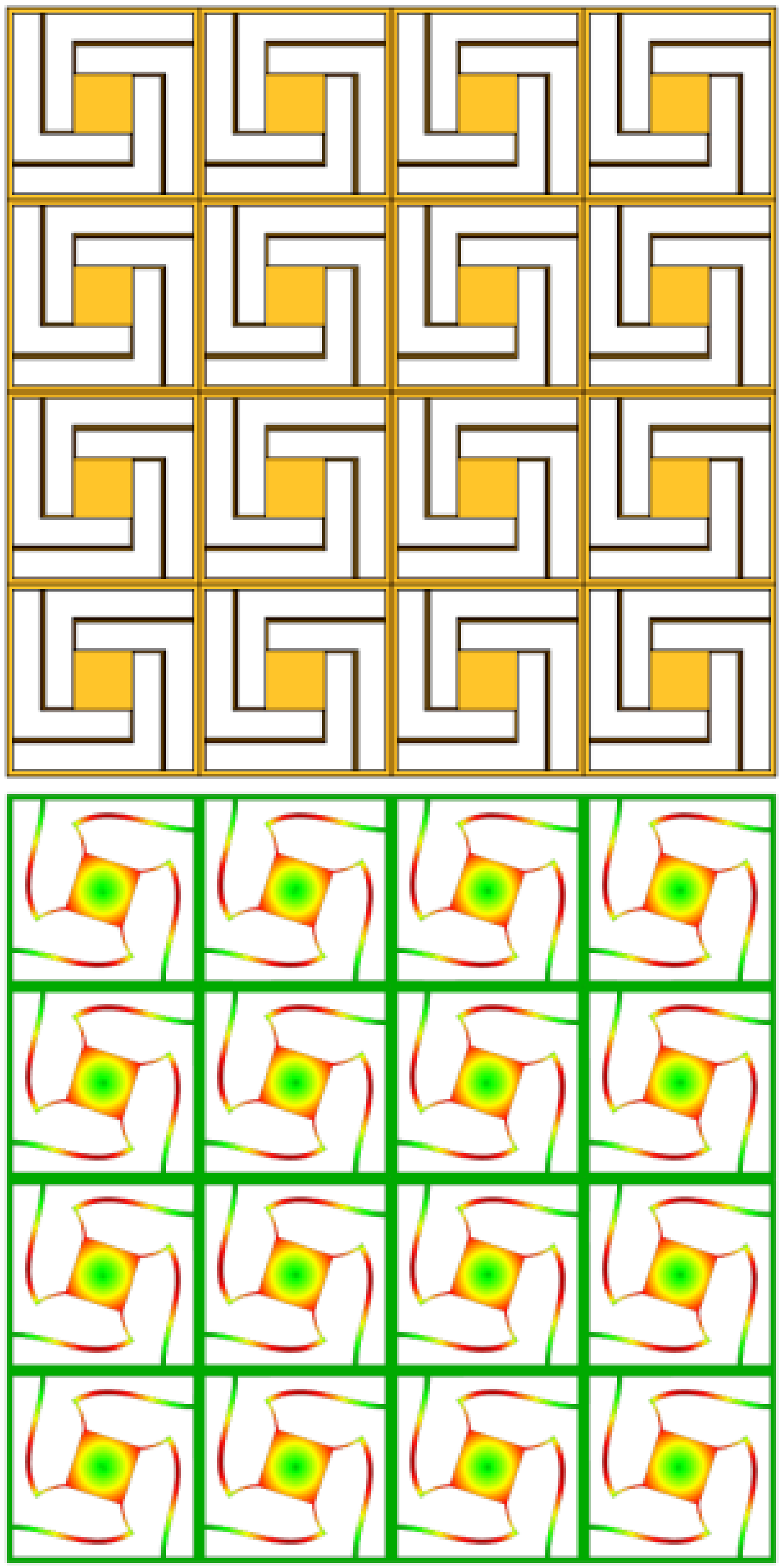

(a)

(b)

**Figure 32**. A finite sample of the proposed periodic structure at **(a)** undeformed state, **(b)** deformed by thermal actuation, $\Delta \boldsymbol{T} = \boldsymbol{200K}$.

Ligaments are modeled as nearly incompressible neo-Hookean materials. When the thermal expansion takes place, a volume change occurs in the structure. The total deformation gradient tensor, $\boldsymbol{F}$ can be decomposed into an elastic, $\boldsymbol{F_e}$ and a thermal, $\boldsymbol{F_{th}}$ gradient tensors; $\boldsymbol{F} = \boldsymbol{F_e}\boldsymbol{F_{th}}$. The total volume ratio is related to the mechanical and thermal volume ratios as; $J = J_e J_{th}$ or $det\boldsymbol{F} = det\boldsymbol{F_e} det\boldsymbol{F_{th}}$.

The thermal strain appears as $\varepsilon_{th} = \alpha(T - T_{ref})$, where $\alpha$ is the thermal expansion coefficient and $T$ and $T_{ref}$ are the current and reference temperatures, respectively. For isotropic materials, the thermal gradient tensor is a diagonal matrix; $\boldsymbol{F_{th}} = \lambda_{th}\boldsymbol{I}$ . Hence, $J_{th} = \lambda_{th}^3$. The hyperelastic model is characterized by a nonlinear energy density function [136-137]:

$$W = \frac{\mu}{2}(\boldsymbol{I_1} - 3) - \mu \log J_e + \frac{\kappa}{2}(\log J_e)^2 \qquad (1)$$

Where $\mu$ and $\kappa$ are the shear and bulk moduli, respectively, and $\boldsymbol{I_1} = \boldsymbol{tr}(\boldsymbol{C_e})$ is the first invariant of the right Cauchy-Green deformation tensor $\boldsymbol{C_e} = \boldsymbol{F_e^T}\boldsymbol{F_e}$. Thus, the first Piola-Kirchhoff stress tensor, $\boldsymbol{P} = \frac{\partial W}{\partial \boldsymbol{F}}$ is:

$$\boldsymbol{P} = \mu\boldsymbol{F_e} + (\kappa J_e(J_e - 1) - \mu)\boldsymbol{F_e^{-T}} \qquad (2)$$

where $\boldsymbol{F_e^{-T}}$ is the transpose matrix of the inverse of $\boldsymbol{F_e}$. We note that $\boldsymbol{F_e} = \boldsymbol{F}\boldsymbol{F_{th}^{-1}}$ and $J_e = \frac{J}{J_{th}}$. Equation governing the incremental motions superimposed on pre-deformed structures in Lagrangian coordinates is:

$$\nabla . \dot{\boldsymbol{P}} = \rho \frac{\partial^2 \dot{x}}{\partial t^2} \qquad (3)$$

where $\dot{\boldsymbol{x}}$ is the incremental displacemet, $\dot{\boldsymbol{P}}$ is the incremental first Piola-Kirchhoff stress tensor, and $\rho$ is the density of the material. The increment of first Piola-Kirchhoff stress tensor, $\dot{\boldsymbol{P}} = \mathbb{L} : \dot{\boldsymbol{F}}$ is a function of the incremental deformation gradient tensor, $\dot{\boldsymbol{F}} = \frac{\partial \dot{x}}{\partial \boldsymbol{X}}$ . The incremental moduli tensor is a fourth-order tensor defined by $\mathbb{L} = \frac{\partial^2 W}{\partial \boldsymbol{F} \partial \boldsymbol{F}}$ .

A solution of the wave equation in the form of plane wave $\dot{\boldsymbol{x}}(\boldsymbol{X}, t) = \dot{\mathbb{x}}(\boldsymbol{X}) e^{-i\omega t}$ is sought, where $\dot{\mathbb{x}}$ is the amplitude vector and $\omega$ is the angular frequency. Thus, the stress can be written in the following form:

$$\dot{\boldsymbol{P}} = \dot{\mathbb{P}} e^{-i\omega t} \tag{4}$$

Therefore, the equation of motion is an eigenvalue problem, as follows:

$$\nabla . \dot{\mathbb{P}} + \rho \omega^2 \dot{\mathbb{x}} = 0 \tag{5}$$

where $\omega$ is the eigenfrequency of the system. Wave propagation in PnCs is investigated through the application of Bloch type boundary conditions on parallel boundaries of the unit cell; the smallest repetitive structural element of the structure. The primitive unit cell and the deformed unit cell are shown in Figure 33. Elastic modulus and Poisson's ratio are input in the model as $E = 7MPa$ and $\upsilon = 0.25$ and $E = 14GPa$ and $\upsilon = 0.25$ for frame and square core. $\lambda = 2GPa$ and $\mu = 10MPa$ are used as the Lame constants for the strip and $\lambda = 0.2GPa$ and $\mu = 1MPa$ for the L-shaped ligament. $\alpha_1 = 25e - 4$

and $\alpha_2 = 1e - 4$ are presumed for thermal expansion coefficient of the strip and L-shaped ligament. The unit cell size is $160mm$ and the size of the core square is $50mm$.

A 2D propagation of infinitesimal harmonic plane waves is considered in a periodic structure subjected to a pre-existing homogeneous deformation gradient tensor. PnCs are characterized by a unit cell that is defined through direct lattice vectors, $a_1$ and $a_2$. These vectors are the periodicity of the lattice in $\boldsymbol{x_1}$ and $\boldsymbol{x_2}$ directions. $a_1 = a_2 = 160mm$ are used as the lattice vectors in the study.

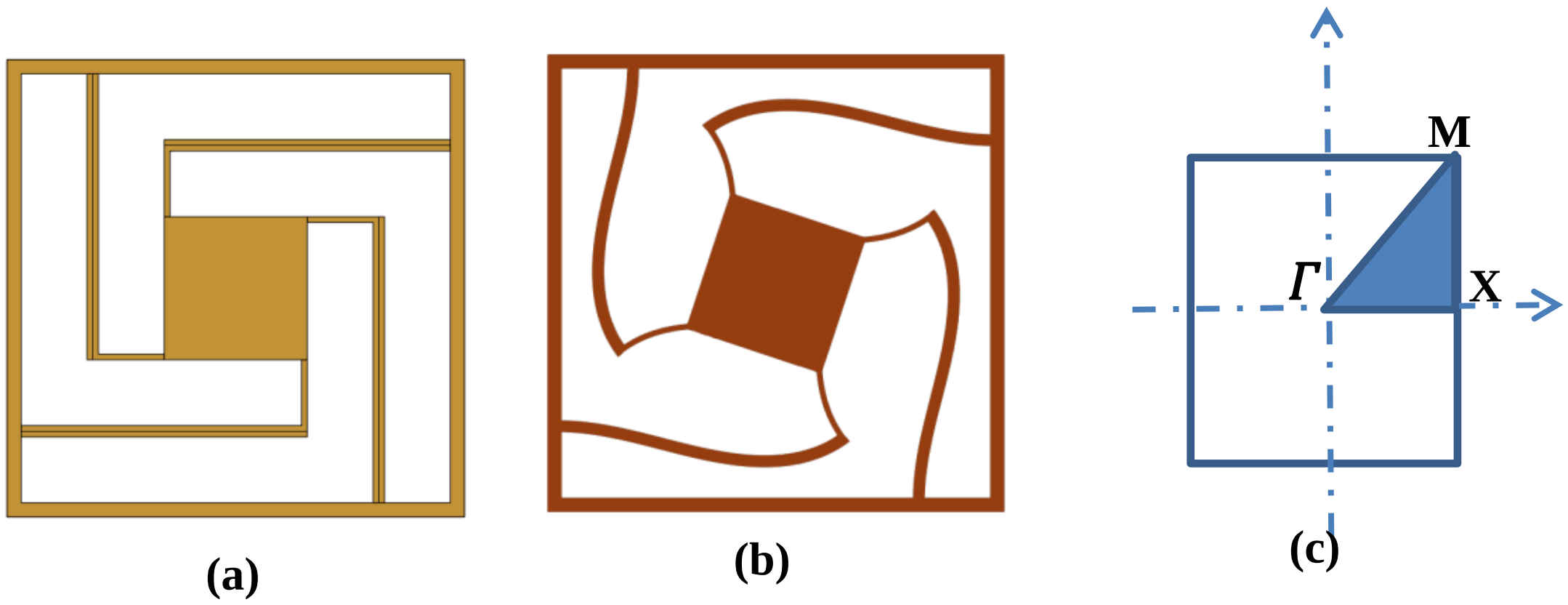

**Figure 33. (a)** The primitive unit cell, **(b)** deformed unit cell, under $\boldsymbol{\Delta T} = \boldsymbol{200K}$ and **(c)** First and irreducible Brillouin zone chosen for the wave propagation analysis.

### 4.2 Results and discussions

The deformed unit cell is shown in Figure 33(b) for $\Delta T = 200K$. Bloch type displacement boundary conditions are applied on the opposite boundaries of the deformed unit cell so that $\dot{\mathrm{x}}^{+}(\boldsymbol{X} + \boldsymbol{r}) = \dot{\mathrm{x}}^{-}(\boldsymbol{X})e^{i(\boldsymbol{k}.\boldsymbol{r})}$, where $\boldsymbol{k}$ is the Bloch wave vector and $\boldsymbol{r}$ denotes the distance vector between parallel boundaries. The superscripts (+) and (–) denote the corresponding opposite boundaries; right (top) and left (bottom) in Figure 33(b), respectively. The band diagram and mode shapes for undeformed structure are shown in Figure 34. 40 modes are shown in the diagram for frequencies larger than $1.5e4\ Hz$.

The band diagram shows narrow gaps of frequencies at $\omega = 1.88e4$, $\omega = 2.48e4$, $\omega = 2.51e4$, $\omega = 3.01e4$ , $\omega = 3.47e4$ and $\omega = 4.51e4$. Psudo-gaps in $\Gamma - \mathrm{X}$, $\mathrm{X} - \mathrm{M}$ and $\mathrm{M} - \Gamma$ directions are shown separately in diagrams. The structure shows partial gaps in the specific directions of symmetries, while complete band-gaps cover very narrow range of frequencies. The band diagram and mode shapes for the deformed structure are shown in Figure 35. 40 modes are shown in the diagram for frequencies larger than $1.5e4\ Hz$. The band diagram shows narrow gaps of frequency at $\omega = 2.49e4$, $\omega = 4.81e4$ and $\omega = 4.89e4$. Psudo-gaps in $\Gamma - \mathrm{X}$, $\mathrm{X} - \mathrm{M}$ and $\mathrm{M} - \Gamma$ directions are also shown in diagrams. The structure shows partial gaps in the specific directions of symmetries, while complete band-gaps occur at very narrow range of frequencies.

Comparing the band diagram results in the undeformed and deformed structure demonstrates that the band-gaps are suppressed by the applied temperature difference. Moreover, the position and width of psudo-gaps in specific directions are changed by the applied temperature difference. Figures 34(b) and 35(b) compare the effect of temperature difference on the mode-shapes of the structure.

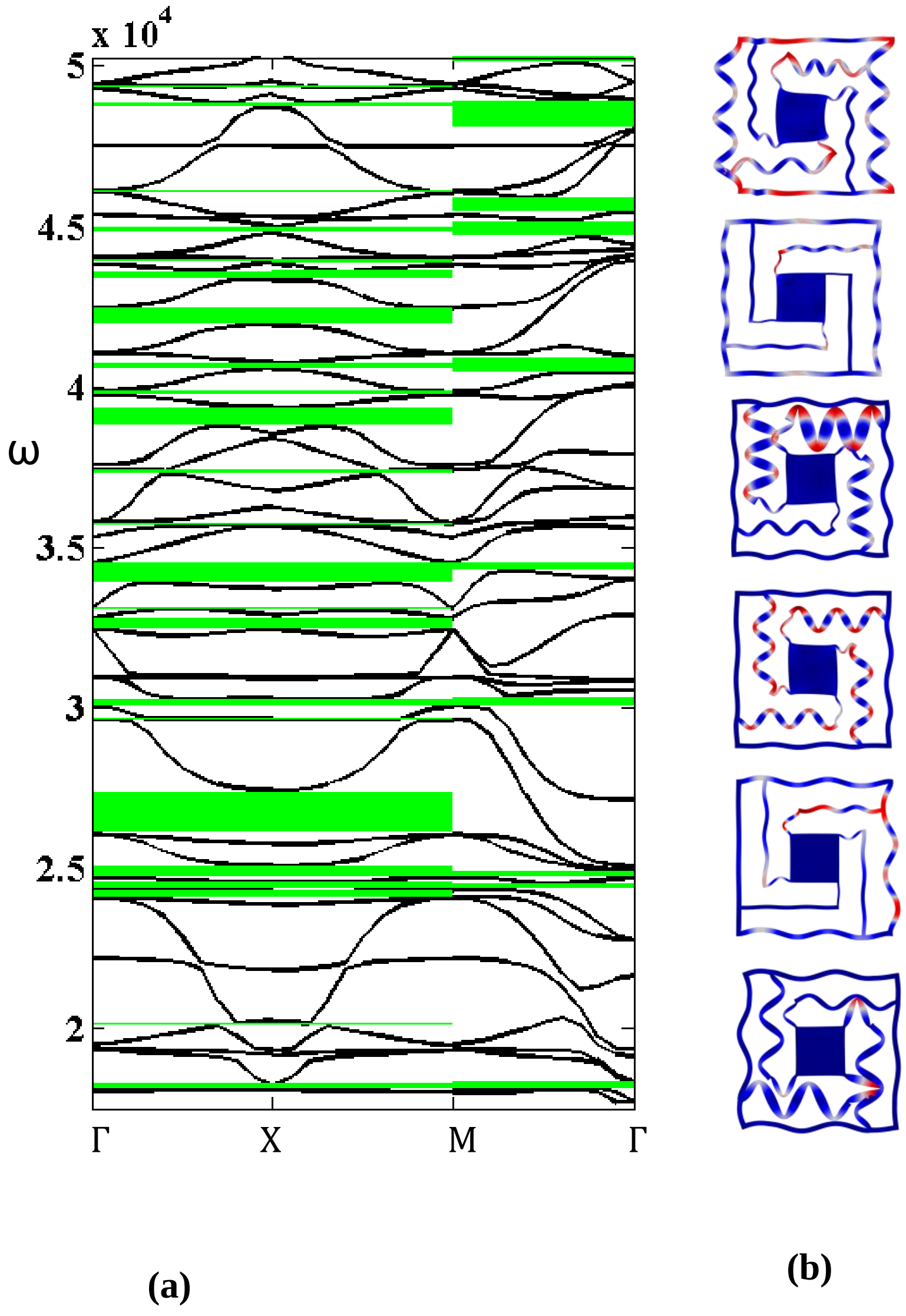

**Figure 34. (a)** Band diagram of undeformed structure and **(b)** the first 6 mode-shapes at *Γ*. Psudo-gaps and complete gaps are shown in color region in the band diagram.

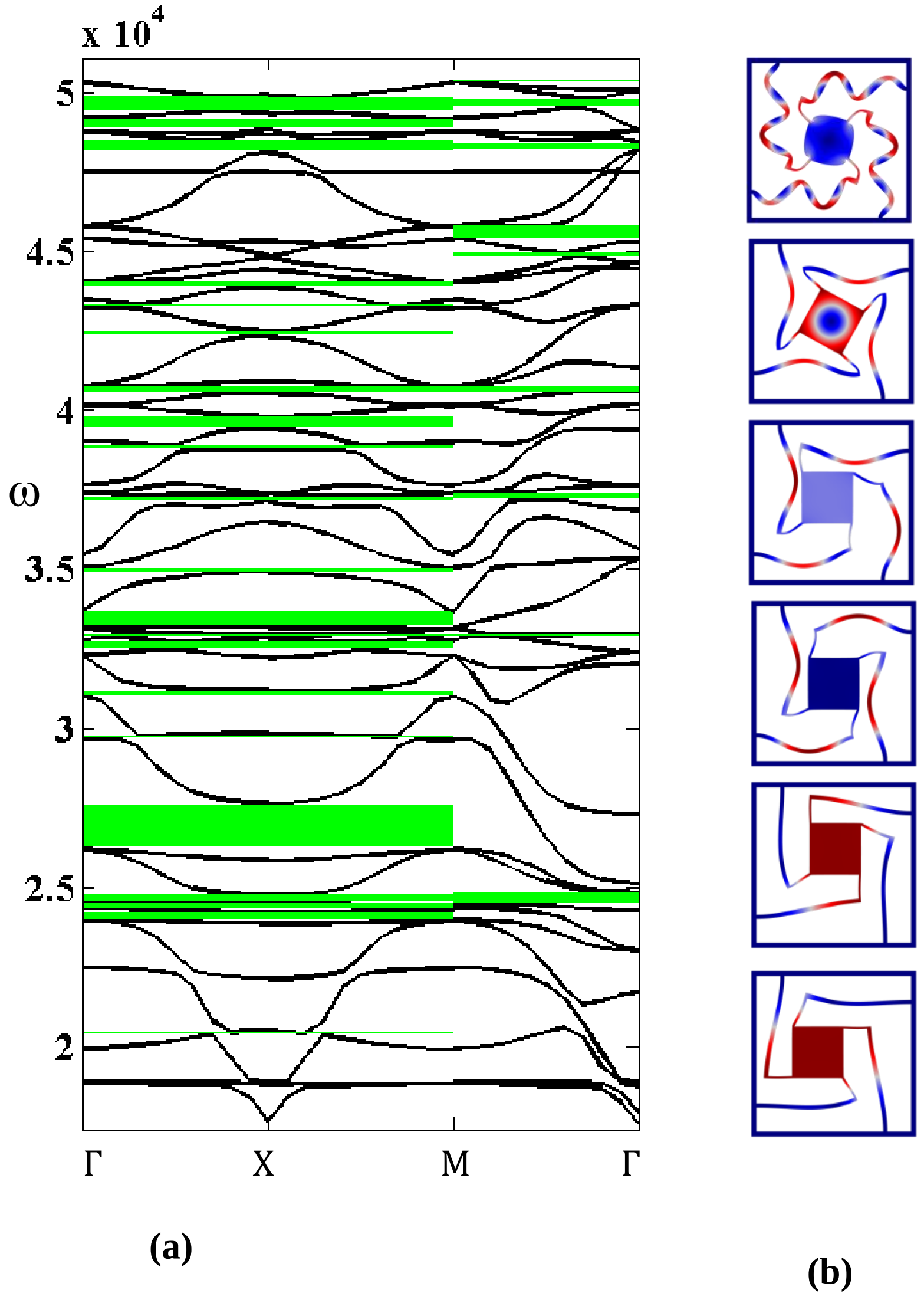

**Figure 35. (a)** Band diagram of deformed structure and (b) the first 6 mode-shapes at *Γ*. Psudo-gaps and complete gaps are shown in color region in the band diagram.

### 4.3 Summary and Conclusion

A thermally tunable PnC is designed and analyzed through FEM modeling. The proposed periodic structure utilizes temperature induced large deformations in bimaterial ligaments of an in plane resonant unit cell to exploit pattern transformation in a periodic structure. The propagation of elastic waves is studied on the prestressed deformed structure. Complete band-gaps are shifted by the applied temperature difference. Moreover, psudo-gaps are transformed in specific directions of symmetry of the unit cell. Band diagram results demonstrate the ability of the thermally tunable periodic structure to be used in the control of elastic wave propagation.

# CHAPTER FIVE

## A New Computational Method for Overall Tangent Moduli of a Soft Magnetoactive Composite Using Periodic Homogenization[1]

A finite element methods based homogenization approach is presented to simulate the nonlinear behavior of magnetoactive composites under a macroscopic deformation and an external magnetic field. The coupled magnetoelastic constitutive law and governing equations are developed in micro-scale for large deformations. Micro-scale formulation is employed on a characteristic volume element, taking into account periodic boundary conditions. Periodic homogenization method is utilized to compute macroscopic properties of the magnetoelastic composite at different mechanical and magnetic loading paths. A new and cost effective numerical scheme is used to develop the magnetoelastic tangent moduli tensors. The sensitivity analysis is proposed to compute the overall tangent moduli tensors of the composite through the finite difference method. The presented approach is useful in characterization of magnetoactive and electroactive composites and $FE^2$ methods. Results are presented for typical equilibrium states.

In this study, a FEM-based homogenization method is employed to compute the effective response of a periodic MEC under applied magnetic fields and large deformations. Due

---

[1] Results of this chapter are published in:

Bayat A and Gordaninejad F 2015 A New Computational Method for Homogenized Tangent Moduli of a Soft Magnetoelastic Composite *Journal of Smart Materials and Structures* **24** 075010.

to high interest in magneto-active and electro-active composites, the characterization of these composites needs to be realized through commercial software. This study differs from the prior work in the proposed numerical approach for the computation of the tangent moduli tensors using the commercial FEM package COMSOL which is capable of multiphysics modeling. The overall tangent moduli are developed based on the sensitivity analysis of deformation gradient tensor and magnetic field vector by utilizing the finite difference method. Constitutive laws of the homogenized material are not derived from a macroscopic energy function, but a CVE is attached to a material point to extract the effective response through the volume average of microscopic quantities. It is assumed that the principle of separation of scales is satisfied for the relative dimensions of the microstructure and fluctuation field in contrast to that of the CVE. Theoretical framework for constitutive laws and coupled governing equations for magnetoelastic continuum is presented following the finite elasticity theory [21-28]. The FEM discretization is carried out on the CVE consisting of a magnetically permeable particle and a hyperelastic matrix considering periodic boundary conditions. The periodic homogenization is employed to extract macroscopic constitutive laws of the nominal stress tensor and magnetic induction vector.

## 5.1 Modeling

A direct micro to macro extraction of the material properties is defined through the FEM-based homogenization approach. A CVE consisting of a magnetically permeable particle and a soft matrix is used for micro-scale. The permeable particle and soft matrix are characterized by a magnetoelastic energy function where the magnetic permeability of the matrix is presumed as that of the free space.

A typical CVE chosen for micro-scale analysis is shown in Figure 36. The MEC is considered to be initially at an undeformed state, denoted by $\mathbb{C}_r$ with boundary $\partial\mathbb{C}_r$ as the reference configuration. The body deforms when subjected to time-dependent magnetic and mechanical loadings. The region occupied by the continuum $\mathbb{C}_\mathrm{t}$, with boundary $\partial\mathbb{C}_\mathrm{t}$, at a given time $t$ is the deformed configuration. Let $\boldsymbol{X}$ and $\boldsymbol{x}$ be the position vectors of the material point at reference and deformed configurations, respectively, where $\boldsymbol{x} = \chi(\boldsymbol{X}, t)$ and $\chi: \mathbb{C}_r \rightarrow \mathbb{C}_t$ is the deformation mapping. The deformation gradient tensor is defined by $\boldsymbol{F} = Grad_{\boldsymbol{X}}\, \boldsymbol{x} = \partial\chi/\partial\boldsymbol{X}$, where $Grad_{\boldsymbol{X}}$ is the gradient operator with respect to material coordinates, $\boldsymbol{X}$. In this study, notations $Grad_{\boldsymbol{X}},\ Div_{\boldsymbol{X}}\ and\ Curl_{\boldsymbol{X}}$ are used for micro-scale differential operators in Lagrangian coordinates**.** A Lagrangian formulation is adopted to develop magnetoelastic relations. The Lagrangian magnetic field and magnetic induction vectors are denoted by $\boldsymbol{H} = \boldsymbol{H}(\boldsymbol{X})$ and $\boldsymbol{B} = \boldsymbol{B}(\boldsymbol{X})$, respectively. It is assumed that the magnetic field is stationary and the non-conducting MEC material is initially at the static configuration and subjected to only magnetic and mechanical interactions. Thus, $\boldsymbol{H}$ and $\boldsymbol{B}$ are independent of time. Maxwell equations of magneto-statics can be written, as follows:

$$Curl_{\boldsymbol{X}}\, \boldsymbol{H} = \boldsymbol{0}, \qquad Div_{\boldsymbol{X}}\, \boldsymbol{B} = 0, \tag{1}$$

It is worth mentioning that Equation (1) is resulted from no electric field, no free charge and no current density assumptions on the continuum. Equation (1) is used to define a scalar potential $\varphi(\boldsymbol{X})$, such that $\boldsymbol{H} = -Grad_{\boldsymbol{X}}\varphi$, since $Curl_{\boldsymbol{X}}\,\boldsymbol{H} = Curl_{\boldsymbol{X}}(-Grad_{\boldsymbol{X}}\varphi) = \boldsymbol{0}.$ Thus, the magneto-static differential equations are solved for the scalar potential.

In the absence of body forces, the equilibrium equation on the micro-scale reads:

$$Div_{\boldsymbol{X}}\,\boldsymbol{T} = 0, \tag{2}$$

where $\boldsymbol{T}$ is the nominal stress tensor defined at reference configuration. Equations (1) and (2) are coupled governing equations of the magneto-elastic continuum. Constitutive relations of the magneto-elastic medium are derived from a nonlinear magnetoelastic energy density, $\Psi = \Psi(\boldsymbol{F}, \boldsymbol{H})$, which is a function of the deformation gradient tensor and magnetic field vector, defined per unit volume at $\mathbb{C}_{\mathrm{r}}$. For a compressible material, constitutive relations for the nominal stress and the magnetic induction are:

$$\boldsymbol{T} = \frac{\partial\Psi}{\partial\boldsymbol{F}}\,, \quad \boldsymbol{B} = -\frac{\partial\Psi}{\partial\boldsymbol{H}} \qquad \text{in } \mathbb{C}_r \tag{3}$$

In macro-scale, the volume occupied by a body in reference (undeformed) configuration is denoted by $\overline{\mathbb{C}_r}$, which is bounded by $\overline{\partial\mathbb{C}_{\mathrm{r}}}$ and notations $\overline{\mathbb{C}_{\mathrm{t}}}$ and $\overline{\partial\mathbb{C}_{\mathrm{t}}}$ are assigned for the corresponding deformed configuration of the continuum. $\overline{\mathbf{X}}$ and $\overline{\mathbf{x}}$ are associated with the macroscopic Lagrangian and Eulerian coordinates, respectively. The macroscopic deformation mapping, $\bar{\chi}$ follows $\overline{\boldsymbol{x}} = \bar{\chi}(\overline{\boldsymbol{X}}, t)$. Accordingly, $Grad_{\overline{\boldsymbol{X}}}$, $Div_{\overline{\boldsymbol{X}}}$ and $Curl_{\overline{\boldsymbol{X}}}$ are used for macro-scale differential operators in Lagrangian coordinates. Hence, $\overline{\boldsymbol{F}} = Grad_{\overline{\boldsymbol{X}}}\,\overline{\boldsymbol{x}} = \partial\bar{\chi}/\partial\overline{\boldsymbol{X}}$ is the macroscopic deformation gradient tensor. Consequently, coupled governing equations of the continuum are:

$$Curl_{\overline{X}}\,\overline{\boldsymbol{H}} = \boldsymbol{0}, \qquad Div_{\overline{X}}\,\overline{\boldsymbol{B}} = 0, \qquad Div_{\overline{X}}\,\overline{\boldsymbol{T}} = 0 \tag{4}$$

where $\overline{\boldsymbol{H}}$, $\overline{\boldsymbol{B}}$ and $\overline{\boldsymbol{T}}$ are macroscopic magnetic field, magnetic induction and nominal stress, respectively. Similarly, macroscopic quantities can be related to a macroscopic energy function $\overline{\Psi} = \overline{\Psi}(\overline{\boldsymbol{F}}, \overline{\boldsymbol{H}})$ through:

$$\overline{\boldsymbol{T}} = \frac{\partial\overline{\Psi}}{\partial\overline{\boldsymbol{F}}}, \quad \overline{\boldsymbol{B}} = -\frac{\partial\overline{\Psi}}{\partial\overline{\boldsymbol{H}}} \qquad \text{in } \overline{\mathbb{C}_r} \tag{5}$$

Derivation of constitutive laws through a macroscopic energy function is beyond the scope of this study. Computation of these macroscopic quantities is performed through surface integrals of corresponding microscopic counterparts across the CVE's boundary. Notations, $< \blacksquare >_{\mathbb{C}_r} = \frac{1}{V}\int_{\mathbb{C}_r} \blacksquare \, dV$ and $\sqsubset \blacksquare \sqsupset_{\partial\mathbb{C}_r} = \frac{1}{V}\oint_{\partial\mathbb{C}_r} \blacksquare \, dS$, are introduced for volume and surface integrals on the body's domain and boundary, respectively, where $V$ is the volume of the domain in the reference configuration. Assuming the continuity of the deformation gradient tensor and the magnetic induction vector on the boundary of the CVE, the surface integral can be equivalently estimated as the volume integral of corresponding properties on the CVE's domain. Using Gauss theorem, macroscopic deformation and nominal stress tensors are given by:

$$\overline{\boldsymbol{F}} = < \boldsymbol{F} >_{CVE} = \sqsubset \boldsymbol{x} \otimes \boldsymbol{n} \sqsupset_{\partial CVE}, \qquad \overline{\boldsymbol{T}} = < \boldsymbol{T} >_{CVE} = \sqsubset \boldsymbol{t} \otimes \boldsymbol{X} \sqsupset_{\partial CVE} \tag{6}$$

where $\boldsymbol{x}$, $\boldsymbol{t}$ and $\boldsymbol{n}$ are the position vector, traction vector, and normal vector on the boundary of the CVE, respectively. Likewise, the corresponding magnetic field and magnetic induction vectors in macro-scale are defined by integral equations:

$$\bar{\boldsymbol{H}} = <\boldsymbol{H}>_{CVE} = \sqsubset \varphi \boldsymbol{n} \sqsupset_{\partial CVE} \; , \qquad \bar{\boldsymbol{B}} = <\boldsymbol{B}>_{CVE} = \sqsubset b\boldsymbol{X} \sqsupset_{\partial CVE} \tag{7}$$

where $\varphi$ and $b = \boldsymbol{B}.\boldsymbol{n}$ are the magnetic potential and magnetic flux on the boundary of the CVE, respectively.

In periodic homogenization, the microscopic position vector can be expressed as a linear function of the macroscopic deformation gradient and a fluctuation field; $\boldsymbol{x}(\boldsymbol{X}) = \bar{\boldsymbol{F}}\boldsymbol{X} + \boldsymbol{g}(\boldsymbol{X})$, where $\boldsymbol{g}(\boldsymbol{X})$ is the fluctuation field, a vector function of the position vector. Similarly, the magnetic potential follows $\varphi(\boldsymbol{X}) = \bar{\boldsymbol{H}}\boldsymbol{X} + h(\boldsymbol{X})$, where $h(\boldsymbol{X})$ is a scalar function of $\boldsymbol{X}$. Boundary conditions of the CVE are derived from the classical Hill-Mandel homogeneity condition. Hill-Mandel condition states that the increment of the macroscopic energy function is equivalent to the volume average of the increment of the microscopic energy function, $\dot{\bar{\Psi}} = <\dot{\Psi}>$. It is shown that periodic boundary conditions for the deformation and magnetic potential on $\partial CVE$ satisfy the Hill-Mandel condition [46-48]. Fluctuation fields are assumed to be periodic on $\partial CVE$; thus, the traction vector and magnetic flux are anti-periodic, as follows:

$$\begin{aligned} &\boldsymbol{g}(\boldsymbol{X}^-) = \boldsymbol{g}(\boldsymbol{X}^+), \qquad h(\boldsymbol{X}^-) = h(\boldsymbol{X}^+), \\ &\boldsymbol{t}(\boldsymbol{X}^-) = -\boldsymbol{t}(\boldsymbol{X}^+), \qquad b(\boldsymbol{X}^-) = -b(\boldsymbol{X}^+), \qquad \text{on } \partial CVE \end{aligned} \tag{8}$$

where superscripts (+) and (–) are associated with nodes on opposite boundaries (right (top) and left (bottom) edges on Figure 36 (b-c)) of the $\partial CVE$. In order to define the boundary value problem (BVP), the macro-scale boundary conditions on $\partial\mathbb{C}_r$, for mechanical and magnetic problem must be specified. Figure 36 shows a schematic of the BVP. The boundary of the homogenous body is decomposed into different sections, where corresponding Dirichlet and Neumann boundary conditions for mechanical and magnetic problems are prescribed. Displacement type boundary conditions, $\bar{\boldsymbol{u}} = \bar{\boldsymbol{u}}_{\boldsymbol{0}}$ on $\partial\bar{\mathbb{C}}_u$ as well as traction boundary conditions $\bar{\boldsymbol{t}} = \bar{\boldsymbol{T}}.\bar{\boldsymbol{n}}$ on $\partial\bar{\mathbb{C}}_\sigma$ are prescribed to define mechanical BVP, where $\bar{\boldsymbol{u}}$, $\bar{\boldsymbol{t}}$ and $\bar{\boldsymbol{n}}$ are the macroscopic displacement, traction and normal to the surface vectors defined on the boundary of the body, respectively. The corresponding magnetic BVP is defined through scalar magnetic potential boundary condition; $\bar{\varphi} = \bar{\varphi}_0$ on $\partial\bar{\mathbb{C}}_\varphi$and magnetic flux boundary condition; $\bar{b} = \bar{\boldsymbol{B}}.\bar{\boldsymbol{n}}$ on $\partial\bar{\mathbb{C}}_b$, where $\bar{\varphi}$ and $\bar{b}$ are the macroscopic magnetic potential and magnetic flux, respectively.

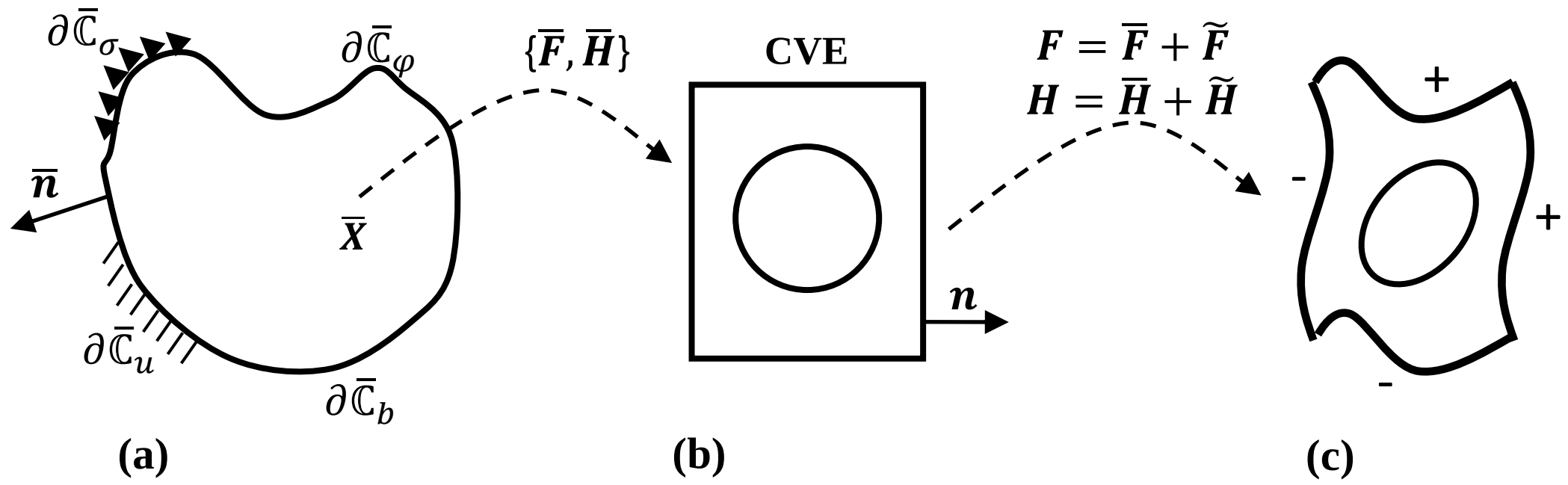

**Figure 36. (a)** The homogenized body and corresponding boundary decomposition in Lagrangian configuration, **(b)** corresponding CVE, attached to $\bar{\boldsymbol{X}}$, selected for homogenization study including a circular permeable particle inside a soft square matrix, **(c)** the deformed CVE in Eulerian configuration.

The macroscopic deformation gradient tensor and magnetic field are obtained from macro-scale BVP which are inputs of the micro-scale problem. The transition from microscopic to macroscopic properties is defined on the micro-scale problem through FEM-based averaging process. Thus, the effective nominal stress and magnetic induction as well as macroscopic (effective) moduli tensors are computed from corresponding microscopic properties on CVE domain.

To develop a finite element model, the variational formulation of the equilibrium and magneto-statics equations is derived. To derive the weak form of governing equations, inner product of Equations $(1)_2$ and (2) with an arbitrary test function, is considered and then integrated over the CVE domain. Let us consider $\delta\boldsymbol{x}$ and $\delta\varphi$ be arbitrary variations of $\boldsymbol{x}$ and $\varphi$, respectively, that satisfy the boundary conditions on $\partial CVE$. Taking the variational form of Equations $(1)_2$ and (2), using integration by part and then the divergence theorem yields:

$$\int_{CVE} \boldsymbol{T}: Grad_{\boldsymbol{X}}\delta\boldsymbol{x}\, dV - \oint_{\partial CVE} \boldsymbol{t}.\delta\boldsymbol{x}\, dS = 0 \tag{9}$$

$$\int_{CVE} \boldsymbol{B}: Grad_{\boldsymbol{X}}\delta\varphi\, dV - \oint_{\partial CVE} b\delta\varphi\, dS = 0 \tag{10}$$

It is noted that the weak form of magneto-statics equation results from a stationary magnetic field condition and a non-conducting MEC medium in the absence of surface current density on $\partial$CVE. Natural boundary conditions in Equations (9) and (10) appear as work terms applied to $\partial CVE$, which arise from the normal traction force and the magnetic flux. Recalling Equation (8), it is evident from Figure 36(b) that boundary integrals $\oint_{\partial CVE} \boldsymbol{t}.\,\delta\boldsymbol{x}\, dS$, in Equation (9) and $\oint_{\partial CVE} b\delta\varphi\, dS$, in Equation (10) vanish, since the normal unit vector, $\boldsymbol{n}$ acts in opposite directions on the parallel boundaries of the deformed CVE. Weak forms in Equations (9) and (10) define the coupled magnetoelastic behavior of the model which is solved utilizing the nonlinear FEM solver. COMSOL Multiphysics is used for numerical simulations which allows for direct implementation of weak expressions.

To predict the overall response of the MEC through homogenization approach, computation of the effective tangent moduli is required. Homogenized tangent moduli tensors contribute to constitutive laws of macro-scale BVP. In contrast to macroscopic quantities; $\overline{\boldsymbol{F}}, \overline{\boldsymbol{T}}, \overline{\boldsymbol{H}}$ and $\overline{\boldsymbol{B}}$, which can be directly computed by Equations (6) and (7), the overall tangent moduli $\frac{\partial \overline{\boldsymbol{T}}}{\partial \overline{\boldsymbol{F}}}, \frac{\partial \overline{\boldsymbol{T}}}{\partial \overline{\boldsymbol{H}}}, \frac{\partial \overline{\boldsymbol{B}}}{\partial \overline{\boldsymbol{F}}}, \frac{\partial \overline{\boldsymbol{B}}}{\partial \overline{\boldsymbol{H}}}$ cannot be obtained through volume averaging of microscopic counterparts. This is due to the fact that there is no explicit relation of macroscopic nominal stress tensor and magnetic induction vector as a function of $\overline{\boldsymbol{F}}$ and $\overline{\boldsymbol{H}}$. Incremental (linearized) constitutive relations of the coupled magnetoelastic BVP can be estimated, as follows:

$$\begin{aligned} \Delta\overline{\boldsymbol{T}} &= \frac{\partial \overline{\boldsymbol{T}}}{\partial \overline{\boldsymbol{F}}} : \Delta\overline{\boldsymbol{F}} + \frac{\partial \overline{\boldsymbol{T}}}{\partial \overline{\boldsymbol{H}}} \cdot \Delta\overline{\boldsymbol{H}} = \overline{\mathbb{A}} : \Delta\overline{\boldsymbol{F}} + \overline{\mathbb{C}} \cdot \Delta\overline{\boldsymbol{H}} \;, \\ \Delta\overline{\boldsymbol{B}} &= \frac{\partial \overline{\boldsymbol{B}}}{\partial \overline{\boldsymbol{F}}} : \Delta\overline{\boldsymbol{F}} + \frac{\partial \overline{\boldsymbol{B}}}{\partial \overline{\boldsymbol{H}}} \cdot \Delta\overline{\boldsymbol{H}} = \overline{\mathbb{C}}^{\boldsymbol{T}} : \Delta\overline{\boldsymbol{F}} + \overline{\mathbb{B}} \cdot \Delta\overline{\boldsymbol{H}} \end{aligned} \tag{11}$$

where $\overline{\mathbb{A}}, \overline{\mathbb{C}}$ and $\overline{\mathbb{B}}$ are macroscopic mechanical, magneto-mechanical and magnetic moduli tensors, respectively. Taking the gradient of displacement vector and magnetic potential defined for periodic homogenization, the deformation gradient and magnetic field can be decomposed into a constant and a fluctuation part $\boldsymbol{F} = \overline{\boldsymbol{F}} + \widetilde{\boldsymbol{F}}$ and $\boldsymbol{H} = \overline{\boldsymbol{H}} + \widetilde{\boldsymbol{H}}$**,** where $\widetilde{\boldsymbol{F}} = Grad_{\boldsymbol{X}}\boldsymbol{g}(\boldsymbol{X})$ and $\widetilde{\boldsymbol{H}} = Grad_{\boldsymbol{X}}\boldsymbol{h}(\boldsymbol{X})$**.** Substituting in moduli tensor relations one has,

$$\overline{\mathbb{A}} =< \frac{\partial \boldsymbol{T}}{\partial \boldsymbol{F}} : \frac{\partial(\overline{\boldsymbol{F}}+\widetilde{\boldsymbol{F}})}{\partial \overline{\boldsymbol{F}}} >_{CVE} =< \mathbb{A} >_{CVE} +< \mathbb{A} : \frac{\partial \widetilde{\boldsymbol{F}}}{\partial \overline{\boldsymbol{F}}} >_{CVE},$$

$$\overline{\mathbb{C}} =< \frac{\partial \boldsymbol{T}}{\partial \boldsymbol{H}} \cdot \frac{\partial(\overline{\boldsymbol{H}}+\widetilde{\boldsymbol{H}})}{\partial \overline{\boldsymbol{H}}} >_{CVE} =< \mathbb{C} >_{CVE} +< \mathbb{C} \cdot \frac{\partial \widetilde{\boldsymbol{H}}}{\partial \overline{\boldsymbol{H}}} >_{CVE} \tag{12}$$

$$\overline{\mathbb{B}} =< \frac{\partial \boldsymbol{B}}{\partial \boldsymbol{H}} \cdot \frac{\partial(\overline{\boldsymbol{H}}+\widetilde{\boldsymbol{H}})}{\partial \overline{\boldsymbol{H}}} >_{CVE} =< \mathbb{B} >_{CVE} +< \mathbb{B} \cdot \frac{\partial \widetilde{\boldsymbol{H}}}{\partial \overline{\boldsymbol{H}}} >_{CVE}$$

where $\mathbb{A}, \mathbb{C}$ and $\mathbb{B}$ are microscopic counterparts of moduli tensors. In Equations (12), the challenge is to compute partial derivatives of fluctuation fields; $\frac{\partial \widetilde{\boldsymbol{F}}}{\partial \overline{\boldsymbol{F}}}$ and $\frac{\partial \widetilde{\boldsymbol{H}}}{\partial \overline{\boldsymbol{H}}}$, since there is no explicit expression of microscopic fluctuation fields in terms of macroscopic variables, $\overline{\boldsymbol{F}}$ and $\overline{\boldsymbol{H}}$. Computation of the sensitivity of $\widetilde{\boldsymbol{F}}$ and $\widetilde{\boldsymbol{H}}$, with respect to their macroscopic counterparts needs to be performed through numerical methods. Sensitivity analysis is a technique developed to measure incremental variations of a single input parameter that affects a particular dependent variable of an objective function, while remaining inputs are kept constant. The objective function is in general a function of the solution to a multiphysics problem $y = \widetilde{\boldsymbol{F}}(\boldsymbol{u}(\boldsymbol{\xi}), \boldsymbol{\xi})$, which is manipulated by the control variables $\boldsymbol{\xi} = \xi_i$. Using a Taylor expansion around the state $\xi_0$., the sensitivity of $y$ to $\xi_i$ is defined by $\frac{\partial y}{\partial \xi_i}$. A finite difference method is used to compute macroscopic tangent moduli and sensitivity of $\widetilde{\boldsymbol{F}}$ and $\widetilde{\boldsymbol{H}}$ with respect to $\overline{\boldsymbol{F}}$ and $\overline{\boldsymbol{H}}$. Details are given in section 3.

## 5.2 Results and Discussion

The FEM-based homogenization approach presented in this study is carried out on a typical two dimensional CVE consisting of a permeable inclusion and a matrix shown in Figure 36(b). The matrix is a square of $1mm$ edge and the radius of circular inclusion is $0.3mm$. The CVE is analyzed under different states of deformation and magnetic field to compute the distribution of microscopic nominal stress and magnetic induction as well as effective counterparts. Both inclusion and matrix are modeled as a compressible neo-Hookean magnetoelastic material. To conduct numerical analysis, a particular form of energy function is required. Due to lack of experimental data on MECs, limited data are available in the literature. In this study, a typical magnetoelastic energy function is considered:

$$\Psi = \frac{\mu}{2}(tr(\boldsymbol{c}) - 3) - \mu\, lnJ + \frac{\lambda}{2}(lnJ)^2 + pJ^{-2/3}(\boldsymbol{cH}.\boldsymbol{H}) \tag{13}$$

where $\boldsymbol{c} = \boldsymbol{F^T F}$ is the right Cauchy–Green deformation tensor, $tr(\boldsymbol{c})$ is the trace of $\boldsymbol{c}$ and $J = det\boldsymbol{F}$. $\mu$ and $\lambda$ are Lamé constants and $p$ is the magnetic permeability. Material properties of the inclusion are chosen as $\mu^i = 80\ GPa$, $\lambda^i = 120\ GPa$ and $p^i = 1250\mu_0$ and those of the matrix are $\mu^m = 4$ MPa, $\lambda^m = 30\ MPa$ and $p^m = \mu_0$, where $\mu_0 = 1.256 \times 10^{-6} NA^{-2}$ is the magnetic permeability of the vacuum [45, 46, 47].

For the magnetoelastic energy function given in the Equation (13), the nominal stress tensor is calculated as:

$$\boldsymbol{T}_{ij} = \mu\boldsymbol{F}_{ij} + [\lambda lnJ - \mu]\boldsymbol{F^{-T}}_{ij} + 2pJ^{-\frac{2}{3}}\left(\boldsymbol{H}_i\boldsymbol{F}_{jk}\boldsymbol{H}_k - \frac{\mathbf{1}}{\mathbf{3}}\boldsymbol{F^{-T}}_{km}\boldsymbol{H}_m(\boldsymbol{c}_{kn}\boldsymbol{H}_n)\boldsymbol{\delta}_{ij}\right) \quad (14)$$

where $\boldsymbol{F^{-T}}$ is the inverse matrix of $\boldsymbol{F^{T}}$ and $\boldsymbol{\delta}_{ij}$is the identity tensor. The magnetic induction vector would be:

$$\boldsymbol{B}_i = [-\frac{\partial\Psi}{\partial\boldsymbol{H}}]_i = -2pJ^{-\frac{2}{3}}\boldsymbol{c}_{ij}\boldsymbol{H}_j \quad (15)$$

The microscopic moduli tensors are derived as:

$$\mathbb{A}_{mnpq} = [\frac{\partial\boldsymbol{T}}{\partial\boldsymbol{F}}]_{mnpq} = [\frac{\partial^2\Psi}{\partial\boldsymbol{F}\partial\boldsymbol{F}}]_{mnpq} = \lambda\boldsymbol{F}^{-1}_{nm}\boldsymbol{F}^{-1}_{qp} + \mu\delta_{mp}\delta_{nq} + [\lambda lnJ - \mu]\boldsymbol{F}^{-1}_{np}\boldsymbol{F}^{-1}_{qm} - \frac{2}{3}pJ^{\frac{-2}{3}}\left(\boldsymbol{H}_i\boldsymbol{H}_j\left(\frac{-2}{3}\boldsymbol{c}_{ij}\boldsymbol{F}^{-1}_{np}\boldsymbol{F}^{-1}_{qm}\right) + 2\boldsymbol{H}_q\boldsymbol{F}_{pj}\boldsymbol{H}_j\boldsymbol{F}^{-T}_{mn}\right)\Big) + pJ^{\frac{-2}{3}}([\left(\frac{-2}{3}\right)\left(2\boldsymbol{H}_q\boldsymbol{F}_{pj}\boldsymbol{H}_j\boldsymbol{F}^{-T}_{mn} + \boldsymbol{H}_i\boldsymbol{H}_j\boldsymbol{c}_{ij}\boldsymbol{F}^{-1}_{np}\boldsymbol{F}^{-1}_{qm}\right)] + \delta_{mp}\boldsymbol{H}_q\boldsymbol{H}_n) \quad (16)$$

$$\mathbb{C}_{\alpha\beta i} = [\frac{\partial\boldsymbol{T}}{\partial\boldsymbol{H}}]_{\alpha\beta i} = [\frac{\partial^2\Psi}{\partial\boldsymbol{H}\partial\boldsymbol{F}}]_{\alpha\beta i} = 2pJ^{\frac{-2}{3}}[(\delta_{\alpha\beta}\boldsymbol{F}_{i\gamma}\boldsymbol{H}_\gamma + \boldsymbol{H}_\alpha\boldsymbol{F}_{i\beta}) - \frac{2}{3}\boldsymbol{c}_{\alpha j}\boldsymbol{H}_j\boldsymbol{F}^{-T}_{i\beta}] \quad (17)$$

$$\mathbb{B}_{\alpha\beta} = [\frac{\partial\boldsymbol{B}}{\partial\boldsymbol{H}}]_{\alpha\beta} = [\frac{\partial^2\Psi}{\partial\boldsymbol{H}\partial\boldsymbol{H}}]_{\alpha\beta} = -2pJ^{\frac{-2}{3}}\boldsymbol{c}_{\alpha\beta} \quad (18)$$

For clarity, distribution of microscopic stress and magnetic induction are investigated for three different cases of magneto-mechanical loadings in $\boldsymbol{x_1} - \boldsymbol{x_2}$ plane:

- A plane strain uniaxial stretch at constant magnetic field: three cases of macroscopic deformation is considered while magnetic field is being kept constant at $\boldsymbol{x_2}$ direction:

$$\overline{\boldsymbol{F}} = \begin{bmatrix} f_{11} & 0 \\ 0 & f_{22} \end{bmatrix} \quad f_{11} = 0.8, 1.1, 1.4 \quad \text{and} \quad \overline{\boldsymbol{H}} = \begin{bmatrix} 0 \\ 8e5 \end{bmatrix} Am^{-1}$$

$f_{22}$ is numerically computed at each loading step, so as $\overline{\boldsymbol{T}}_{22} = 0$.

- A plane strain pure shear at constant magnetic field: three cases of macroscopic deformation is considered while magnetic field is being kept constant at $\boldsymbol{x_1}$ direction:

$$\overline{\boldsymbol{F}} = \begin{bmatrix} 1 & f_{12} \\ f_{12} & 1 \end{bmatrix} \quad f_{12} = -0.3, 0.1, 0.2 \quad \text{and} \quad \overline{\boldsymbol{H}} = \begin{bmatrix} 8e5 \\ 0 \end{bmatrix} Am^{-1}$$

- A plane strain equally biaxial stretch at different levels of magnetic field: three cases of magnetic loading is considered at $\boldsymbol{x_1}$ direction while macroscopic deformation is kept constant:

$$\overline{\boldsymbol{F}} = \begin{bmatrix} 0.9 & 0 \\ 0 & 0.9 \end{bmatrix} \quad \text{and} \quad \overline{\boldsymbol{H}} = \begin{bmatrix} h_1 \\ 0 \end{bmatrix} \quad h_1 = 1e5, 5e5, 1e6\ Am^{-1}$$

Simulation results for uniaxial loading case are shown in Figure 37. All contour plots presented are normalized by the corresponding volumetric average of the quantity concerned. Figures 37(a-c) show the distribution of the microscopic nominal stress component, $\boldsymbol{T}_{22}$, for uniaxial stretch at constant magnetic field $\overline{H}_2 = 8e5\ Am^{-1}$. The macroscopic stress changes as $-12.5\ MPa, 1.48\ MPa$ and $7.22\ MPa$, as the stretch is increased by $0.8, 1.1$ and $1.4$, respectively. The microscopic stress is a nonlinear function of the deformation gradient tensor and magnetic field. Distribution of the microscopic stress is different for compressive and tensile stretches. Smaller stress occurs in the matrix compared to the inclusion, as it is expected due to stiffer material properties of the inclusion.

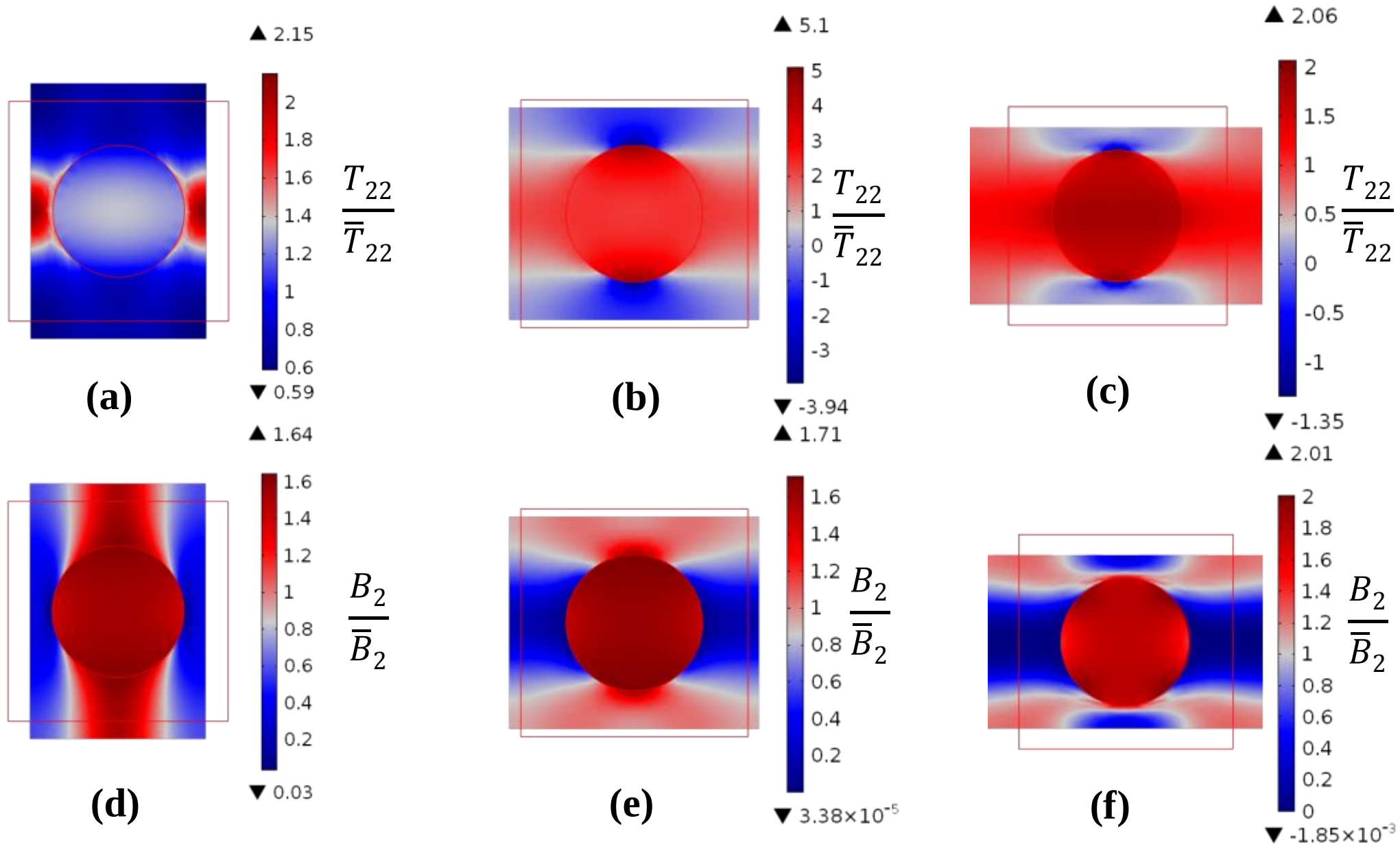

**Figure 37.** Contour plots of normalized microscopic distribution of $\boldsymbol{T_{22}}$ for uniaxial loading case at **(a)** 0.8, **(b)** 1.1 and **(c)** 1.4 stretches. Normalized microscopic distribution of $\boldsymbol{B_2}$ for uniaxial loading case at **(d)** 0.8, **(e)** 1.1 and **(f)** 1.4 stretches.

Distribution of the $B_2$ component of the microscopic magnetic induction, normalized by $\bar{B}_2$ , is shown in Figure 37(d-f). Due to high contrast between magnetic permeability of the inclusion and matrix, high magnetic induction occurs at the inclusion area. Effective magnetic induction varies by $5.76\,T, 3.12\,T$ and $2.68\,T$, as the stretch is increased by $0.8, 1.1$ and 1.4, respectively. This increase is due to the fact that the magnetic induction vector is a nonlinear function of stretch. Moreover, nonuniform deformation distribution in the matrix area causes a higher magnetic induction at top and bottom edges of the CVE.

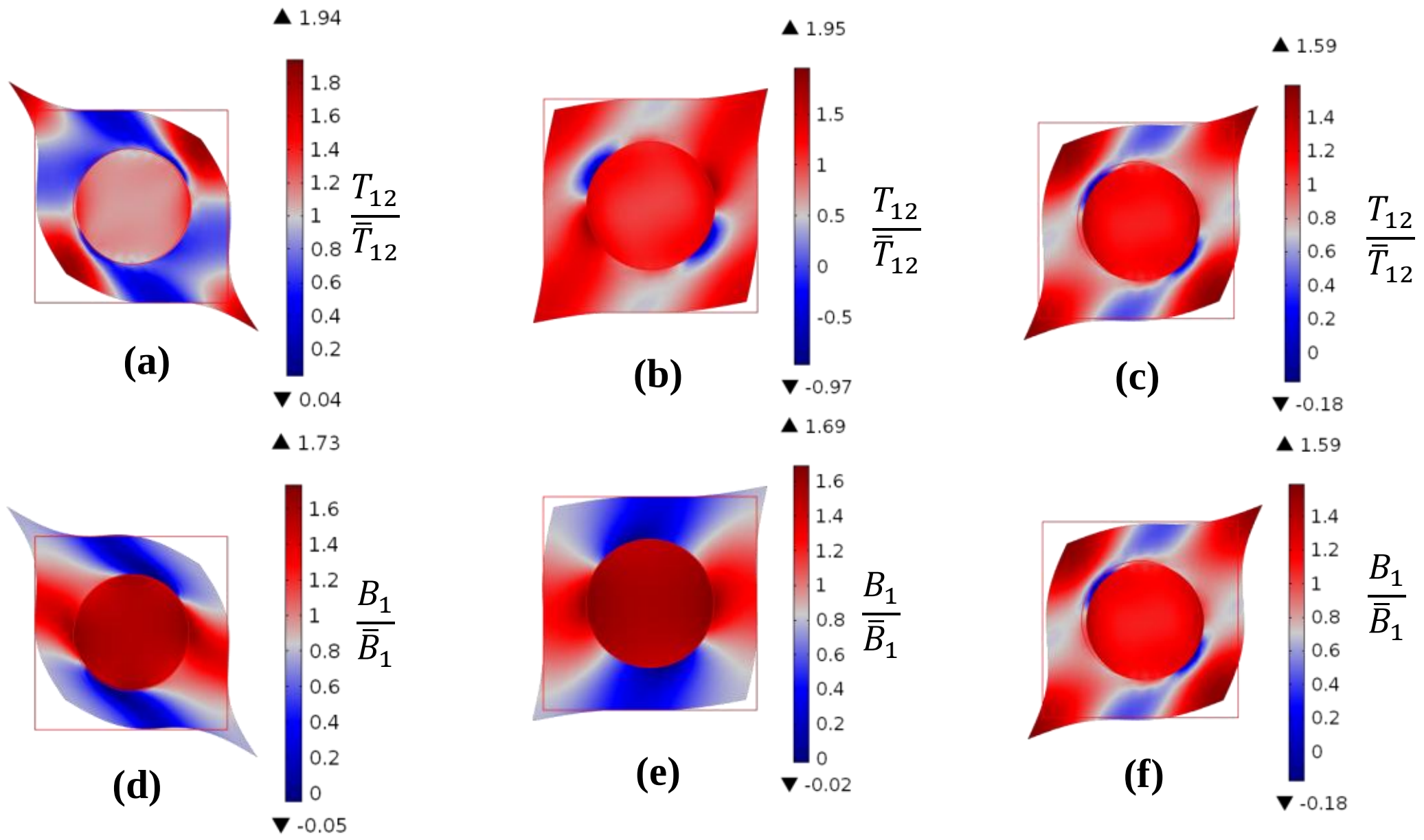

**Figure 38.** Contour plots of normalized microscopic distribution of $\boldsymbol{T_{12}}$ for pure shear loading case at **(a)** -0.3, **(b)** 0.1 and **(c)** 0.2 shear stretches. Normalized microscopic distribution of $\boldsymbol{B_1}$ for pure shear loading case at **(d)** -0.3, **(e)** 0.1 and **(f)** 0.2 shear stretches. The red boundary lines represent the $\boldsymbol{\partial CVE}$ in undeformed configuration. All deformed plots are scaled to 1.

Figures 38(a-c) and 38(d-f) show the contour plots of the microscopic stress and magnetic induction for case of pure shear loading, respectively. Macroscopic magnetic field is applied in $\boldsymbol{x_1}$ direction. The $T_{12}$ component of stress tensor and $B_1$ component of the magnetic induction vector are shown for analysis. The macroscopic stress takes the values $-6.37\ MPa, 1.58\ MPa$ and $3.58\ MPa$ , as the shear stretch changes by $-0.3, 0.1$ and 0.2, respectively. Similar to the case of uniaxial loading case, the highest stress and magnetic induction occurs at the inclusion, due to stiffer mechanical properties and higher magnetic permeability of the inclusion. The macroscopic magnetic induction changes by $-4.83\ T, -3.74\ T$ and $-4.12\ T$, as the shear stretch increases by $-0.3, 0.1$ and 0.2, respectively.

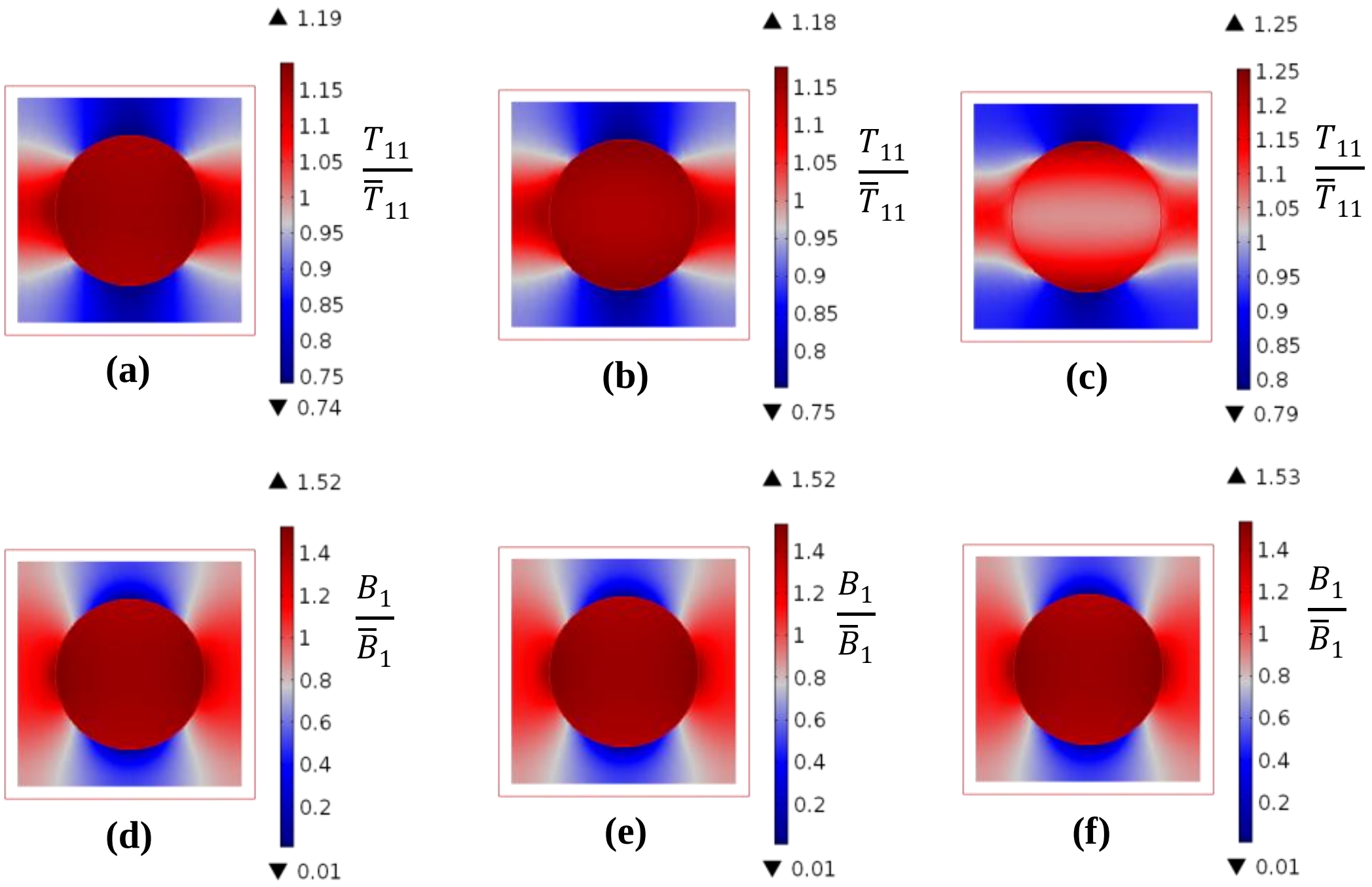

**Figure 39.** Contour plots of normalized microscopic distribution of $\boldsymbol{T_{11}}$ for equally biaxial loading case at **(a)** $\boldsymbol{1e5Am^{-1}}$, **(b)** $5\boldsymbol{e5Am^{-1}}$ and **(c)** $\boldsymbol{1e6Am^{-1}}$ magnetic field. Normalized microscopic distribution of $\boldsymbol{B_1}$ for equally biaxial loading case at **(d)** $\boldsymbol{1e5Am^{-1}}$, **(e)** $5\boldsymbol{e5Am^{-1}}$ and **(f)** $\boldsymbol{1e6Am^{-1}}$ magnetic field.

Figures 39(a-c) and 39(d-f) show the contour plots of the microscopic stress and magnetic induction for equally biaxial loading case, respectively. Macroscopic magnetic field is applied as $1e5, 5e5$ and $1e6\ Am^{-1}$ in $\boldsymbol{x_1}$ direction. The $T_{11}$ component of stress tensor and $B_1$ component of the magnetic induction vector are shown for comparison. The macroscopic stress takes the values $-13.07\ MPa, -13.03\ MPa$ and $-12.9\ MPa$, as the magnetic field is increased by $1e5, 5e5$ and $1e6\ Am^{-1}$, respectively. The highest stress and magnetic induction occurs at the inclusion, due to stiffer mechanical properties and higher magnetic permeability of the inclusion. The macroscopic magnetic induction changes by $-0.4\ T, -1.8\ T$ and $-3.4\ T$, as the magnetic field is increased by $1e5, 5e5$ and $1e6\ Am^{-1}$, respectively.

In all cases of magneto-mechanical loadings presented, the relative distribution of the microscopic stress components mainly depends on the deformation state. This arises obviously due to the dominant geometric nonlinearity term compared to the magnetic nonlinearity term, in the microscopic stress formulation at considered magnetic field.

One might note that periodic boundary conditions are not appeared in the deformed unit cells shown in Figures 37-39. This is because of the fact that periodic boundary conditions are applied on the fluctuation fields but the deformed unit cells displays the microscopic position vector $\boldsymbol{x}(\boldsymbol{X}) = \overline{\boldsymbol{F}}\boldsymbol{X} + \boldsymbol{\mathcal{g}}(\boldsymbol{X})$. The fluctuation fields on the boundaries are shown in Figure 40 for the pure shear loading case. Figure 40 corresponds to the Figure 38 which depicts the deformed unit cell at pure shear loading case.

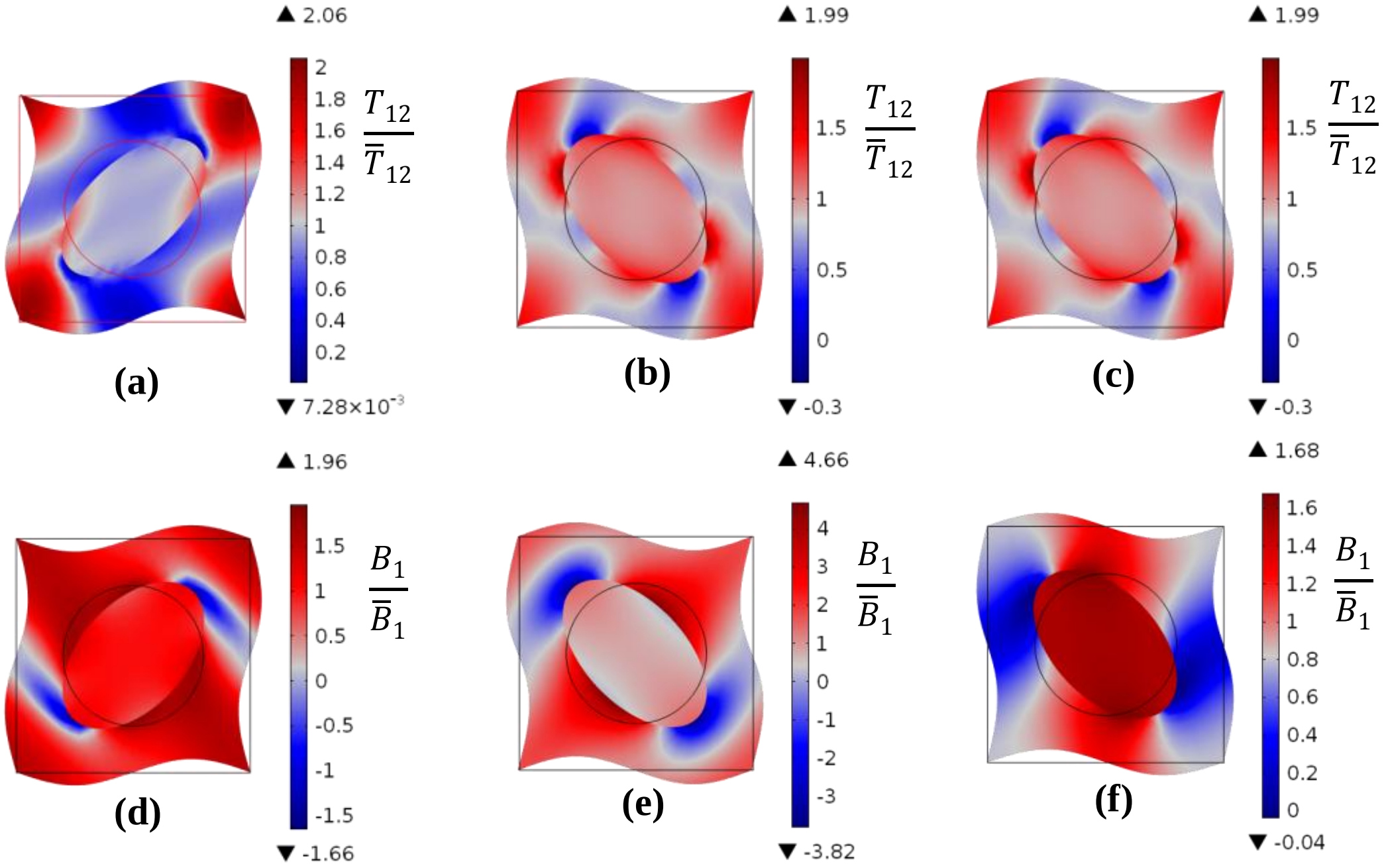

**Figure 40.** Contour plots showing the periodic boundary conditions on the fluctuation fields. The normalized microscopic distribution of $\boldsymbol{T_{12}}$ for pure shear loading case at **(a)** -0.3, **(b)** 0.1 and **(c)** 0.2 shear stretches. Normalized microscopic distribution of $\boldsymbol{B_1}$ for pure shear loading case at **(d)** -0.3, **(e)** 0.1 and **(f)** 0.2 shear stretches. The red boundary lines represent the $\boldsymbol{\partial CVE}$ in undeformed configuration. All deformed plots are scaled to 1. The Figure corresponds to Figure 38 which sketches the deformed unit cells based on of microscopic position vector.

A parametric study is carried out to understand the effect of the increasing magnetic field on the macroscopic quantities. Figure 41 shows the plots for macroscopic stress and magnetic induction components with respect to the stretch at different levels of macroscopic magnetic field. For comparison, magnetic field is applied in $\boldsymbol{x_1}$ direction for all loading cases. The model is parametrically swept on both stretch and magnetic field component for some distinct equilibrium states.

Results for components of the stress tensor for the uniaxial, equally biaxial and pure shear loadings are shown in Figures 41(a), 41(c) and 41(e), respectively. In Figure 41(a), the average stress $\bar{T}_{11}$ turns into compressive one and is highly increased when the magnetic field reaches $\bar{H}_1 = 2e6\ Am^{-1}$. The quadratic magnetic field-dependent term in the stress function becomes a negative term and shows significant effect at higher magnitude of the applied magnetic field, while the influence of the deformation only dependent terms is dominant at lower magnetic fields. In Figure 41(c), the effect of the magnetic term tends to switch the stress component $\bar{T}_{11}$ to compressive stress. Analogous to the uniaxial loading case, the deformation dependent term of the stress dominates at low magnetic fields. Figure 41(e) demonstrates the effect of macroscopic magnetic field on the $\bar{T}_{12}$ stress component for pure shear loading case. The macroscopic magnetic field has strong effect on the shear stress component as is expected from Equation (15).

Corresponding parametric study results for $\bar{B}_1$ component of the effective magnetic induction are depicted in Figures 41(b), 41(d) and 41(f) for uniaxial, equally biaxial and pure shear loading cases, respectively. From Equation (16), for a constant deformation state, the magnetic induction has a linear relation with the applied magnetic field for both matrix and inclusion domains. In all loading cases, magnetic induction increases by the increasing stretch.

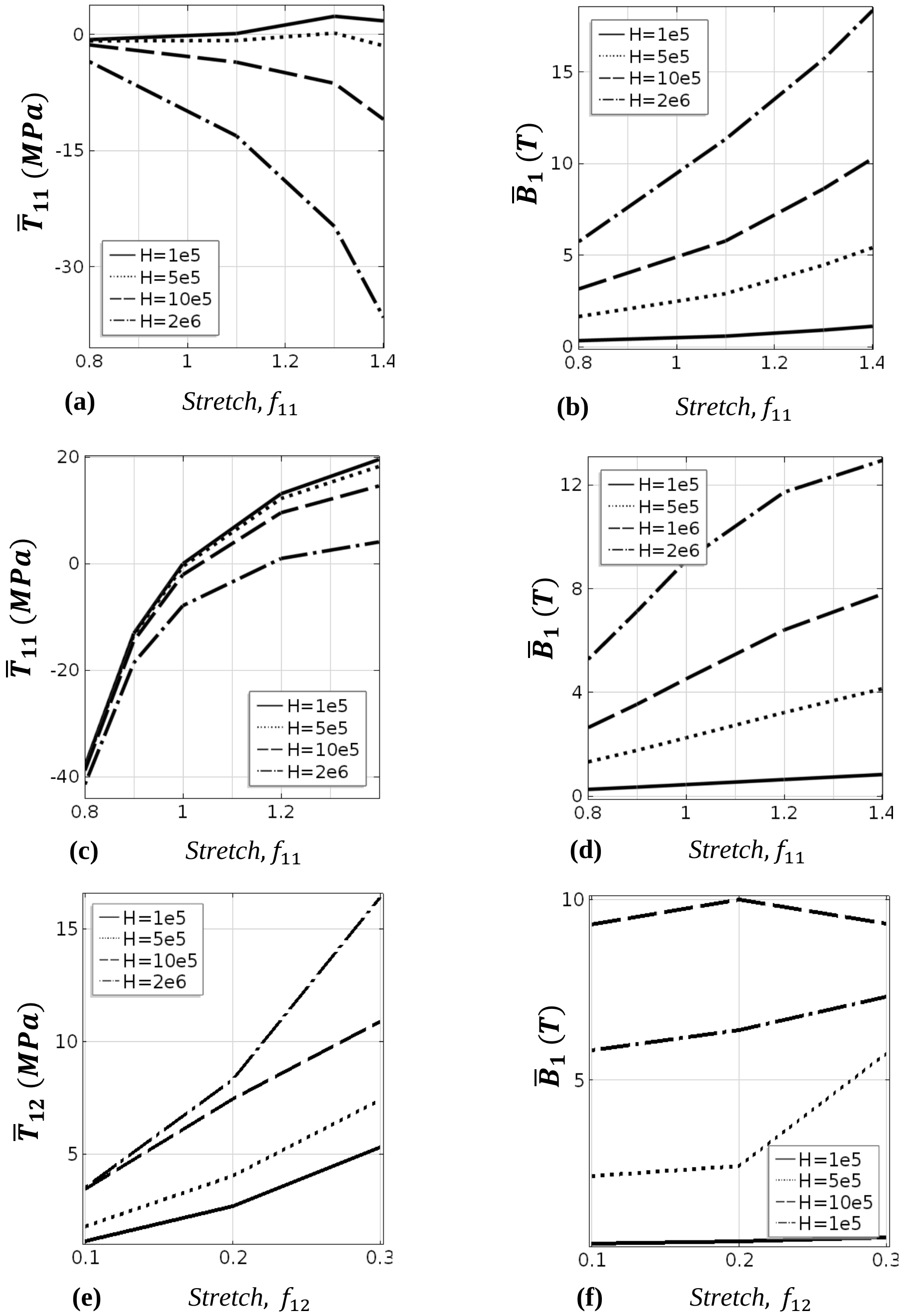

**Figure 41.** Parametric analysis resulting from numerical study representing **(a)** $\overline{\boldsymbol{T}}_{\mathbf{11}}$ vs. stretch **(b)** $\overline{\boldsymbol{B}}_{\mathbf{1}}$ vs. stretch for uniaxial loading case, **(c)** $\overline{\boldsymbol{T}}_{\mathbf{11}}$ vs. stretch (d) $\overline{\boldsymbol{B}}_{\mathbf{1}}$ vs. stretch for equally biaxial loading case and **(e)** $\overline{\boldsymbol{T}}_{\mathbf{12}}$ vs. stretch **(f)** $\overline{\boldsymbol{B}}_{\mathbf{1}}$ vs. stretch for pure shear loading case at different levels of macroscopic magnetic field.

### 5.3 Computation of macroscopic moduli tensors

Computation of the first term in right hand side of Equation (12) requires integration of components of microscopic moduli tensors given by Equations (17-19), over CVE's domain. Calculation of the second term of integral Equations (12) requires the computation of $\frac{\partial \widetilde{\boldsymbol{F}}}{\partial \overline{\boldsymbol{F}}}$ and $\frac{\partial \widetilde{\boldsymbol{H}}}{\partial \overline{\boldsymbol{H}}}$. For small perturbations, a suitable approach is the Taylor expansion of fluctuation functions around macroscopic quantities, $\overline{\boldsymbol{F}}$ and $\overline{\boldsymbol{H}}$. Thus, $\frac{\partial \widetilde{\boldsymbol{F}}}{\partial \overline{\boldsymbol{F}}}$ and $\frac{\partial \widetilde{\boldsymbol{H}}}{\partial \overline{\boldsymbol{H}}}$ are first-order sensitivity of $\widetilde{\boldsymbol{F}}$ and $\widetilde{\boldsymbol{H}}$, with respect to their macroscopic counterparts, respectively.

The finite difference method is used to find the first order sensitivity for two independent cases of loading paths, deformation induced loading and magnetic induced loading. Practically, for an arbitrary deformation loading path, an incremental loading is performed on the model by sweeping the uniaxial stretch at gradually decreasing incremental steps. This is performed through parametric sweep study of the software. At each increment, the model is run for small perturbations, e.g. $\pm 1\%$, of four components of macroscopic deformation gradient tensor, while all other parameters of the model are kept constant. The four components of $\widetilde{\boldsymbol{F}}_{ij}$ are then computed for each independent perturbation and the 4th- order sensitivity tensor is estimated through $[\frac{\partial \widetilde{\boldsymbol{F}}}{\partial \overline{\boldsymbol{F}}}]_{ijkl} = \frac{[\Delta \widetilde{\boldsymbol{F}}_{ij}]}{[\Delta \overline{\boldsymbol{F}}_{kl}]} = \frac{\widetilde{\boldsymbol{F}}_{ij}^{+} - \widetilde{\boldsymbol{F}}_{ij}^{-}}{\overline{\boldsymbol{F}}_{kl}^{+} - \overline{\boldsymbol{F}}_{kl}^{-}}$. Analogously, for magnetic loading case, an incremental parametric sweep is applied on the macroscopic magnetic field. At each incremental level, the model is run for small variation, e.g. $\pm 1\%$, of two components of the macroscopic magnetic field vector, while the rest of parameters are kept constant. Two components of $\widetilde{\boldsymbol{H}}_{i}$ are computed for each independent perturbation component and the 2nd –order magnetic sensitivity tensor is approximated by $[\frac{\partial \widetilde{\boldsymbol{H}}}{\partial \overline{\boldsymbol{H}}}]_{ij} = \frac{[\Delta \widetilde{\boldsymbol{H}}_i]}{[\Delta \overline{\boldsymbol{H}}_j]} = \frac{\widetilde{\boldsymbol{H}}_i^+ - \widetilde{\boldsymbol{H}}_i^-}{\overline{\boldsymbol{H}}_j^+ - \overline{\boldsymbol{H}}_j^-}$. In Equations (12), single and double contractions are denoted by $(\cdot)$ and $(:)$, respectively.

In this study, LiveLink™ *for* MATLAB® interface is used for the computation of the macroscopic moduli tensors, which connects COMSOL Multiphysics to MATLAB scripting environment. LiveLink™ *for* MATLAB® is a Java® based interface that increases the FEM modeling ability by using MATLAB commands and functions to set up the model and physics from scripts, control the model and analyze the results. Microscopic moduli tensors are calculated with direct implementation of Equations (17-19), as local variables in the model. The model is created and saved as `.mph` file in the MATLAB directory. Then the model is imported in the MATLAB script for further processing.

In terms of MATLAB script implementation, a loop is used for each loading path to perform the incremental sweep on the parameter concerned. At every increment step, components of the deformation gradient tensor are perturbed by 0.002 stretch, while $1000Am^{-1}$ is chosen for the perturbation of components of the magnetic field vector. Components of the microscopic moduli tensors and fluctuation fields are computed and integrated on CVE's domain according to the Equation (12). Once the FEM model is created and saved in the MATLAB directory, it is called in the MATLAB script for evaluation of the homogenized moduli tensors. Main syntaxes used in the code are:

1. `model = mphload('modelname')`
2. `model.param.set('parameter','value');`

```
3. model.study('std1').run;
4. [v1,v2,...] = mphint2(model,{'q1','q2',... },'surface')
5. data = mpheval(model,{'q1','q2',... },'selection',1)
```

The first command loads the `.mph` file (COMSOL files extension) which is already saved in the MATLAB working directory. The second command assigns the quantity in `'value'` to the desired parameter in the model. The third command runs the model. The 4th command evaluates the surface integration of the string expressions `'qi's` on the CVE's domain. The 5th command evaluates the string expressions `'qi's` as a field value at each node points. All the data and integration evaluations returns in matrix format and stored for plotting and further processing.

Results for selected components of overall mechanical tangent moduli tensors for a typical uniaxial loading path are depicted in Figure 42. $f_{11}$ component of the deformation gradient tensor is incrementally increased from 0.8 to 1.35 at a constant macroscopic magnetic field, $\bar{H}_1 = 8e6\ Am^{-1}$. From Equation (17), the components of $\mathbb{A}_{mnpq}$ are highly nonlinear and complex functions of the deformation gradient tensor. As it is shown in Figure 42(a), the components $\bar{\mathbb{A}}_{1212}$and $\bar{\mathbb{A}}_{1221}$ are of order of $10^{10}$, since they are highly dependent on $\bar{H}_1$ component of the macroscopic magnetic field. While $\bar{H}_1$-dependent terms are not dominant in $\bar{\mathbb{A}}_{1111}$and $\bar{\mathbb{A}}_{1122}$ components, as demonstrated in Figure 42(b). Consequently, $\bar{\mathbb{A}}_{1212}$, $\bar{\mathbb{A}}_{1221}$ and their symmetric counterparts of the mechanical moduli tensors are dominant terms in the stress tensor.

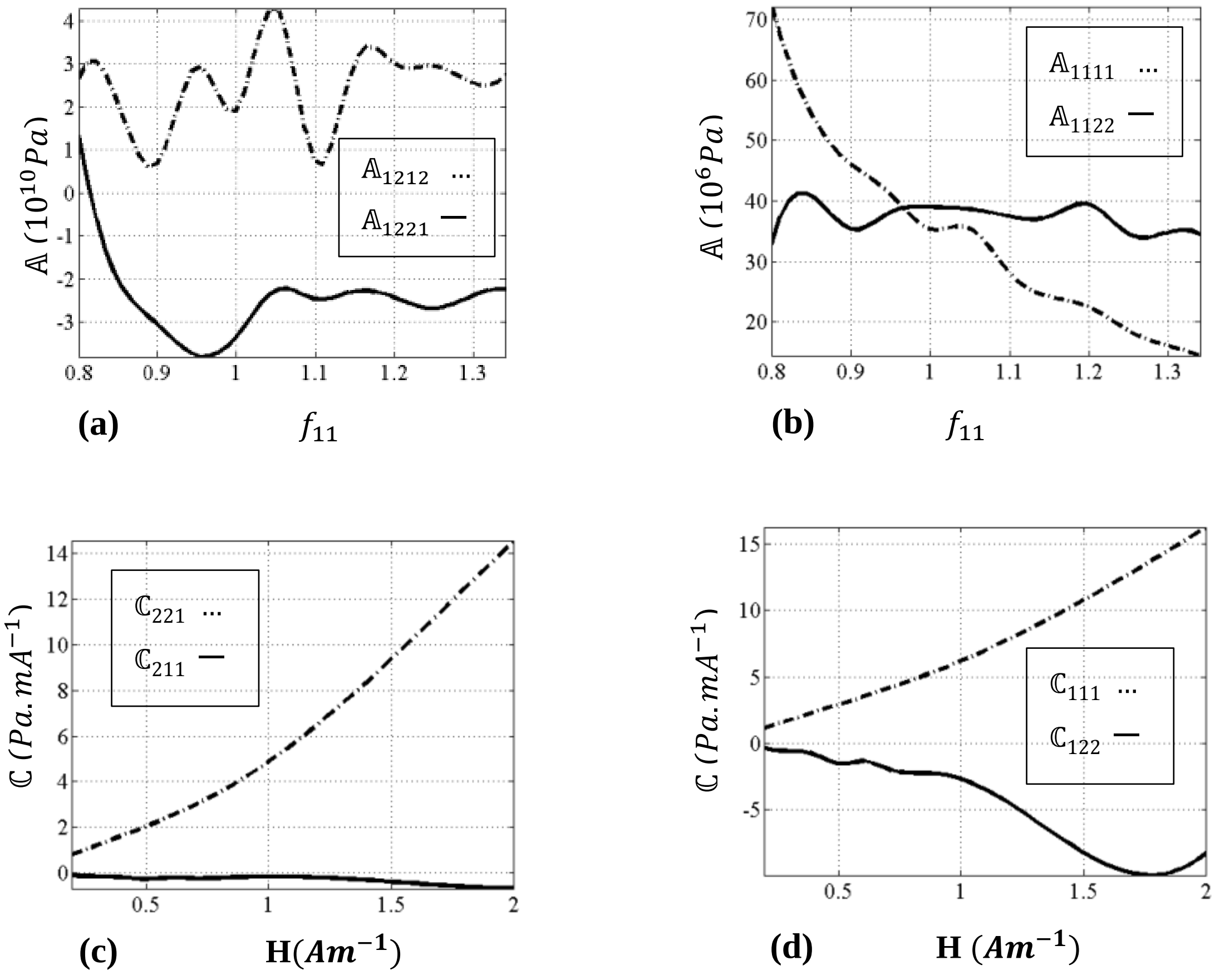

**Figure 42**. **(a,b)** Components of homogenized mechanical moduli tensor, **(c,d)** Components of the homogenized coupling magneto-mechanical moduli tensor.

Results for selected components of the coupling moduli tensor for a uniaxial magnetic loading path are shown in Figure 42(c-d). $\overline{H}_1$ is incrementally increased form $0.2\text{e}6\ \text{Am}^{-1}$ to $2\text{e}6\ \text{Am}^{-1}$at a constant macroscopic deformation state $\overline{\boldsymbol{F}} = [1.3\ \ 0; 0\ \ 0.84]$. From Equation (18), components of $\mathbb{C}_{\alpha\beta i}$ moduli are linear functions of $\overline{H}_1$. The nonlinearity observed in the Figures 42(c,d) stems from the effect of sensitivity tensor as given by Equation (12). The magnitude of $\bar{\mathbb{C}}_{\alpha\beta i}$ tends to increase by the increasing magnetic field. Moreover, as it is deduced from Equation (19), components of $\mathbb{B}_{ij}$tensors are independent of the macroscopic magnetic field. The nonzero components of macroscopic counterparts are nearly computed as $\bar{\mathbb{B}}_{11} = \bar{\mathbb{B}}_{22} = 18 \times 10^{-10}$ (not shown in the Figure). It has to be mentioned that tangent moduli tensors are computed through a set of perturbation tests conducted along a loading path about a reference macroscopic equilibrium state $\{\bar{F}, \bar{H}\}$. Hence, the overall properties presented here depend on the selected uniaxial loading path and the equilibrium state.

### 5.4 Summary and Conclusions

A numerical tool is demonstrated to compute the homogenized properties of the magneto-active composite structures through a commercial FEM package. The proposed algorithm provides a computational approach to study the micro to macro transition of the mechanical and magnetic properties of the MEC structures. The presented approach differs from all prior work in two aspects. The methods presented in references [48-51] for computation of effective tangent moduli tensors are carried out through a C++ based in-house FE code and are of high computational cost. In this study, the FEM package -COMSOL Multiphysics- is employed for all numerical simulations which allows for direct implementation of weak forms of governing equations. It also takes the advantage of MATLAB scripting environment for parametric study and control of the model's variables and physics. More importantly, the finite difference method proposed in this study offers a computationally cost-effective methodology to evaluate homogenized tangent moduli tensors for different loading paths.

A neo-Hookean type magnetoelastic energy function is proposed to demonstrate the nonlinear coupling behavior of the matrix and magnetic inclusion of the composite. A CVE with periodic boundary condition is selected to extract the microscopic distribution of stress and magnetic induction. Macroscopic properties are evaluated through volumetric averaging of microscopic counterparts. Macroscopic properties of the composite are extracted from the microscopic counterparts, through the homogenization procedure. No effective (macroscopic) energy function is assumed in this process. For both uniaxial and biaxial loading cases considered here, macroscopic stress results confirm that the magnetic field dependent terms of stress are dominant at high macroscopic magnetic fields. Moreover, magnetic induction increases by increasing stretch and magnetic field. At constant magnetic field, both stress and magnetic induction increase by increasing the stretch.

In conclusion, homogenization is an essential mechanism to compute the effective properties of the magneto-active composites, especially when finding an effective constitutive law is very difficult for complicated composites. Homogenized tangent moduli tensors are useful and necessary characteristics of the magneto-active composites for evaluating the overall response of the composite, $FE^2$ modeling and macro-scale instability analysis.

# CHAPTER SIX

## Homogenization Approach in Random Magnetoelastic Composites[1]

Homogenization is a numerical approach used as a tool to study the overall response of the composite and heterogeneous materials presumed to be statistically homogenous. In this study, a FEM-based homogenization method is employed to compute the effective response of a random MEC under applied magnetic fields and large deformations. Spatially random distribution of identically circular inclusions inside a soft homogenous matrix is investigated. FEM-based averaging process is combined with Monte-Carlo method (MCM) to generate ensembles of randomly distributed MECs. The ensemble is utilized as a statistical volume element (SVE) in a scale-dependent statistical algorithm to approach the desired characteristic volume element (CVE) size.

It is assumed that the principle of separation of scales is satisfied for the relative dimensions of the microstructure and fluctuation fields in contrast to that of the CVE. The overall tangent moduli tensors are developed based on the sensitivity analysis of deformation gradient tensor and magnetic field vector by utilizing the finite difference method. Theoretical framework for constitutive laws and coupled governing equations for magnetoelastic continuum is presented following the finite elasticity theory [21-28]. The FE discretization is carried out on SVEs consisting of randomly distributed magnetically permeable particles within a hyperelastic matrix. The random homogenization is employed to extract macroscopic constitutive laws of the nominal stress tensor and magnetic induction vector.

[1] Results of this chapter are submitted to the Journal of Computational mechanics.

### 6.1 Modeling

A direct micro-meso-macro extraction of material properties is defined through the FEM-based homogenization approach. A SVE consisting of magnetically permeable particles within a soft matrix is used as ensembles in meso-scale. The magnetic particles and the soft matrix are characterized by a magnetoelastic energy function where the magnetic permeability of the matrix is presumed as that of the free space.

A typical SVE, chosen for meso-scale analysis is shown in Figure 43(b). The magnetoelastic composite (MEC) is considered to be initially at an undeformed state, denoted by $\mathbb{C}_r$ with boundary $\partial\mathbb{C}_r$ as the reference configuration. The body deforms when subjected to time-dependent magnetic and mechanical loadings. The region occupied by the continuum $\mathbb{C}_\mathrm{t}$, with boundary $\partial\mathbb{C}_\mathrm{t}$, at a given time $t$ is the deformed configuration. Let $\boldsymbol{X}$ and $\boldsymbol{x}$ be the position vectors of the material point at reference and deformed configurations, respectively, where $\boldsymbol{x} = \chi(\boldsymbol{X}, t)$ and $\chi: \mathbb{C}_r \rightarrow \mathbb{C}_t$ is the deformation mapping. The deformation gradient tensor is defined by $\boldsymbol{F} = Grad_{\boldsymbol{X}}\, \boldsymbol{x} = \partial\chi/\partial\boldsymbol{X}$, where $Grad_{\boldsymbol{X}}$ is the gradient operator with respect to material coordinates, $\boldsymbol{X}$.

In this study, notations $Grad_{\boldsymbol{X}},\ Div_{\boldsymbol{X}}\ and\ Curl_{\boldsymbol{X}}$ are used for micro-scale differential operators in Lagrangian coordinates. A Lagrangian formulation is adopted to develop magnetoelastic relations. The Lagrangian magnetic field and magnetic induction vectors are denoted by $\boldsymbol{H} = \boldsymbol{H}(\boldsymbol{X})$ and $\boldsymbol{B} = \boldsymbol{B}(\boldsymbol{X})$, respectively. It is assumed that the magnetic field is stationary and the non-conducting MEC material is initially at the static configuration and subjected to only magnetic and mechanical interactions. Thus, $\boldsymbol{H}$ and $\boldsymbol{B}$ are independent of time. Maxwell equations of magneto-statics can be written, as follows:

$$Curl_{\boldsymbol{X}}\, \boldsymbol{H} = \boldsymbol{0}, \qquad Div_{\boldsymbol{X}}\, \boldsymbol{B} = 0, \tag{1}$$

It is worth mentioning that Equation (1) is resulted from no electric field, no free charge and no current density assumptions on the continuum. Equation (1) is used to define a scalar potential $\varphi(\boldsymbol{X})$, such that $\boldsymbol{H} = -Grad_{\boldsymbol{X}}\varphi$, since $Curl_{\boldsymbol{X}}\, \boldsymbol{H} = Curl_{\boldsymbol{X}}(-Grad_{\boldsymbol{X}}\varphi) = \boldsymbol{0}$. Thus, the magneto-statics differential equations are solved for the scalar potential.

In the absence of body forces, the equilibrium equation on the micro-scale reads:

$$Div_{\boldsymbol{X}}\, \boldsymbol{T} = 0, \tag{2}$$

where $\boldsymbol{T}$ is the nominal stress tensor defined at reference configuration. Equations (1) and (2) are coupled governing equations of the magneto-elastic continuum. Constitutive relations of the magneto-elastic medium are derived from a nonlinear magnetoelastic energy density, $\Psi = \Psi(\boldsymbol{F}, \boldsymbol{H})$, which is a function of the deformation gradient tensor and magnetic field vector, defined per unit volume at $\mathbb{C}_{\mathrm{r}}$. For a compressible material, constitutive relations for the nominal stress and the magnetic induction are:

$$\boldsymbol{T} = \frac{\partial \Psi}{\partial \boldsymbol{F}}, \quad \boldsymbol{B} = -\frac{\partial \Psi}{\partial \boldsymbol{H}} \qquad \text{in } \mathbb{C}_r \tag{3}$$

In macro-scale, the volume occupied by a body in reference (undeformed) configuration is denoted by $\overline{\mathbb{C}_r}$, which is bounded by $\overline{\partial\mathbb{C}_r}$ and notations $\overline{\mathbb{C}_t}$ and $\overline{\partial\mathbb{C}_t}$ are assigned for the corresponding deformed configuration of the continuum. $\bar{\mathbf{X}}$ and $\bar{\mathbf{x}}$ are associated with the macroscopic Lagrangian and Eulerian coordinates, respectively. The macroscopic deformation mapping, $\bar{\chi}$ follows $\bar{\boldsymbol{x}} = \bar{\chi}(\bar{\boldsymbol{X}}, t)$. Accordingly, $Grad_{\bar{X}}$, $Div_{\bar{X}}$ and $Curl_{\bar{X}}$ are used for macro-scale differential operators in Lagrangian coordinates. Hence, $\bar{\boldsymbol{F}} = Grad_{\bar{X}}\,\bar{\boldsymbol{x}} = \partial\bar{\chi}/\partial\bar{\boldsymbol{X}}$ is the macroscopic deformation gradient tensor. Consequently, coupled governing equations of the continuum are:

$$Curl_{\bar{X}}\,\bar{\boldsymbol{H}} = \mathbf{0}, \qquad Div_{\bar{X}}\,\bar{\boldsymbol{B}} = 0, \qquad Div_{\bar{X}}\,\bar{\boldsymbol{T}} = 0 \tag{4}$$

where $\bar{\boldsymbol{H}}$, $\bar{\boldsymbol{B}}$ and $\bar{\boldsymbol{T}}$ are macroscopic magnetic field vector, magnetic induction vector and nominal stress tensor, respectively. Similarly, macroscopic quantities can be related to a macroscopic energy function $\bar{\Psi} = \bar{\Psi}(\bar{\boldsymbol{F}}, \bar{\boldsymbol{H}})$ through:

$$\bar{\boldsymbol{T}} = \frac{\partial\bar{\Psi}}{\partial\bar{\boldsymbol{F}}}, \quad \bar{\boldsymbol{B}} = -\frac{\partial\bar{\Psi}}{\partial\bar{\boldsymbol{H}}} \qquad \text{in } \overline{\mathbb{C}_r} \tag{5}$$

Derivation of constitutive laws through a macroscopic energy function is beyond the scope of this study. Computation of macroscopic quantities is performed through surface integrals of corresponding microscopic counterparts across the ensemble's boundary. Notations, $< \blacksquare >_{\mathbb{C}_r} = \frac{1}{V}\int_{\mathbb{C}_r} \blacksquare\, dV$ and $\sqsubset \blacksquare \sqsupset_{\partial\mathbb{C}_r} = \frac{1}{V}\oint_{\partial\mathbb{C}_r} \blacksquare\, dS$, are introduced for volume and surface integrals on the body's domain and boundary, respectively, where $V$ is the volume of the domain in the reference configuration. Assuming the continuity of the deformation gradient tensor and the magnetic induction vector on the boundary of the SVE, the surface integral can be equivalently estimated as the volume integral of corresponding properties on the SVE's domain. Using Gauss theorem, macroscopic deformation and nominal stress tensors are given by:

$$\overline{\boldsymbol{F}} =< \boldsymbol{F} >_{\boldsymbol{SVE}} = \sqsubset \boldsymbol{x} \otimes \boldsymbol{n} \sqsupset_{\boldsymbol{\partial SVE}} \ , \qquad \overline{\boldsymbol{T}} =< \boldsymbol{T} >_{\boldsymbol{SVE}} = \sqsubset \boldsymbol{t} \otimes \boldsymbol{X} \sqsupset_{\boldsymbol{\partial SVE}} \qquad (6)$$

where $\boldsymbol{x}$, $\boldsymbol{t}$ and $\boldsymbol{n}$ are the position vector, traction vector, and normal vector on the boundary of the SVE, respectively. Likewise, the corresponding magnetic field and magnetic induction vectors in macro-scale are defined by integral equations:

$$\overline{\boldsymbol{H}} =< \boldsymbol{H} >_{\boldsymbol{SVE}} = \sqsubset \varphi \boldsymbol{n} \sqsupset_{\boldsymbol{\partial SVE}} \ , \qquad \overline{\boldsymbol{B}} =< \boldsymbol{B} >_{\boldsymbol{SVE}} = \sqsubset b \boldsymbol{X} \sqsupset_{\boldsymbol{\partial SVE}} \qquad (7)$$

where $\varphi$ and $b = \boldsymbol{B}.\boldsymbol{n}$ are the magnetic potential and magnetic flux on the boundary of the SVE, respectively.

The microscopic position vector can be expressed as a linear function of the macroscopic deformation gradient and a fluctuation field; $\boldsymbol{x}(\boldsymbol{X}) = \overline{\boldsymbol{F}}\boldsymbol{X} + \boldsymbol{\mathscr{g}}(\boldsymbol{X})$, where $\boldsymbol{\mathscr{g}}(\boldsymbol{X})$ is the fluctuation field, a vector function of the position vector. Similarly, the magnetic potential follows $\varphi(\boldsymbol{X}) = \overline{\boldsymbol{H}}\boldsymbol{X} + \mathscr{h}(\boldsymbol{X})$, where $\mathscr{h}(\boldsymbol{X})$ is a scalar function of $\boldsymbol{X}$. Boundary conditions of the SVE are derived from the classical Hill-Mandel homogeneity condition. Hill-Mandel condition, $\delta\overline{\Psi} =< \delta\Psi >$ states that the increment of the macroscopic energy function is equivalent to the volume average of the increment of the microscopic energy function,.

$$\begin{aligned}
&< \delta\Psi >_{\mathrm{SVE}} =< \boldsymbol{T}: \delta\boldsymbol{F} >_{\mathrm{SVE}} -< \boldsymbol{B} \cdot \delta\boldsymbol{H} >_{\mathrm{SVE}} =< \boldsymbol{T} >_{\mathrm{SVE}}: \delta\overline{\boldsymbol{F}} -< \boldsymbol{B} >_{\mathrm{SVE}} \cdot \delta\overline{\boldsymbol{H}} +< \\
&\boldsymbol{T}: Grad_{\boldsymbol{X}}\delta\boldsymbol{\mathscr{g}} + \delta\boldsymbol{\mathscr{g}}.Div_{\boldsymbol{X}}\boldsymbol{T} >_{\mathrm{SVE}} -< \boldsymbol{B} \cdot Grad_{\boldsymbol{X}}\delta\boldsymbol{\mathscr{g}} + \delta\mathscr{h}.Div_{\boldsymbol{X}}\boldsymbol{H} >_{\mathrm{SVE}} \\
&= \overline{\boldsymbol{T}}: \delta\overline{\boldsymbol{F}} - \overline{\boldsymbol{B}} \cdot \delta\overline{\boldsymbol{H}} + \sqsubset \boldsymbol{t} \cdot \delta\boldsymbol{\mathscr{g}} \sqsupset_{\boldsymbol{\partial SVE}} - \sqsubset b\delta\mathscr{h} \sqsupset_{\boldsymbol{\partial SVE}} \\
&= \delta\overline{\Psi} + + \sqsubset \boldsymbol{t} \cdot \delta\boldsymbol{\mathscr{g}} \sqsupset_{\boldsymbol{\partial SVE}} - \sqsubset b\delta\mathscr{h} \sqsupset_{\boldsymbol{\partial SVE}} \ ,
\end{aligned} \qquad (8)$$

To satisfy the Hill's condition, both the mechanical and magnetic boundary integrals in Equation (8) are required to vanish through defining appropriate boundary conditions. Two types of boundary conditions are considered here:

- Linear displacement boundary conditions (LD-BC) with zero mean fluctuation fields; $\bar{\boldsymbol{g}} = 0$ and $\bar{\boldsymbol{h}} = 0$.
- Periodic boundary condition (PF-BC) for fluctuation fields; $\boldsymbol{g}(\boldsymbol{X})$ and $\boldsymbol{h}(\boldsymbol{X})$ and anti-periodic condition for $\boldsymbol{t}$ and $b$ [46-48]:

$$\begin{aligned} &\boldsymbol{g}(\boldsymbol{X}^-) = \boldsymbol{g}(\boldsymbol{X}^+), \qquad \boldsymbol{h}(\boldsymbol{X}^-) = \boldsymbol{h}(\boldsymbol{X}^+), \\ &\boldsymbol{t}(\boldsymbol{X}^-) = -\boldsymbol{t}(\boldsymbol{X}^+), \qquad b(\boldsymbol{X}^-) = -b(\boldsymbol{X}^+), \qquad \text{on } \partial SVE \end{aligned} \tag{9}$$

where superscripts (+) and (–) are associated with nodes on opposite boundaries (right (top) and left (bottom) edges on Figure 43 (b)) of the $\partial SVE$. One may argue that the use of periodic boundary conditions is not suitable for a random heterogeneous structure. It has been shown that adopting periodic boundary conditions can estimate the effective properties, even when the structure is characterized with random distribution of the inclusions [53-61].

In order to define the boundary value problem (BVP), macro-scale boundary conditions on $\partial\mathbb{C}_r$ for mechanical and magnetic problem must be specified. Figure 43(a) shows a schematic of the BVP. The boundary of the homogenous body is decomposed into different sections, where corresponding Dirichlet and Neumann boundary conditions for mechanical and magnetic problems are prescribed. Displacement type boundary conditions, $\bar{\boldsymbol{u}} = \bar{\boldsymbol{u}}_{\mathbf{0}}$ on $\partial\bar{\mathbb{C}}_u$ as well as traction boundary conditions $\bar{\boldsymbol{t}} = \bar{\boldsymbol{T}}.\bar{\boldsymbol{n}}$ on $\partial\bar{\mathbb{C}}_\sigma$ are prescribed to define mechanical BVP, where $\bar{\boldsymbol{u}}$, $\bar{\boldsymbol{t}}$ and $\bar{\boldsymbol{n}}$ are the macroscopic displacement, traction and normal to the surface vectors defined on the boundary of the body, respectively. The corresponding magnetic BVP is defined through scalar magnetic potential boundary condition; $\bar{\varphi} = \bar{\varphi}_0$ on $\partial\bar{\mathbb{C}}_\varphi$ and magnetic flux boundary condition; $\bar{b} = \bar{\boldsymbol{B}}.\bar{\boldsymbol{n}}$ on $\partial\bar{\mathbb{C}}_b$, where $\bar{\varphi}$ and $\bar{b}$ are the macroscopic magnetic potential and magnetic flux, respectively.

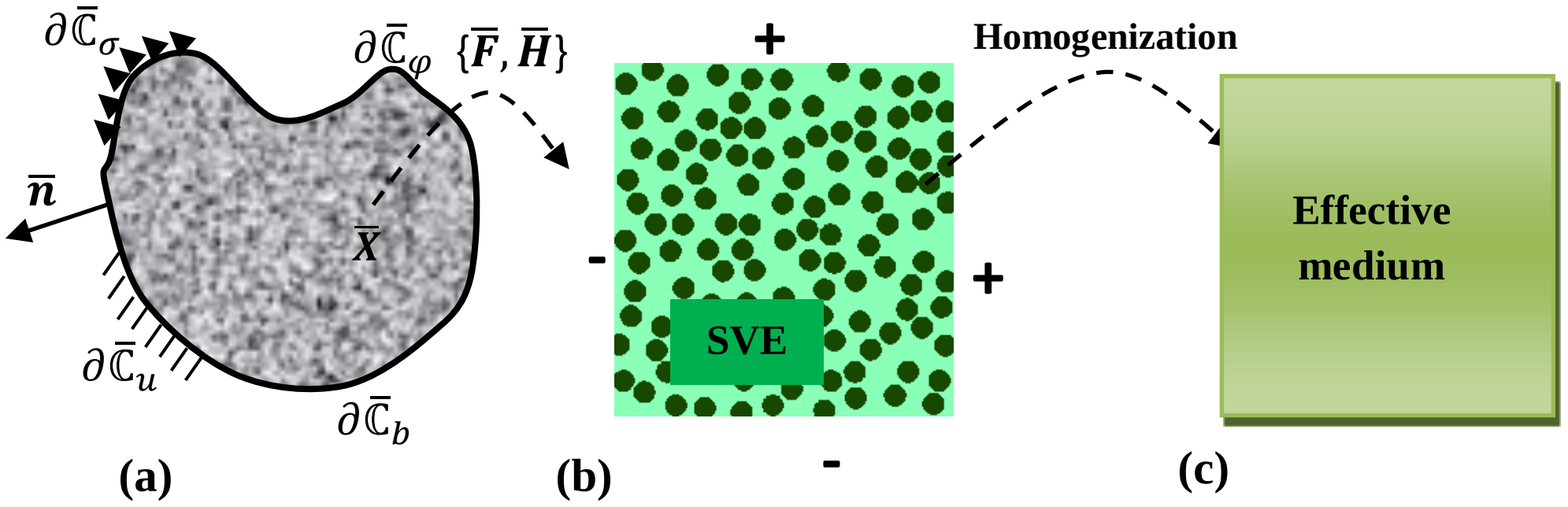

**Figure 43.** **(a)** The homogenized body and corresponding boundary decomposition in Lagrangian configuration, **(b)** corresponding SVE, attached to $\bar{\boldsymbol{X}}$, selected for statistical analysis including randomly distributed circular permeable particle inside a soft matrix **(c)** the effective medium modeled using homogenization approach.

The macroscopic deformation gradient tensor and magnetic field are obtained from macro-scale BVP which are inputs of the micro-scale problem. The transition from microscopic to macroscopic properties is defined on the micro-scale problem through FEM-based averaging process. Thus, the effective nominal stress and magnetic induction as well as macroscopic (effective) moduli tensors are computed from corresponding microscopic properties on the SVE domain.

To develop a finite element model, variational formulation of the equilibrium and magneto-statics equations is derived. To derive the weak form of governing equations, inner product of Equations $(1)_2$ and (2) with an arbitrary test function, is considered and then integrated over the SVE domain. Let us consider $\delta\boldsymbol{x}$ and $\delta\varphi$ be arbitrary variations of $\boldsymbol{x}$ and $\varphi$, respectively, that satisfy the boundary conditions on $\partial SVE$. Taking the variational form of Equations $(1)_2$ and (2), using integration by part and then the divergence theorem yields:

$$\int_{SVE} \boldsymbol{T} : Grad_{\boldsymbol{X}} \delta\boldsymbol{x} \, dV - \oint_{\partial SVE} \boldsymbol{t} . \delta\boldsymbol{x} \, dS = 0 \tag{10}$$

$$\int_{SVE} \boldsymbol{B} : Grad_{\boldsymbol{X}} \delta\varphi \, dV - \oint_{\partial SVE} b\delta\varphi \, dS = 0 \tag{11}$$

It is noted that the weak form of magneto-statics equation results from a stationary magnetic field condition and a non-conducting MEC medium in the absence of surface current density on $\partial$SVE. Natural boundary conditions in Equations (10) and (11) appear as work terms applied to $\partial SVE$, which arise from the normal traction force and the magnetic flux. For linear displacement boundary conditions, the traction $\boldsymbol{t}$ and magnetic flux $b$ are calculated from the applied displacement field and magnetic potential on the boundaries of the ensemble. The FEM solver automatically computes the surface tractions from the displacement type boundary conditions, and applies them as work terms in the variational formulations.

For periodic boundary conditions, recalling Equation (9), it is evident from Figure 43(b) that boundary integrals $\oint_{\partial SVE} \boldsymbol{t} . \delta\boldsymbol{x} \, dS$, in Equation (10) and $\oint_{\partial SVE} b\delta\varphi \, dS$, in Equation (11) vanish, since the normal unit vector, $\boldsymbol{n}$ acts in opposite directions on the parallel boundaries of the deformed SVE. Weak forms in Equations (10) and (11) define the coupled magnetoelastic behavior of the model which are solved utilizing the nonlinear FEM solver. COMSOL Multiphysics is used for numerical simulations which allows for direct implementation of weak expressions.

To predict the overall response of the MEC through homogenization approach, computation of the effective tangent moduli is required. Homogenized tangent moduli tensors contribute to constitutive laws of macro-scale BVP. In contrast to macroscopic quantities; $\overline{\boldsymbol{F}}, \overline{\boldsymbol{T}}, \overline{\boldsymbol{H}}$ and $\overline{\boldsymbol{B}}$, which can be directly computed by Equations (6) and (7), the overall tangent moduli $\frac{\partial \overline{\boldsymbol{T}}}{\partial \overline{\boldsymbol{F}}}, \frac{\partial \overline{\boldsymbol{T}}}{\partial \overline{\boldsymbol{H}}}, \frac{\partial \overline{\boldsymbol{B}}}{\partial \overline{\boldsymbol{F}}}, \frac{\partial \overline{\boldsymbol{B}}}{\partial \overline{\boldsymbol{H}}}$ cannot be obtained through volume averaging of microscopic counterparts. This is due to the fact that there is no explicit relation of macroscopic nominal stress tensor and magnetic induction vector as a function of $\overline{\boldsymbol{F}}$ and $\overline{\boldsymbol{H}}$. Incremental (linearized) constitutive relations of the coupled magnetoelastic BVP can be estimated, as follows:

$$\begin{aligned} \Delta\overline{\boldsymbol{T}} &= \frac{\partial \overline{\boldsymbol{T}}}{\partial \overline{\boldsymbol{F}}} : \Delta\overline{\boldsymbol{F}} + \frac{\partial \overline{\boldsymbol{T}}}{\partial \overline{\boldsymbol{H}}} \cdot \Delta\overline{\boldsymbol{H}} = \overline{\mathbb{A}} : \Delta\overline{\boldsymbol{F}} + \overline{\mathbb{C}} \cdot \Delta\overline{\boldsymbol{H}} \ , \\ \Delta\overline{\boldsymbol{B}} &= \frac{\partial \overline{\boldsymbol{B}}}{\partial \overline{\boldsymbol{F}}} : \Delta\overline{\boldsymbol{F}} + \frac{\partial \overline{\boldsymbol{B}}}{\partial \overline{\boldsymbol{H}}} \cdot \Delta\overline{\boldsymbol{H}} = \overline{\mathbb{C}}^{\boldsymbol{T}} : \Delta\overline{\boldsymbol{F}} + \overline{\mathbb{B}} \cdot \Delta\overline{\boldsymbol{H}} \end{aligned} \tag{12}$$

where $\overline{\mathbb{A}}, \overline{\mathbb{C}}$ and $\overline{\mathbb{B}}$ are macroscopic mechanical, magneto-mechanical and magnetic moduli tensors, respectively. Taking the gradient of displacement vector and magnetic potential defined for periodic homogenization, the deformation gradient and magnetic field can be decomposed into a constant and a fluctuation part $\boldsymbol{F} = \overline{\boldsymbol{F}} + \widetilde{\boldsymbol{F}}$ and $\boldsymbol{H} = \overline{\boldsymbol{H}} + \widetilde{\boldsymbol{H}}$, where $\widetilde{\boldsymbol{F}} = Grad_{\boldsymbol{X}} \boldsymbol{g}(\boldsymbol{X})$ and $\widetilde{\boldsymbol{H}} = Grad_{\boldsymbol{X}} \boldsymbol{h}(\boldsymbol{X})$. Substituting in moduli tensor relations one has,

(13)

$$\bar{\mathbb{A}} =< \frac{\partial \boldsymbol{T}}{\partial \boldsymbol{F}} : \frac{\partial (\bar{\boldsymbol{F}}+\tilde{\boldsymbol{F}})}{\partial \bar{\boldsymbol{F}}} >_{SVE} =< A >_{SVE} +< A : \frac{\partial \tilde{\boldsymbol{F}}}{\partial \bar{\boldsymbol{F}}} >_{SVE},$$

$$\bar{\mathbb{C}} =< \frac{\partial \boldsymbol{T}}{\partial \boldsymbol{H}} \cdot \frac{\partial (\bar{\boldsymbol{H}}+\tilde{\boldsymbol{H}})}{\partial \bar{\boldsymbol{H}}} >_{SVE} =< C >_{SVE} +< C \cdot \frac{\partial \tilde{\boldsymbol{H}}}{\partial \bar{\boldsymbol{H}}} >_{SVE}$$

$$\bar{\mathbb{B}} =< \frac{\partial \boldsymbol{B}}{\partial \boldsymbol{H}} \cdot \frac{\partial (\bar{\boldsymbol{H}}+\tilde{\boldsymbol{H}})}{\partial \bar{\boldsymbol{H}}} >_{SVE} =< B >_{SVE} +< B \cdot \frac{\partial \tilde{\boldsymbol{H}}}{\partial \bar{\boldsymbol{H}}} >_{SVE}$$

where $\mathbb{A}, \mathbb{C}$ and $\mathbb{B}$ are microscopic counterparts of moduli tensors. In Equations (13), the challenge is to compute partial derivatives of fluctuation fields; $\frac{\partial \tilde{\boldsymbol{F}}}{\partial \bar{\boldsymbol{F}}}$ and $\frac{\partial \tilde{\boldsymbol{H}}}{\partial \bar{\boldsymbol{H}}}$, since there is no explicit expression of microscopic fluctuation fields in terms of macroscopic variables, $\bar{\boldsymbol{F}}$ and $\bar{\boldsymbol{H}}$. Computation of the sensitivity of $\tilde{\boldsymbol{F}}$ and $\tilde{\boldsymbol{H}}$, with respect to their macroscopic counterparts needs to be performed through numerical methods presented in chapter two.

### 6.2 Random homogenization framework

In this study, the homogenized response of a 2D magnetoactive heterogeneous material is sought through random homogenization process. A scale dependent statistical approach is employed to perform the convergence analysis on the ensembles of the random MEC following the approaches documented in [53-61].

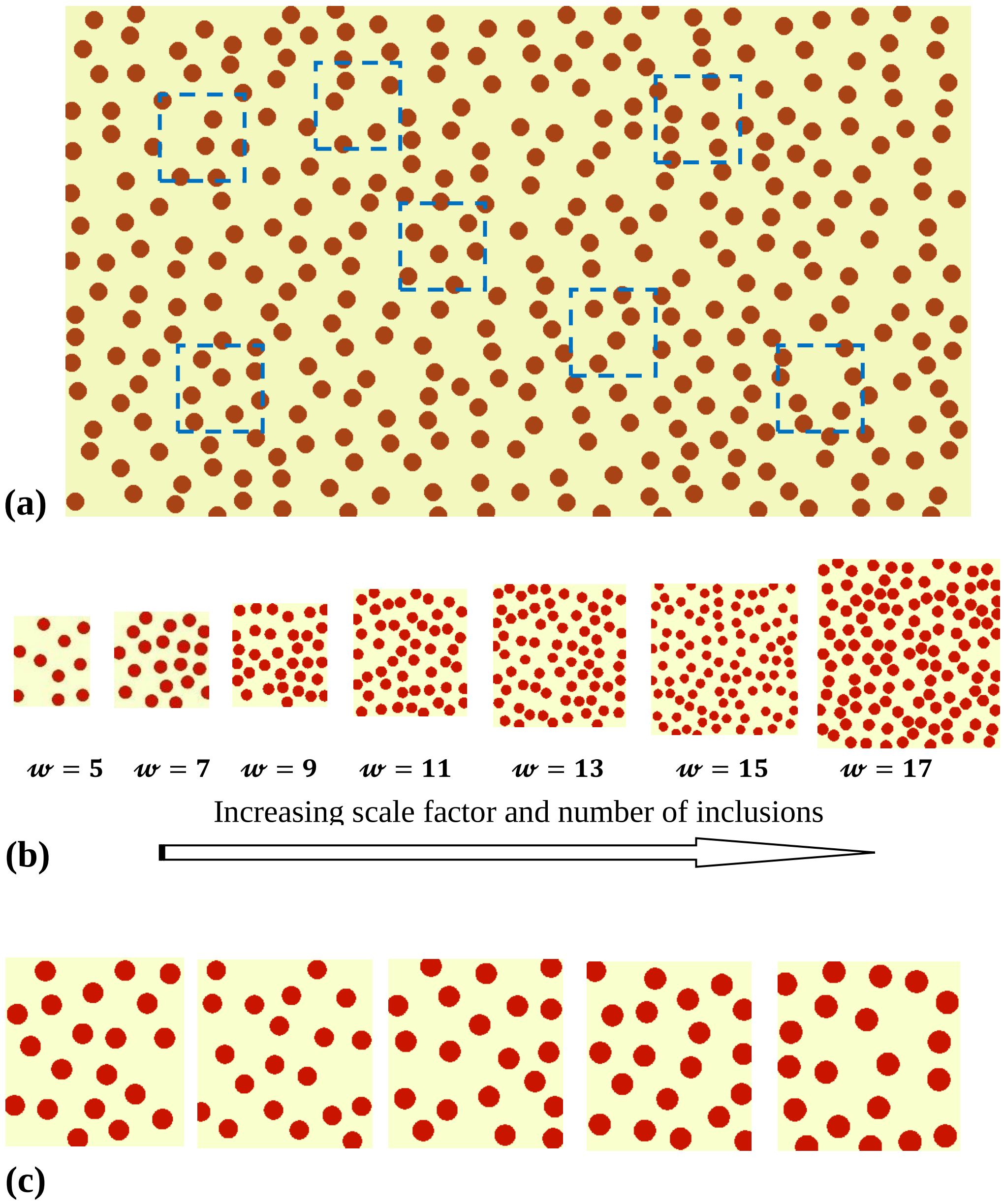

**Figure 44.** Schematic of the statistical algorithm used in the study. **(a)** selection of random meso-scale ensembles of the heterogeneous MEC for a typical test window size, **(b)** increasing the size of the test window and number of particles towards the convergence window and **(c)** five typical realizations for $\boldsymbol{w} = \mathbf{7}$. All ensembles are generated for $\boldsymbol{A_f} = \mathbf{0.35},\ \boldsymbol{d} = \mathbf{10}\ \textbf{and}\ \boldsymbol{d_c} = \mathbf{1}$.

The magnetoactive material is a two-phase composite consists of randomly distributed hard and permeable inclusions in a soft non-magnetic matrix. It is assumed that the Hill's general homogenization limit between the size of heterogeneity, $d$ and that of CVE, $L$ is satisfied; $d \ll L$.

The effective behavior of composite material depends on properties of microstructure's constituents. SVE is an area of the random heterogeneous material that is chosen to calculate the overall properties. In random homogenization, the problem is to find the appropriate size of the SVE which can represent the heterogeneous material properties, called CVE. Here, the goal is to develop a FEM-based random homogenization approach which uses the Monte-Carlo method to generate random SVEs and capture the effective properties.

Figure 44 shows the schematic of statistical approach and generation of the meso-scale SVE ensembles in the random MEC. The statistical approach is equivalent to moving limited number of square-shaped frames at different points of the heterogeneous structure, where the size of the square frames is successively increased step by step till the desired CVE size is approached (Figure 44(b)). Typical realizations of the random structure for $\mathcal{w} = 7$ are shown in Figure 44(c).

The computational algorithm utilized in the simulation is depicted in Figure 45. First, constant parameters of simulations: $A_f$, $\mu^i$, $\lambda^i$, $\mu^m$, $\lambda^m$, $d$, $d_c$, $Tol$ are input in the model. An area fraction $A_f$, is assigned to find the number of inclusions inside the square ensemble. $A_f$ is defined as the ratio of total area of inclusions to the area of the SVE. Material properties of the matrix and inclusions, $A_f$, particle diameter, $d$, and distance between particles, $d_c$ are held constant in simulations. Next, a window size, $L$ is assigned to generate the realization, $R_{w}$, where $w = \frac{L}{d}$ is a non-dimensional scale factor, discretely increased in a loop till the convergence point is achieved. Each $R_{w}$ is generated through random positioning of circular inclusions in the square matrix, using a so called hard-core Poisson point field, so that the circles are prevented from overlapping, by assigning a minimum distance $d_c$ between the boundaries of inclusions. For each ensemble $R_{w}$, the variational formulations (10) and (11) are solved to find the components of the constitutive laws and moduli tensors. Steps 1-3 are repeated in a loop for each $w$ and for maximum of 45 independent simulations. For each $w$, the number of realizations, *N* is determined through $\frac{1.96\sigma}{\bar{x}\sqrt{N}} \leq Tol$, where $\sigma = \sqrt{\frac{1}{N-1}\sum_{i=1}^{N}(x_i - \bar{x})^2}$ is the standard deviation, $\bar{x}$ is the average of the selected component of the macroscopic moduli tensors and *Tol* is the assigned tolerance [54]. Dispersion of the data is checked through calculation of *N* at each step.

$\overline{\mathbb{A}} = \frac{\overline{\mathbb{A}}_{1111}+\overline{\mathbb{A}}_{2222}}{2}$ and $\overline{\mathbb{B}} = \frac{\overline{\mathbb{B}}_{11}+\overline{\mathbb{B}}_{22}}{2}$, are defined as the selected components of the moduli tensors being monitored for statistical accuracy. The algorithm stops when the statistical accuracy is achieved and the corresponding ensemble size, $w^*$ is qualified as the desired CVE. The CVE is defined as the minimum window size for which the number of required realizations *N*, is less than 5 for both parameters $\overline{\mathbb{A}}$ and $\overline{\mathbb{B}}$. In other words, the average components of moduli tensors remain in the tolerance interval and the minimum dispersion in $\overline{\mathbb{A}}$ and $\overline{\mathbb{B}}$ are observed.

- Assign inputs: $A_f$, $\mu^i$, $\lambda^i$, $\mu^m$, $\lambda^m$, $d$, $d_c$, *Tol.*
- *Loop* $w$*: 5*$\longmapsto\infty$. Subsequently increase the size of SVE till the convergence is achieved.
  - *Loop* $t$:1$\longmapsto N$.

    Calculate number of particles, $= 4A_f L^2/(\pi(d+d_c)^2)$ .
    Calculate size of SVE square $L = wd$.
    Generate spatially random distribution of *n* particles in square ensembles.
    Solve the variational formulations (10) and (11).
    Compute constitutive laws and overall moduli tensors over each SVE.
    Compute $\overline{\mathbb{A}} = \frac{\overline{\mathbb{A}}_{1111}+\overline{\mathbb{A}}_{2222}}{2}$ and $\overline{\mathbb{B}} = \frac{\overline{\mathbb{B}}_{11}+\overline{\mathbb{B}}_{22}}{2}$.
    Calculate number of realizations, *N* from $\frac{1.96\sigma}{\bar{x}\sqrt{N}} \leq Tol$ for each parameters $\overline{\mathbb{A}}$ and $\overline{\mathbb{B}}$.
    If $N < 5$

    *End Loop t*

*End Loop* $w$

**Figure 45.** Computational algorithm used in the statistical approach.

### 6.3 Results and Discussion

The FEM-based homogenization approach presented in this study is carried out on a typical two dimensional SVE consisting of permeable inclusions and a matrix shown in Figure 43(b). The SVE is analyzed under a deformation and magnetic field loading state to compute the distribution of microscopic nominal stress and magnetic induction as well as effective counterparts. Both inclusion and matrix are modeled as a compressible neo-Hookean magnetoelastic material. To conduct numerical analysis, a particular form of energy function is required. Due to lack of experimental data on MECs, limited data are available in the literature. In this study, a typical magnetoelastic energy function is considered:

$$\Psi = \frac{\mu}{2}(tr(\boldsymbol{c}) - 3) - \mu\, lnJ + \frac{\lambda}{2}(lnJ)^2 + pJ^{-2/3}(\boldsymbol{cH}.\boldsymbol{H}) \tag{13}$$

where $\boldsymbol{c} = \boldsymbol{F^T F}$ is the right Cauchy–Green deformation tensor, $tr(\boldsymbol{c})$ is the trace of $\boldsymbol{c}$ and $J = det\boldsymbol{F}$. $\mu$ and $\lambda$ are Lamé constants and $p$ is the magnetic permeability. Material properties of the inclusion are chosen as $\mu^i = 80\ GPa$, $\lambda^i = 120\ GPa$ and $p^i = 1250\mu_0$ and those of the matrix are $\mu^m = 4$ MPa, $\lambda^m = 30\ MPa$ and $p^m = \mu_0$, where $\mu_0 = 1.256 \times 10^{-6} NA^{-2}$ is the magnetic permeability of the vacuum [45, 46, 47].

The nominal stress tensor is calculated as:

$$\boldsymbol{T}_{ij} = \mu \boldsymbol{F}_{ij} + [\lambda lnJ - \mu]\boldsymbol{F^{-T}}_{ij} + 2pJ^{-\frac{2}{3}}\left(\boldsymbol{H}_i \boldsymbol{F}_{jk} \boldsymbol{H}_k - \frac{\boldsymbol{1}}{\boldsymbol{3}}\boldsymbol{F^{-T}}_{km}\boldsymbol{H}_m(\boldsymbol{c}_{kn}\boldsymbol{H}_n)\boldsymbol{\delta}_{ij}\right) \quad (14)$$

where $\boldsymbol{F^{-T}}$ is the inverse matrix of $\boldsymbol{F^T}$ and $\boldsymbol{\delta}_{ij}$is the identity tensor. The magnetic induction vector would be:

$$\boldsymbol{B}_i = [-\frac{\partial \Psi}{\partial \boldsymbol{H}}]_i = -2pJ^{-\frac{2}{3}}\boldsymbol{c}_{ij}\boldsymbol{H}_j \quad (15)$$

The microscopic moduli tensors are derived as:

$$\mathbb{A}_{mnpq} = [\frac{\partial \boldsymbol{T}}{\partial \boldsymbol{F}}]_{mnpq} = [\frac{\partial^2 \Psi}{\partial \boldsymbol{F} \partial \boldsymbol{F}}]_{mnpq} = \lambda \boldsymbol{F}^{-1}_{nm}\boldsymbol{F}^{-1}_{qp} + \mu \delta_{mp}\delta_{nq} + [\lambda lnJ - \mu]\boldsymbol{F}^{-1}_{np}\boldsymbol{F}^{-1}_{qm} - \frac{2}{3}pJ^{\frac{-2}{3}}\left(\boldsymbol{H}_i\boldsymbol{H}_j\left(\frac{-2}{3}\boldsymbol{c}_{ij}\boldsymbol{F}^{-1}_{np}\boldsymbol{F}^{-1}_{qm}\right) + 2\boldsymbol{H}_q\boldsymbol{F}_{pj}\boldsymbol{H}_j\boldsymbol{F}^{-T}_{mn}\right)\right) + pJ^{\frac{-2}{3}}([\left(\frac{-2}{3}\right)\left(2\boldsymbol{H}_q\boldsymbol{F}_{pj}\boldsymbol{H}_j\boldsymbol{F}^{-T}_{mn} + \boldsymbol{H}_i\boldsymbol{H}_j\boldsymbol{c}_{ij}\boldsymbol{F}^{-1}_{np}\boldsymbol{F}^{-1}_{qm}\right)] + \delta_{mp}\boldsymbol{H}_q\boldsymbol{H}_n) \quad (16)$$

$$\mathbb{C}_{\alpha\beta i} = [\frac{\partial \boldsymbol{T}}{\partial \boldsymbol{H}}]_{\alpha\beta i} = [\frac{\partial^2 \Psi}{\partial \boldsymbol{H} \partial \boldsymbol{F}}]_{\alpha\beta i} = 2pJ^{\frac{-2}{3}}[(\delta_{\alpha\beta}\boldsymbol{F}_{i\gamma}\boldsymbol{H}_\gamma + \boldsymbol{H}_\alpha\boldsymbol{F}_{i\beta}) - \frac{2}{3}\boldsymbol{c}_{\alpha j}\boldsymbol{H}_j\boldsymbol{F}^{-T}_{i\beta}] \quad (17)$$

$$\mathbb{B}_{\alpha\beta} = [\frac{\partial \boldsymbol{B}}{\partial \boldsymbol{H}}]_{\alpha\beta} = [\frac{\partial^2 \Psi}{\partial \boldsymbol{H} \partial \boldsymbol{H}}]_{\alpha\beta} = -2pJ^{\frac{-2}{3}}\boldsymbol{c}_{\alpha\beta} \quad (18)$$

A plane strain pure shear at constant magnetic field is considered while magnetic field is kept constant at $\boldsymbol{x_2}$ direction:

$\bar{\boldsymbol{F}} = \begin{bmatrix} 1 & 0.3 \\ 0.3 & 1 \end{bmatrix}$ and $\bar{\boldsymbol{H}} = \begin{bmatrix} 0 \\ 5e5 \end{bmatrix} Am^{-1}$

$A_f = 0.35,\ d = 10,\ d_c = 1$ and $Tol = 0.01$ are assumed constant in the simulations. Input parameters as well as magneto-mechanical loading state are kept fixed for all results presented in this study. Two cases of boundary conditions; LD-BC and PD-BC are considered. The ensemble's size is successively increased in a loop from 9 to the convergence point. FEM simulations are performed using COMSOL Multiphysics. The statistical algorithm is run through a computer code. LiveLink™ *for* MATLAB® interface is used for computer programming, which connects COMSOL Multiphysics to MATLAB scripting environment. Ensemble size, number of inclusions and random position of inclusions are calculated and the SVE is modeled in the COMSOL graphical environment.

Weak expressions (10) and (11) are directly input in the FEM model. Constitutive laws and micro-scale moduli components are input in the model from Equations (14)-(18). The homogenized tangent moduli tensors are computed through the sensitivity approach documented in the previous chapter. The model is run for each $\mathcal{w}$ following the approach explained in section 3.2. The convergence plots are shown in Figures 46 and 47 for moduli components, $\bar{\mathbb{A}}$ and $\bar{\mathbb{B}}$ and different boundary conditions, respectively. $\bar{\mathbb{A}}$ and $\bar{\mathbb{B}}$ are normalized by their corresponding average value at the convergence point. Figure 46(a) plots the statistical convergence plot for component $\bar{\mathbb{A}}$ versus the number of simulations for different values of $\mathcal{w}$ and LD-BC. The number of necessary realizations, $N$ is calculated from $\frac{1.96\sigma}{\bar{x}\sqrt{N}} \leq Tol$ for each $\mathcal{w}$ and at each simulation loop. The convergence is achieved for $\mathcal{w}^* = 23$.

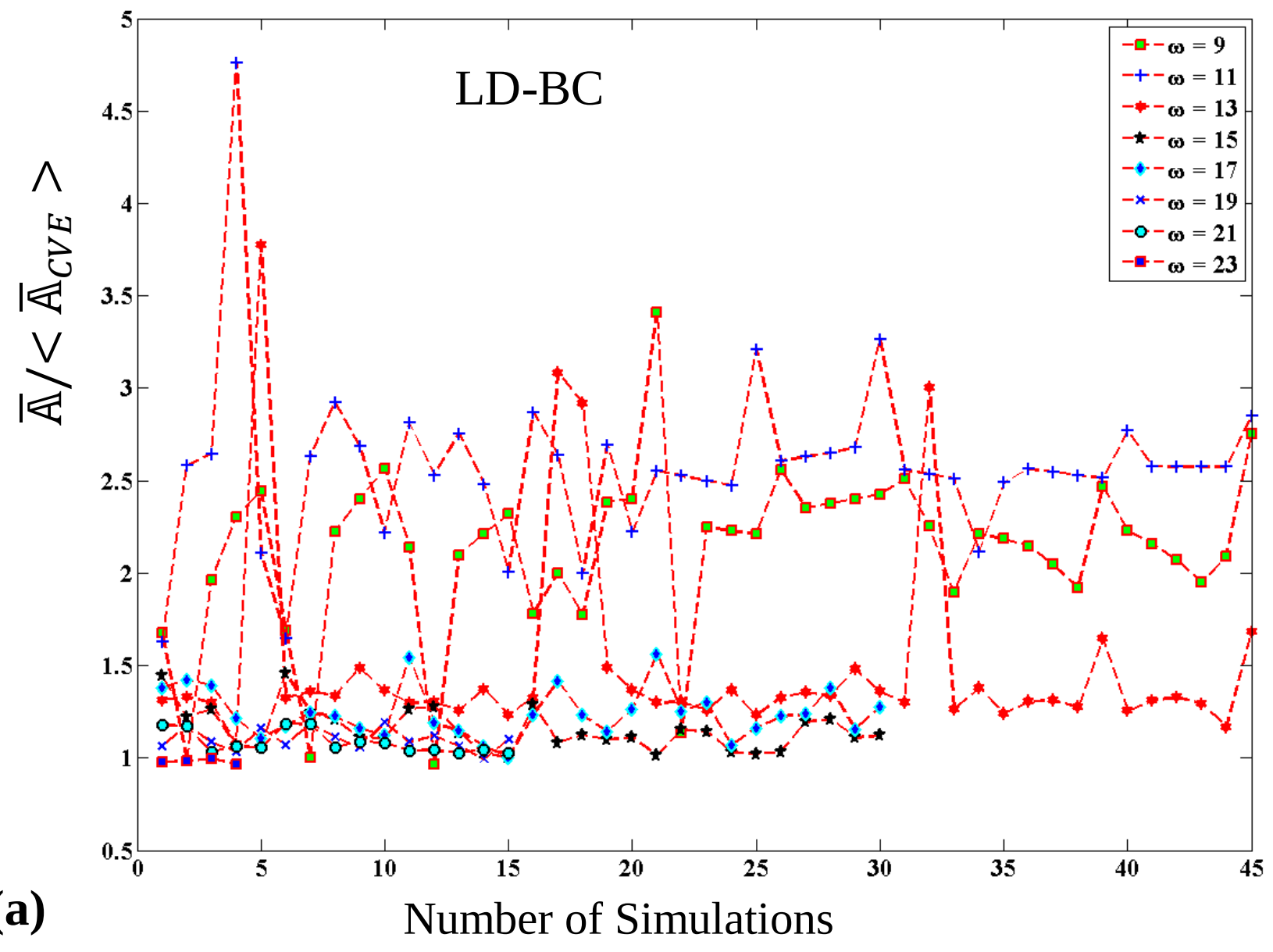

(a)

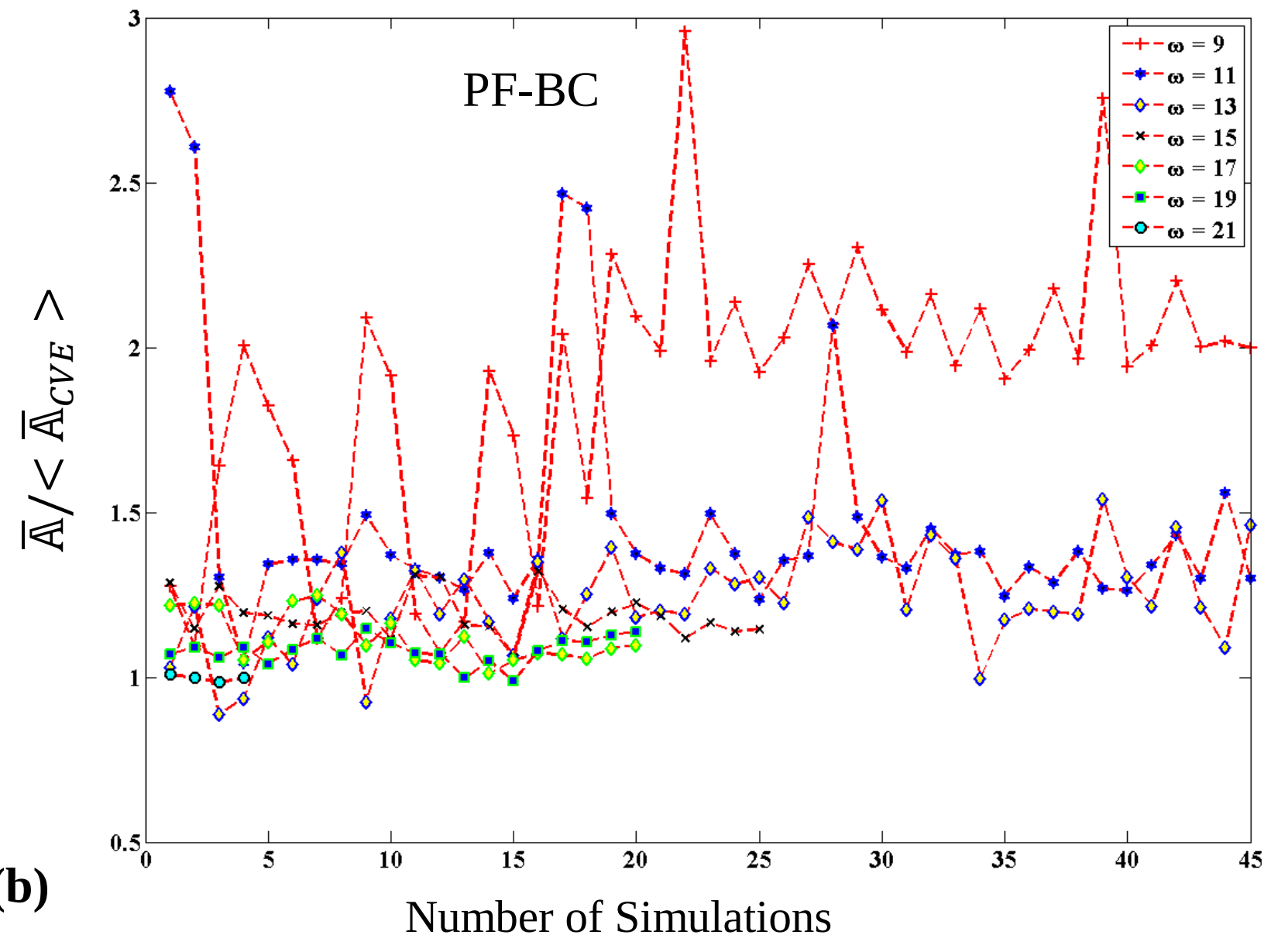

(b)

**Figure 46.** Moduli component $\bar{\mathbb{A}}$ **,** results from the statistical analysis for **(a)** LD-BC and **(b)** PF-BC versus number of simulations.

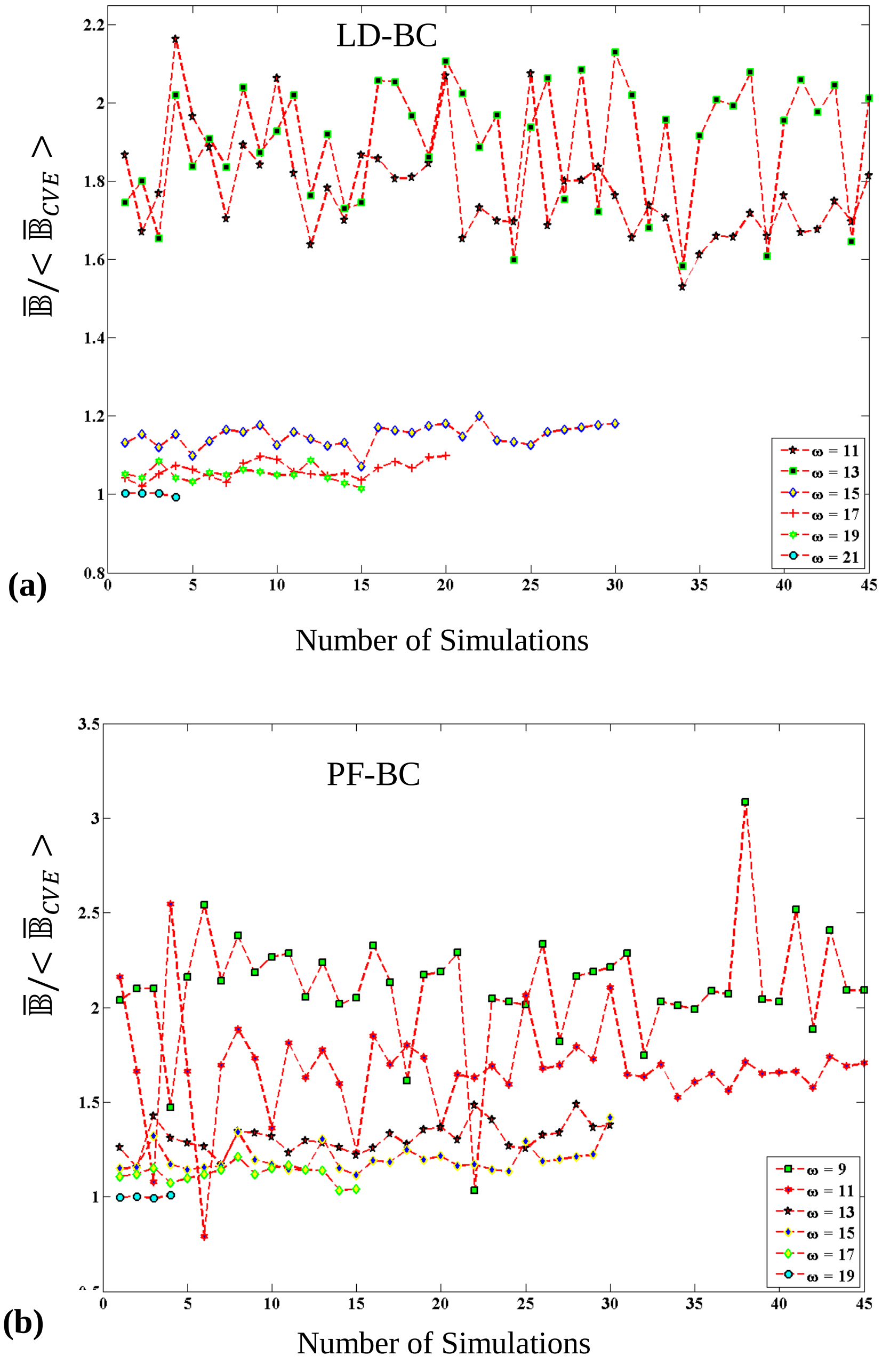

**Figure 47.** Moduli component $\bar{\mathbb{B}}$ **,** results from the statistical analysis for **(a)** LD-BC and **(b)** PF-BC versus number of simulations.

In all convergence plots presented here, the simulations are continued for $w < w^*$ to demonstrate the scattering of the statistical data for different test windows. Figure 46 (b) demonstrates that when PF-BC is adopted, the convergence is occurred at $w^* = 21$. Figure 47 reports the convergence plots for $\overline{\mathbb{B}}$ component of magnetic modulus for LD-BC and LF-BC cases. For both cases of boundary conditions, convergence is reached at $w^* = 21$.

It is observed that in all convergence plots, by increasing the size of SVEs and number of particles, less dispersion is observed in the magnitude of $\overline{\mathbb{A}}$ and $\overline{\mathbb{B}}$ and the convergence is achieved with less number of simulations. It is emphasized that the ensemble size achieved in the statistical approach is only valid for the particular form of the inputs, loading state and boundary conditions. Different size and convergence plots may result for various input and loading conditions. Figure 48 plots the coefficient of variation, $C_v$ versus the scale factor, $w$ corresponds to the convergence plots in Figures 46 and 47. $C_v = \frac{\sigma}{\bar{x}}$ is defined as the ratio between standard deviation to the mean of the statistical data and is a measure of the statistical scattering. Figure 48 confirms that by increasing the size of ensembles, less scattering is observed in the moduli components and $C_v$ approaches to zero.

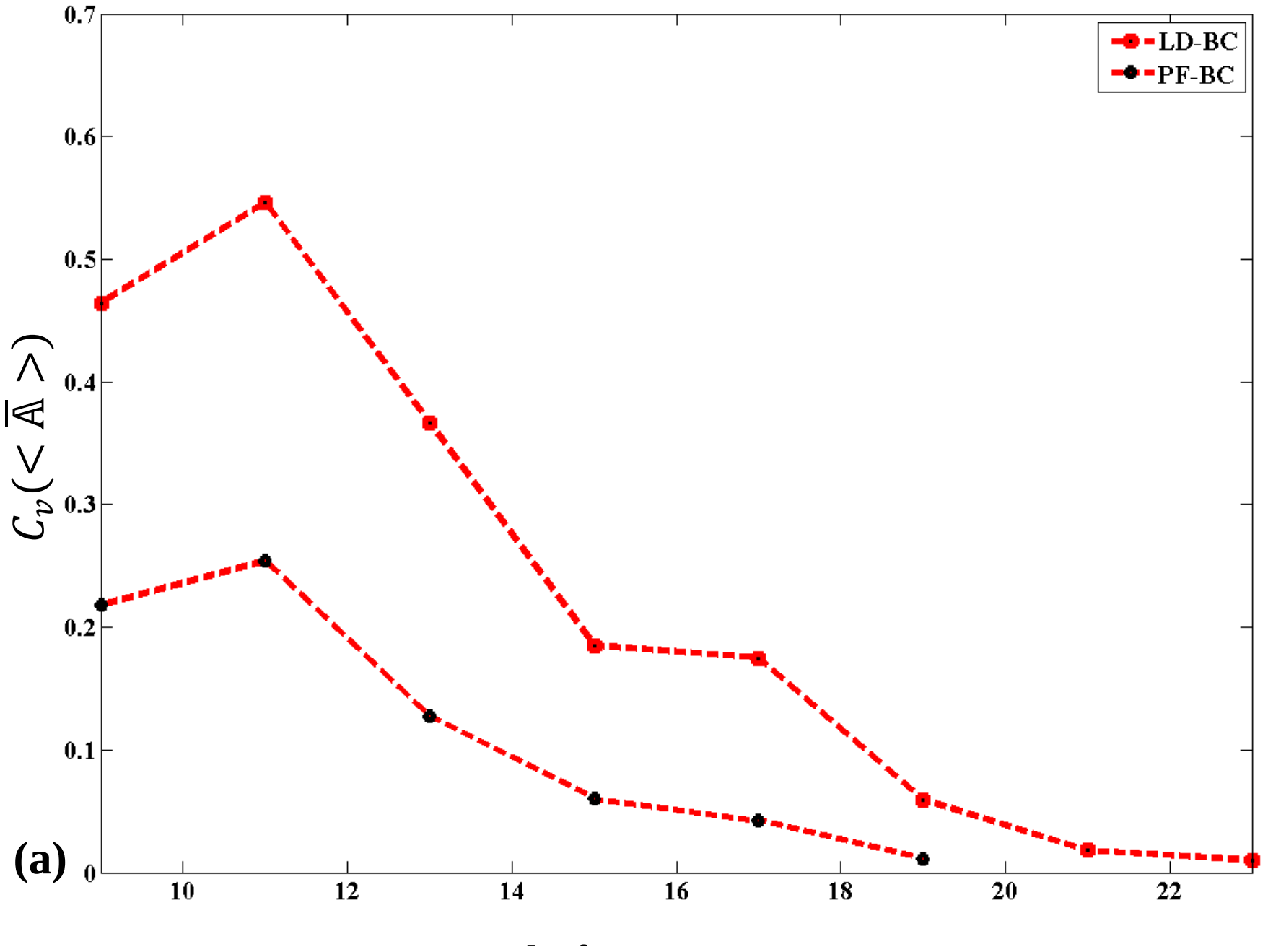

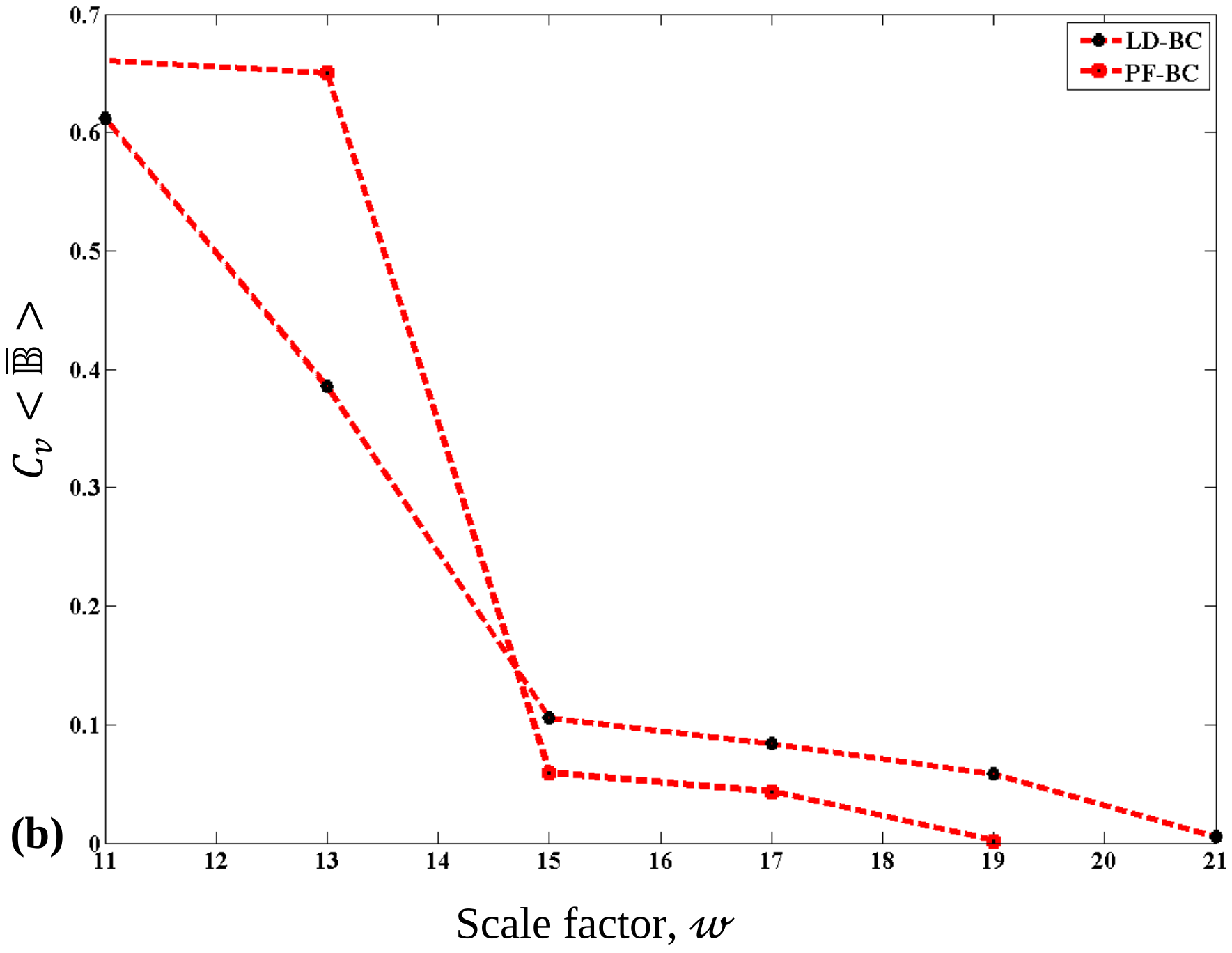

**Figure 48.** Coefficient of variation of **(a)** $< \bar{\mathbb{A}} >$ and **(b)** $< \bar{\mathbb{B}} >$ versus scale factor, $\boldsymbol{w}$.

Figure 49 compares the average of moduli components, $\bar{\mathbb{A}}$ and $\bar{\mathbb{B}}$ versus scale factor, $w$ for both cases of boundary conditions. For $w > 13$, the moduli component $\bar{\mathbb{A}}$ shows the same convergence trend for both cases of LD-BC and PF-BC. The convergence trend for $\bar{\mathbb{B}}$ is shown in Figure 49(b). It is noted that the two types of boundary conditions used in this study, do not define the hierarchies bounds for the corresponding moduli parameters $\bar{\mathbb{A}}$ and $\bar{\mathbb{B}}$ and this plot cannot be used for identification of the CVE size. The CVE is determined statistically through the algorithm demonstrated on Figure 45.

Figure 50 shows the distribution of the microscopic nominal stress component $T_{12}$ for two typical SVE size resulted from different boundary conditions. Solution of the BVP for a meso-scale ensemble with $w = 11$, and 57 number of inclusions for case of LD-BC is shown in Figure 50(a). Result for the same SVE and PF-BC is shown in Figure 50(b). Figure 50(c-d) reports the simulation results for SVE with $w = 13$, and 80 number of inclusions for case of LD-BC and PF-BC, respectively. All contour plots presented here are normalized by the corresponding volumetric average of the quantity concerned. One can notice how the use of PF-BC results in periodic displacement boundary conditions.

For SVEs with $w = 11$ shown in Figures 50(a) and 50(b) the macroscopic stress component $\bar{T}_{12}$ takes the values $9.70\,MPa$ and $7.50\,MPa$, respectively. For ensembles with $w = 13$ in Figures 50(c) and 50(d) $\bar{T}_{12}$ takes the values $4.10\,MPa$ and $2.70\,MPa$ respectively.

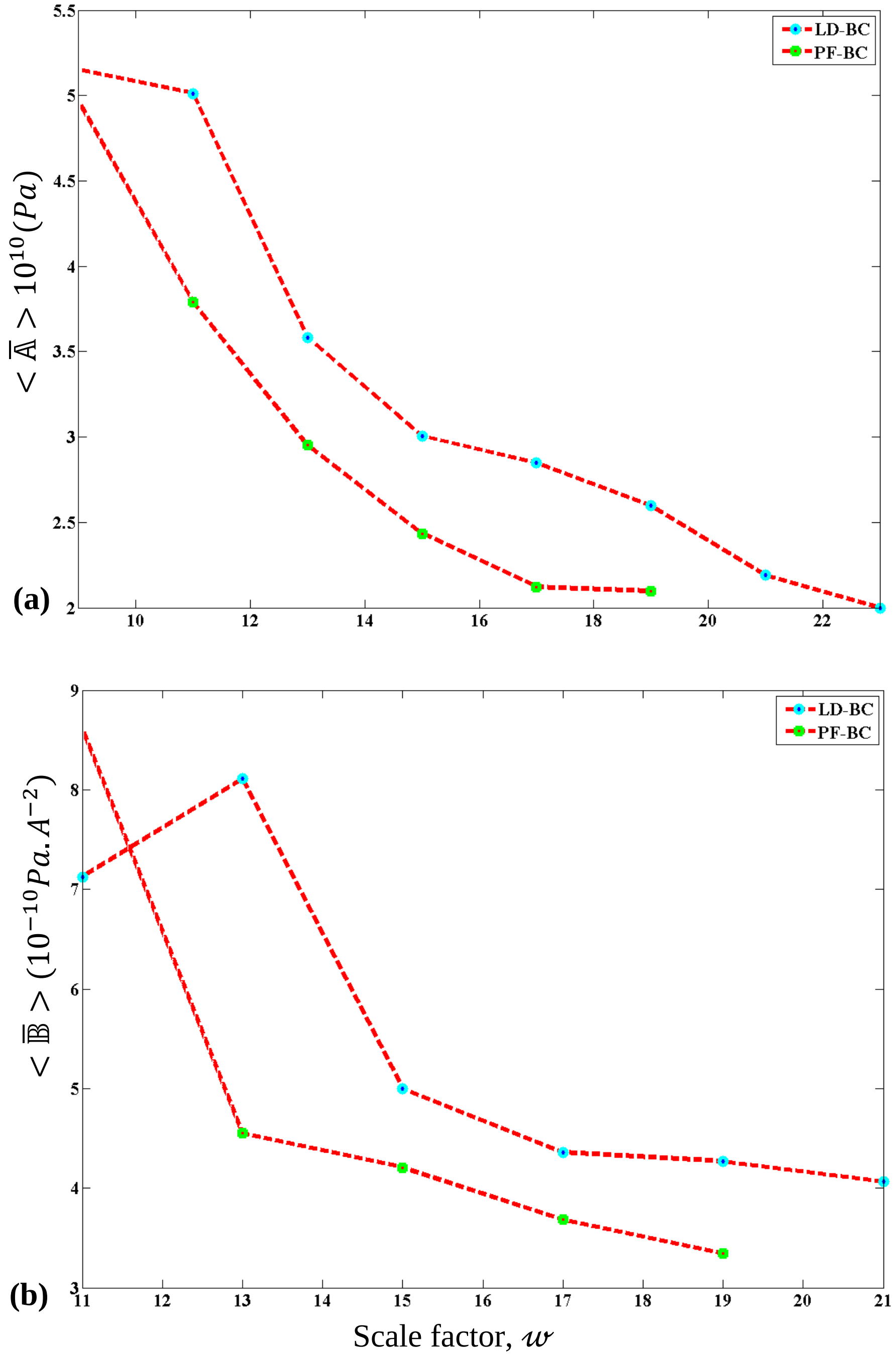

**Figure 49.** Average values of **(a)** $< \bar{\mathbb{A}} >$ and **(b)** $< \bar{\mathbb{B}} >$ versus scale factor, $\boldsymbol{w}$.

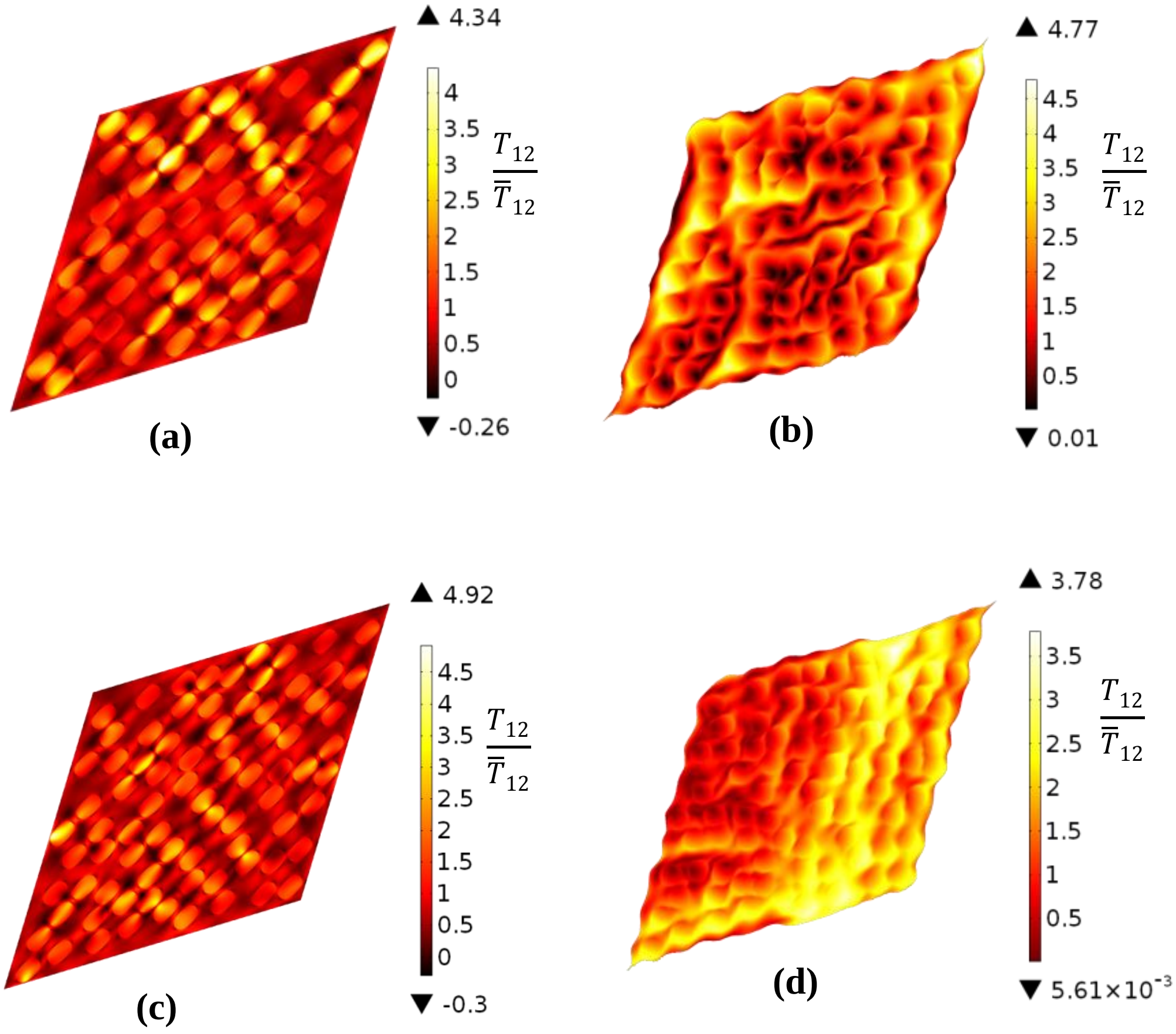

**Figure 50.** FEM results for $\boldsymbol{T_{12}}$ component of stress tensor for SVEs with **(a)** $\boldsymbol{w} = \mathbf{11}$, with LD-BC, **(b)** $\boldsymbol{w} = \mathbf{11}$, with PF-BC, **(c)** $\boldsymbol{w} = \mathbf{13}$, with LD-BC and **(d)** $\boldsymbol{w} = \mathbf{13}$, with PF-BC. All plots are normalized by their average value of $\boldsymbol{T_{12}}$ component.

Figure 51 reports the microscopic distribution of $B_2$ component of magnetic induction, run on the same ensembles described in Figure 50. LD-BC is adopted on the SVEs with $w = 11$ and $w = 13$ as shown in Figure 51(a) and 51(c), respectively. PF-BC is adopted on the SVEs with $w = 11$ and $w = 13$ as shown in Figure 51(b) and 51(d), respectively. Plots are normalized by the corresponding volumetric average component, $\bar{B}_2$. For SVEs with $w = 11$, pictured in Figures 51(a) and 51(c) the macroscopic magnetic induction, $\bar{B}_2$ varies as $3.30\,T$ and $3.74\,T$, respectively. For ensembles with $w = 13$ in Figures 51(c) and 51(d), $\bar{B}_2$ takes the values $3.10\,MPa$ and $2.70\,MPa$ respectively.

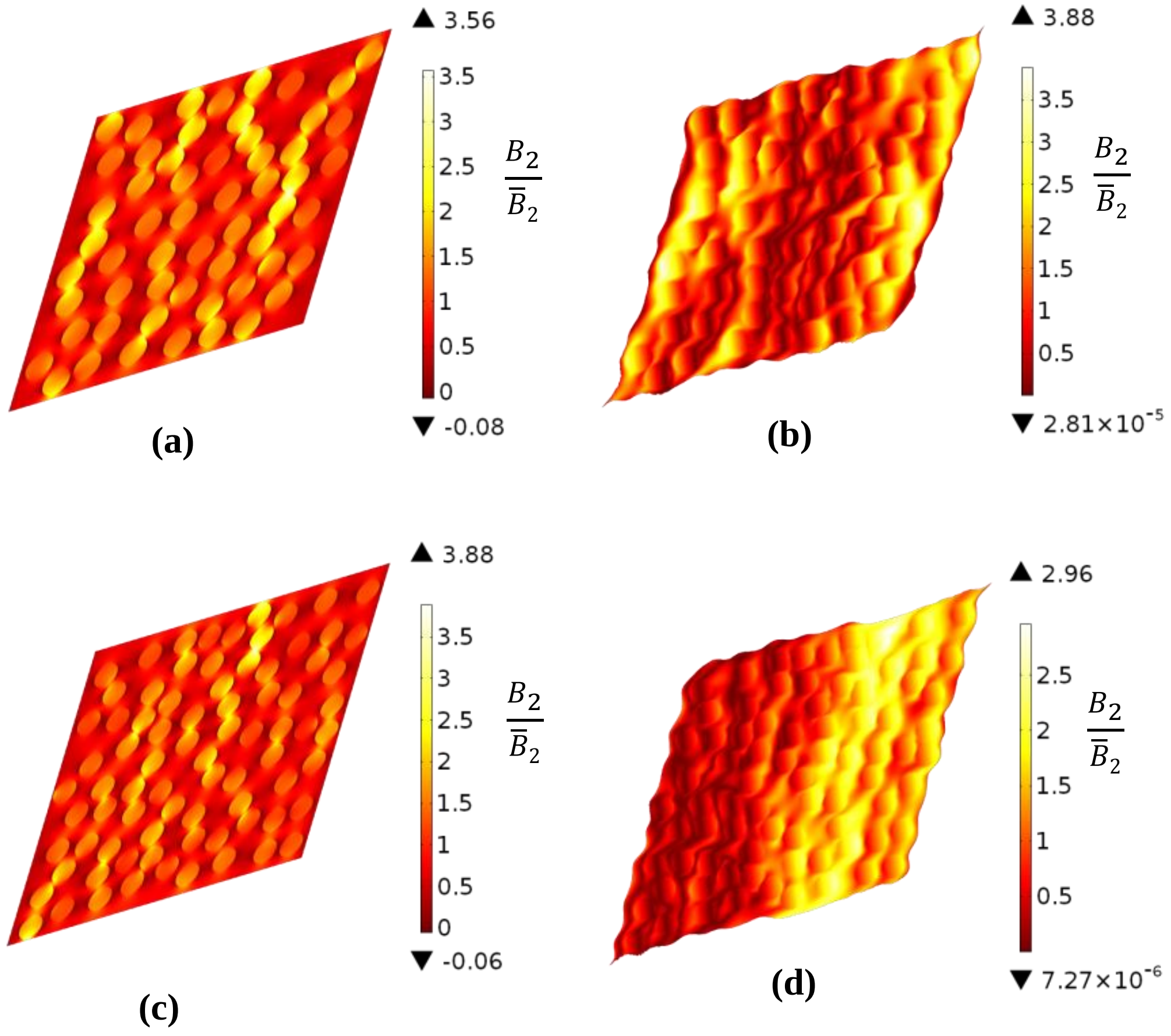

**Figure 51.** FEM results for $\boldsymbol{B_2}$ component of stress tensor for SVEs with **(a)** $\boldsymbol{w = 11}$, with LD-BC, **(b)** $\boldsymbol{w = 11}$, with PF-BC, **(c)** $\boldsymbol{w = 13}$, with LD-BC and **(d)** $\boldsymbol{w = 13}$, with PF-BC. All plots are normalized by their average value of $\boldsymbol{B_2}$ component.

### 6.4 Summary and conclusion

Analytical and FEM approaches are combined with a scale-dependent statistical method to identify a random MEC behavior under magneto-mechanical loadings. Governing equations for magnetoelastic media are presented. BVP is defined on the statistical meso-scale volume elements of the random MEC. SVEs are generated through a computer code with random distribution of particles where particles are prevented from overlapping. The focus is on identifying the minimum CVE size for the random MEC. The CVE size is computed for a particular loading state through a statistical algorithm which combines the FEM-based homogenization approach with Monte-Carlo method.

Results show that for low values of scale factor, $\mathcal{w}$, the random positioning of inclusions in each realization has high influence on the effective response of the MEC, for both cases of boundary conditions. By increasing the scale factor and number of particles (for a constant area fraction) the statistical dispersion and the necessary number of realizations significantly reduces. It is observed that the CVE size $\mathcal{w} = 23$ and $\mathcal{w} = 21$ are obtained for the $\overline{\mathbb{A}}$ and $\overline{\mathbb{B}}$ components respectively. The resultant CVE for the multi-physics system should be identified at the window size where both magneto-mechanical moduli components, $\overline{\mathbb{A}}$ and $\overline{\mathbb{B}}$ are converged. Results show that for PF-BC case, convergence is achieved at a smaller SVE size. FEM results for typical SVEs show that the highest stress and magnetic induction occurs at the inclusions area, due to stiffer mechanical properties and higher magnetic permeability.

# CHAPTER SEVEN

## Concluding Remarks and Future Work

The research presented in this dissertation has resulted in the following contributions:

- High-amplitude wrinkle formation is employed to propose a one-dimensional phononic crystal slab consists of a thin film bonded to a thick compliant substrate. Buckling induced surface instability is employed to generate surface periodic scatterers to control elastic wave propagation in the low thickness composite slab. Simulation results show that the periodic wrinkly structure can be used as a transformative phononic crystal which can switch band diagram of the structure in a reversible manner.
- Dynamic response of a tunable phononic crystal consisting of a porous hyperelastic magnetoelastic elastomer subjected to a macroscopic deformation and an external magnetic field is investigated through considering a magnetoelastic energy function and nonlinear moduli tensors for the medium. The band diagram of the structure is tuned by combined effects of microstructural pattern change and magnetic field.
- A thermally tunable phononic crystal is introduced which utilizes pattern change to control band diagram of the structure.
- A numerical scheme is demonstrated to compute the homogenized properties of the periodic magnetoactive composite structures.

- The finite difference method proposed in this study offers a computationally effective methodology to evaluate homogenized tangent moduli tensors for different loading paths for periodic magnetoactive composites.
- An algorithm for determination of characteristic volume element size and effective properties of random magnetoactive composites is presented which utilizes the FEM-based homogenization approach with Monte-Carlo method.

Recommendations for future work:

- Response of new tunable phononic crystals needs to be studied through combination of control parameters, for example thermal and magnetic effects.
- New pattern transformation paradigms need to be sought in order to propose new periodic structures in controlling elastic wave propagation.
- Dynamic instability in periodic structures has not been explored. Dynamic instability can be occurred in finite amplitude wave propagation in the periodic structures.
- The effect of macroscopic instability in periodic structures needs to be explored.
- The proposed homogenization approach presented in this work needs to be extended to study multi-scale analysis in the periodic and random magnetoelastic composites.
- The proposed random homogenization approach presented in this work needs to be further extended to study different material models and loading conditions. For each case different boundary conditions may be required. The approach needs to be utilized in an optimization process to identify the optimum matrix properties, the particle shape, distribution and percentage in the design of magnetoactive composites.

# REFERENCES

[1]. Kankanala S V and Triantafyllidis N 2004 On finitely strained magnetorheological elastomers *J. Mech. Phys. Solids* **52** 2869-2908

[2]. Ramberg E 2012 *Electrodynamics* (New York: Burlington: Elsevier Science) 88-96.

[3]. Bustamante R 2010 Transversely isotropic nonlinear magneto-active elastomers *Acta Mech.* **210** 183–214.

[4]. Bustamante R, Dorfmann A and Ogden R 2006 Universal relations in isotropic nonlinear magnetoelasticity *Q. J. Mech. Appl. Math.* **59** 435–450.

[5]. Chen L, Gong X L, and Li W H 2007 Microstructures and viscoelastic properties of anisotropic magnetorheological elastomers *Smart Mater. Struct.* **16** 2645–2650.

[6]. Chen T, Nan C-W and Weng G J 2003 Exact connections between effective magnetostriction and effective elastic moduli of fibrous composites and polycrystals *J. Appl. Phys.* **94** 491-495.

[7]. Danas K, Kankanala S V and Triantafyllidis N 2012 Experiments and modeling of iron-particle-filled magnetorheological elastomers *J. Mech. Phys. Solids* **60** 120-138.

[8]. Dorfmann A and Ogden R W 2005 Some problems in nonlinear magnetoelasticity *Z. Angew. Math. Phys.* **56** 718–745.

[9]. Dorfmann A and Ogden R W 2010 Nonlinear electroelastostatics: incremental equations and stability *Int. J. Eng. Sci.* **48** 1–14.

[10]. Farshad M and Benine A 2004 Magnetoactive elastomer composites *Polym. Test* **23** 347-353.

[11]. Jolly M R Carlson J D and Munoz B C 1996 A model of the behaviour of magnetorheological materials *Smart Mater. Struct.* **5** 607–614.

[12]. Rudykh S and Bertoldi K 2013 Stability of anisotropic magnetorheological elastomers in finite deformations: a micromechanical approach *J. Mech. Phys. Solids* **61** 949–967.

[13]. Yang J, Gong X, Deng H, Qin L and Xuan Sh 2012 Investigation on the mechanism of damping behavior of magnetorheological elastomers *Smart Mater. Struct*. **21** 125015(1-11).

[14]. Ginder J M, Nicholes M E, Elie L D and Tardiff J L 1999 Magnetorheological elastomers: properties and applications *Proc. SPIE, Smart Structures and Materials* 1999: Smart Materials Technologies, Newport Beach, CA, Vol. 3675 131–138.

[15]. Brigadnov I A and Dorfmann A 2003 Mathematical modeling of magneto-sensitive elastomers *Int. J. Solids Struct*. **40** 4659–4674.

[16]. Dorfmann A and Ogden R W 2004 Nonlinear magnetoelastic deformations *Q. J. Mech. Appl. Math.* **57** 599–622.

[17]. Jolly M R, Carlson J D and Munoz B C 1996 The magnetoviscoelastic response of elastomer composites consisting of ferrous particles embedded in a polymer Matrix *Smart Mater. Struct* **7** 613–621.

[18]. Dorfmann A, Ogden R W and Saccomandi G 2006 Universal relations for nonlinear magnetoelastic solids *Int. J. NonLin. Mech.* **39** 1699–1708

[19]. Bustamante R, Dorfmann A and Ogden R W 2011 Numerical solution of finite geometry boundary-value problems in nonlinear magnetoelasticity *Int. J. Solids Struct.* **48** 874–883.

[20]. Saxena P 2012 *On wave propagation in finitely deformed magnetoelastic solids* PhD thesis University of Glasgow, Glasgow.

[21]. Ogden R W 2009 Incremental elastic motions superimposed on a finite deformation in the presence of an electromagnetic field *Int. J. NonLin. Mech.* **44** 570-580.

[22]. Otténio M, Destrade M and Ogden R W 2008 Incremental magnetoelastic deformations with application to surface instability *J. Elasticity* **90** 19–42

[23]. Galipeau E and Ponte Castañeda P 2013 A finite-strain constitutive model for magnetorheological elastomers: magnetic torques and fiber rotations. *J. Mech. Phys. Solids* **61** 1065-1090.

[24]. Destrade M and Scott N H 2004 Surface waves in a deformed isotropic hyperelastic material subject to an isotropic internal constraint *Wave Motion* **40** 347–357.

[25]. Dorfmann A L and Ogden R W 2014 nonlinear theory of electroelastic and magnetoelastic interactions Springer US, 91-112.

[26]. Lee J S and Its E N 1992 Propagation of rayleigh waves in magneto-elastic media *J. Appl. Mech.* **59** 812–818.

[27]. Hefni I A Ghaleb A F and Maugin G A 1995 One dimensional bulk waves in a nonlinear magnetoelastic conductor of finite electric conductivity *Int. J. Eng. Sci.* **33** 2067-2084.

[28]. Destrade M and Ogden R W 2011 On magneto-acoustic waves in finitely deformed elastic solids *Math. Mech. Solids* **16** 594–604.

[29]. Kochmann D M and Venturini G N 2013 Homogenized mechanical properties of auxetic composite materials in finite-strain elasticity *Smart Mater. Struct.* **22** 084004(7pp).

[30]. Kanouté P, Boso D P, Chaboche J L and Schrefler B A 2009 Multiscale methods for composites: a review *Arch. Comput. Meth. Eng.* **16** 31–75.

[31]. Kouznetsova V, Geers M G D and Brekelmans W A M 2002 Multi-scale constitutive modeling of heterogeneous materials with a gradient-enhanced computational homogenization scheme *Int. J. Num. Methods Eng.* **54** 1235–1260.

[32]. Miehe C 2002 Strain-driven homogenization of inelastic microstructures and composites based on an incremental variational formulation *Int. J. Num. Methods Eng.* **55** 1285–1322.

[33]. Miehe C and Koch A 2002 Computational micro-to-macro transitions of discretized microstructures undergoing small strains *Arch. Appl. Mech.* **72** 300–317.

[34]. Temizer I and Wriggers P 2008 On the computation of the macroscopic tangent for multiscale volumetric homogenization problems *Comput. Meth. Appl. Mech. Eng.* **198** 495–510.

[35]. Nemat-Nasser S 1999 Averaging theorems in finite deformation plasticity *Mech. Mater.* **31** 493-523.

[36]. Smit R, Brekelmans W and Meijer H 1998 Prediction of the mechanical behavior of nonlinear heterogeneous systems by multi-level finite element modeling *Comput. Meth. Appl. Mech. Eng.* **155** 181–192.

[37]. Wang D, Chen J-S and Sun L 2003 Homogenization of magnetostrictive particle filled elastomers using an interface-enriched reproducing kernel particle method *Finite Elem. Anal. Des.* **39** 765–782.

[38]. Yin H M, Sun L Z and Chen J S 2002 Micromechanics-based hyperelastic constitutive modeling of magnetostrictive particle-filled elastomers *Mech. Mater.* **34** 505–516.

[39]. Bravo-Castillero J, Rodríguez-Ramos R, Mechkour H, Otero J A, Hernández J, Sixto- Camacho L M, Guinovart-Díaz R and Sabina F J 2009 Homogenization and effective properties of thermomagnetoelectroelastic composites *J. Mech. Mater. Struct.* **4** 819–36.

[40]. Li J and Dunn M 1998 Micromechanics of magnetoelectroelastic composite materials: average fields and effective behavior *J. Intell. Mater. Syst. Struct.* **9** 404–16.

[41]. Bravo-Castillero J, Rodriguez-Ramos R, Mechkour H, Otero J and Sabina F 2008 Homogenization of magneto-electro-elastic multilaminated materials *Quart. J. Mech. Appl. Math.* **61** 311–32.

[42]. Berger H, Gabbert U, Koppe H, Rodriguez-Ramos R, Bravo-Castillero J, Guinovart-Diaz R, Otero J A and Maugin G A 2003 Finite element and asymptotic homogenization methods applied to smart material composites *Comput. Mech.* **33** 61–67.

[43]. Costanzo F, Gray G L and Andia P C 2005 On the definitions of effective stress and deformation gradient for use in MD: Hill's macro-homogeneity and the viral theorem *Int. J. Eng. Sci.* **43** 533–555.

[44]. Ponte Castañeda P and Galipeau E 2011 Homogenization-based constitutive models for magneto-rheological elastomers at finite strain *J. Mech. Phys. Solids* **59** 194–215.

[45]. Chatzigeorgiou G, Javili A and Steinmann P 2014 Unified magnetomechanical homogenization framework with application to magnetorheological elastomers *Math. Mech. Solids* **19** 193-211.

[46]. Labusch M, Etier M, Lupascu D C, Schröder J and Keip M-A 2014 Product properties of a two-phase magneto-electric composite: Synthesis and numerical modeling *Comput. Mech.* **54** 71–83.

[47]. Javili A, Chatzigeorgiou G and Steinmann P 2013 Computational homogenization in magneto-mechanics. *Int. J. Solids Struct.* **50** 4197–4216.

[48]. Schröder J 2009 Derivation of the localization and homogenization conditions for electro-mechanically coupled problems *Comput. Mater. Sci.* **46** 595–599

[49]. Schröder J and Keip M A 2012 Two-scale homogenization of electromechanically coupled boundary value problems *Comput. Mech.* **50** 229–244.

[50]. Schröder J 2014 A numerical two-scale homogenization scheme: the $FE^2$-method, in: J. Schröder K Hackl (Eds.), Plasticity and Beyond, in: CISM International Centre for Mechanical Sciences, vol. 550, Springer Vienna, 1–64.

[51]. Keip M A, Steinmann P and Schröder J 2014 Two scale computational homogenization of electro-elasticity at finite strains *Comput. Methods Appl. Mech. Eng.* **278** 62- 79.

[52]. Miehe C 2003 Computational micro-to-macro transitions for discretized micro-structures of heterogeneous materials at finite strains based on the minimization of averaged incremental energy *Comput. Methods Appl. Mech. Engrg.* **192** 559–591.

[53]. Trovalusci P, De Bellis M L, Ostoja-Starzewski M and Murrali 2014 A particulate random composites homogenized as micropolar materials *Meccanica* **49** 2719–2727.

[54]. Trovalusci P, De Bellis M L, Ostoja-Starzewski M and Murrali A 2015 Scale-dependent homogenization of random composites as micropolar continua *Eur. J. Mech. A. Solids* **49** 396-407

[55]. Ma J, Zhang J, Li L, Wriggers P and Sahraee S 2014 Random homogenization analysis for heterogeneous materials with full randomness and correlation in microstructure based on finite element method and Monte-Carlo method *Comput. Mech.* **54** 1395–1414.

[56]. Gitman I M, Askes H and Sluys L J 2007 Representative volume: Existence and size determination *Eng. Fract. Mech.* **74** 2518‑2534.

[57]. Temizer I and Zohdi T I 2007 A numerical method for homogenization in non-linear elasticity *Comput. Mech***.** **40** 281–298.

[58]. Terada K, Horib M, Kyoyac T and Kikuchi N 2000 Simulation of the multi-scale convergence in computational homogenization approaches *Int. J. Solids Struct.* **37** 2285-2311.

[59]. Khisaeva Z F and Ostoja-Starzewski M 2007 Scale effects in infinitesimal and finite thermoelasticity of random composites *J. Therm. Stresses* **30** 587–603.

[60]. Geers M G D, Kouznetsova V G and Brekelmans W A M 2010 Multi-scale computational homogenization: Trends and challenges *J. Comput. Appl. Math.* **234** 2175-2182.

[61]. Temizer I and Wriggers P 2011 An adaptive multiscale resolution strategy for the finite deformation analysis of microheterogeneous structures *Comput. Methods Appl. Mech. Engrg.* **200** 2639–2661.

[62]. Temizer I and Wriggers P 2011 Homogenization in finite thermoelasticity *J. Mech. Phys. Solids* **59** 344–372.

[63]. Francfort G A 1983 Homogenization and linear thermoelasticity SIAM *J. Math. Anal*. **14** 696–708.

[64]. Ostoja-Starzewski M, Du X, Khisaeva Z F and Li W 2007 Comparisons of the size of the representative volume element in elastic, plastic, thermoelastic, and permeable random microstructures *Int. J. Multiscale Comput. Eng.* **5** 73-82.

[65]. Vel S S and Goupee A J 2010 Multiscale thermoelastic analysis of random heterogeneous materials Part I: Microstructure characterization and homogenization of material properties *Comp. Mat. Sci.* **48** 22–38.

[66]. Zhang H W, Yang D S, Zhang S and Zheng Y G 2014 Multiscale nonlinear thermoelastic analysis of heterogeneous multiphase materials with temperature-dependent properties *Finite Elem. Anal. Des.* **88** 97–117.

[67]. Ethiraj G 2014 *Computational modeling of ferromagnetics and magnetorheological elastomers* PhD thesis University of Stuttgart, Stuttgart.

[68]. Castañeda P P and Siboni M H 2012 A finite-strain constitutive theory for electro-active polymer composites via homogenization *Int. J. Non Linear Mech.* **47** 293–306.

[69]. Ponte Castaneda P and Galipeau E 2011, Homogenization-based constitutive models for magnetorheological elastomers at finite strain *J. Mech. Phys. Solids* **59** 194-215.

[70]. Galipeau E, Rudykh S, deBotton G and Castañeda P 2014 Magnetoactive elastomers with periodic and random microstructures *Int. J. Solids Struct.* **51** 3012–3024.

[71]. Elachi C 1976, Waves in active and passive periodic structures a review *Proc. IEEE* **64** 1666 -1698.

[72]. Liu Z, Chan C T and Sheng P 2000 Elastic wave scattering by periodic structures of spherical objects theory and experiment *Phys. Rev. B* **62** 2446-2458.

[73]. Olsson R H and El-Kady I 2009 Microfabricated phononic crystal devices and applications *Meas. Sci. Technol.* **20** 012002-012015.

[74]. Kittle C 1986, Introduction to solid state physics 8th Ed Wiley New York 27-53.

[75]. Sigalas M M and Economou E N 1993 Band structure of elastic waves in two dimensional systems *Solid State Commun.* **86** 141–143.

[76]. Wang G, Wen X, Wen J, Shao L and Liu Y 2004 Two dimensional locally resonant phononic crystals with binary structures *Phys. Rev. Lett.* **93** 154302(1-4).

[77]. Åberg M and Gudmundson P 1997 The usage of standard finite element codes for computation of dispersion relations in materials with periodic microstructure *J. Acoust. Soc. Am.* **102** 2007-2013.

[78]. Langlet P, Hladky-Hennion A C and Decarpigny J N 1995 Analysis of the propagation of plane acoustic waves in passive periodic materials using the finite element method *J. Acoust. Soc. Am*. **98** 2792-2800.

[79]. Liu Z 2000 Locally resonant sonic materials *Science* **289** 1734-1736.

[80]. Sanchez-Perez J V, Rubio C, Martinez-Sala R, Sanchez-Grandia R and Gomez V 2002 Acoustic barriers based on periodic arrays of scatterers *Appl. Phys. Lett.* **81** 5240-5242.

[81]. Vasseur J O, Deymier P A, Khelif A, Lambin P, Djafari-Rouhani B, Akjouj A, Dobrzynski L, Fettouhi N and Zemmouri J 2002 Phononic crystal with low filling fraction and absolute acoustic band gap in the audible frequency range: a theoretical and experimental study *Phys. Rev. E* **65** 056608(1-6).

[82]. Goffaux C and Vigneron J P 2001 Theoretical study of a tunable phononic band gap system *Phys. Rev. B* **64** 075118(1-5).

[83]. Yeh J 2007 Control analysis of the tunable phononic crystal with electrorheological material *Physica B Condensed Matter,* **400** 137-144.

[84]. Wang Y, Li F, Kishimoto K, Wang Y, Huang W and Jiang X 2009 Elastic wave band gaps in magnetoelectroelastic phononic crystals *Wave Motion* **46** 47-56.

[85]. Khelif A, Achaoui Y, Benchabane S, Laude V and Aoubiza B 2010 Locally resonant surface acoustic wave band gaps in a two-dimensional phononic crystal of pillars on a surface *Phys. Rev. B* **81** 214303.

[86]. Gonella S and Ruzzene M 2008 Analysis of in-plane wave propagation in hexagonal and re-entrant lattices *J. Sound Vib.* **312** 125–139.

[87]. Collet M, Ouisse M, Ruzzene M and Ichchou M N 2011 Floquet–Bloch decomposition for the computation of dispersion of two-dimensional periodic, damped mechanical systems *Int. J. Solids Struct.* **48** 2837–2848.

[88]. Robillard J, BouMatar O, Vasseur J O, Deymier P A, Stippinger M, Hladky-Hennion A, Pennec Y and Djafari-Rouhani B 2009 Tunable magnetoelastic phononic crystals *Appl. Phys. Lett.* **95** 124104(1-3).

[89]. Sukhovich A, Jing L and Page J H 2008 Negative refraction and focusing of ultrasound in two-dimensional phononic crystals *Phys. Rev. B* **77** 014301(1-9).

[90]. Goffaux C and Vigneron J P 2001 Theoretical study of a tunable phononic band gap system *Phys. Rev. B* **64** 075118(1-5).

[91]. Deymier P A 2013 Acoustic metamaterials and phononic crystals Springer Berlin 253–372.

[92]. Xu Z, Wu F and Guo Z 2013 Shear-wave band gaps tuned in two-dimensional phononic crystals with magnetorheological material *Solid State Commun.* **154** 43–45.

[93]. Bou Matar O, Robillard J F, Vasseur J O, Hladky-Hennion A-C, Deymier P A Pernod P and Preobrazhensky V 2012 Band gap tunability of magneto-elastic phononic crystal *J. Appl. Phys.* **111** 054901(1-14).

[94]. Ding R, Su X, Zhang J and Gao Y 2014 Tunability of longitudinal wave band gaps in one dimensional phononic crystal with magnetostrictive material *J. Appl. Phys*. **115** 074104(1-8).

[95]. Sukhovich A, Jing L and Page J H 2008 Negative refraction and focusing of ultrasound in two-dimensional phononic crystals *Phys. Rev. B* **77** 014301(1-9).

[96]. Otsuka P H, Nanri K, Matsuda O, Tomoda M, Profunser D M, Veres I A, Danworaphong S, Khelif A, Benchabane S, Laude V and Wright O B 2013 Broadband evolution of phononic-crystal-waveguide eigenstates in real- and k-spaces *Sci. Rep.* **3** 3351.

[97]. Wang P, Shim J and Bertoldi K 2013 Effects of geometric and material nonlinearities on tunable band gaps and low-frequency directionality of phononic crystals *Phys. Rev. B* **88** 014304

[98]. Bertoldi K and Boyce M C 2008 Wave propagation and instabilities in monolithic and periodically structured elastomeric materials undergoing large deformations *Phys. Rev. B* **78** 184107(1-16).

[99]. Babaee S, Wang P and Bertoldi K 2015 Three-dimensional adaptive soft phononic crystals *J. Appl. Phys.* **117** 244903.

[100]. Farhan R *Lecture notes ECE 407*- Spring 2009 - Cornell University.

[101]. Geryak R and Tsukruk V V 2014 Reconfigurable and actuating structures from soft materials *Soft Matter* **10** 1246.

[102]. Bertoldi K, Boyce M C, Deschanel S, Prange S M and Mullin T 2008 Mechanics of deformation-triggered pattern transformations and superelastic behavior in periodic elastomeric structures *J. Mech. Phys. Solids* **56** 2642‑2668.

[103]. Kamat P V, Thomas K G, Barazzouk S, Girishkumar G, Vinodgopal K and Meisel D 2004 Self-assembled linear bundles of single wall carbon nanotubes and their alignment and deposition as a film in a dc field *J. Am. Chem. Soc*. **126** 10757.

[104]. Fuhrer R, Athanassiou E K, Luechinger N A and Stark W J 2009 Crosslinking metal nanoparticles into the polymer backbone of hydrogels enables preparation of soft, magnetic field-driven actuators with muscle-like flexibility *Small* **5** 383.

[105]. Ionov L 2011 Soft microorigami: self-folding polymer films *Soft Matter* **7** 6786-6791.

[106]. Babaee S, Shim J, Weaver J C, Chen E R, Patel N and Bertoldi K 2015 3D soft metamaterials with negative Poisson's ratio *Adv. Mater* **25** 5044-5049.

[107]. Goncu F, Luding S and Bertoldi K 2012 Exploiting pattern transformation to tune phononic band gaps in a two-dimensional granular crystal *J. Acoust. Soc. Am.* **131** EL475.

[108]. Rudykh S and Boyce M C 2014 Transforming wave propagation in layered media via instability-induced interfacial wrinkling *Phys. Rev. Lett.* **112** 034301.

[109]. Cao Y and Hutchinson J W 2012 Wrinkling phenomena in neo-Hookean film/substrate bilayers *J. A Mech.* **79** 031019.

[110]. Saha S K and Culpepper M L 2012 Predicting the quality of one-dimensional periodic micro and nano structures fabricated via wrinkling *in Proceedings of the ASME 2012 International Mechanical Engineering Congress and Exposition,* Houston TX USA.

[111]. Schweikart A and Fery A 2009 Controlled wrinkling as a novel method for the fabrication of patterned surfaces *Microchim. Acta* **165** 249.

[112]. Li Y, Kaynia N, Rudykh S and Boyce M C 2013 Wrinkling of interfacial layers in stratified composites *Adv. Eng. Mater.* **15** 921.

[113]. Zhao Y, Cao Y, W. Hong W, Wadee M K and Fen X-Q 2015 Towards a quantitative understanding of period-doubling wrinkling patterns occurring in film/substrate bilayer systems *Proc. R. Soc. A* **471** 20140695.

[114]. Cai S, Breid D, Crosby A J, Suo Z and Hutchinson J W 2011 Periodic patterns and energy states of buckled films on compliant substrates *J. Mech. Phys. Solids* **59** 1094.

[115]. Huang Z, Hong W and Suo Z 2004 Evolution of wrinkles in hard films on soft substrates *Phys. Rev. E* **70** 030601.

[116]. Damil N, Potier-Ferry M and Hu H 2014 Membrane wrinkling revisited from a multi-scale point of view *Adv. Model. Simul. Eng. Sci.* **1** 6.

[117]. Taylor M, Bertoldi K and Steigmann D J 2014 Spatial resolution of wrinkle patterns in thin elastic sheets at finite strain *J. Mech. Phys. Solids* **62** 163.

[118]. Cerda E and Mahadevan L 2003 Geometry and physics of wrinkling *Phys. Rev. Lett.* **90** 074302.

[119]. Bowden N, Brittain S, Evans A, Hutchinson J W and Whitesides G 1998 Spontaneous formation of ordered structures in thin films of metals supported on an elastomeric polymer *Nature* **393** 146.

[120]. Chan E P, Karp J M, and Langer R S 2011 A self-pinning adhesive based on responsive surface wrinkles *J. Polym. Sci. B.* **49** 40.

[121]. Chen C M and Yang S 2012 Wrinkling instabilities in polymer films and their applications *Polym. Int.* **61** 1041.

[122]. Chandra D, Yang S and Lin PC 2007 Strain responsive concave and convex microlens arrays *Appl. Phys. Lett.* **91** 251912.

[123]. Chan E P and Crosby A J 2006 Fabricating microlens arrays by surface wrinkling *Adv. Mater.* **18** 3238.

[124]. Khang DY, Jiang H, Huang Y and Rogers J A 2006 A stretchable form of single-crystal silicon for high-performance electronics on rubber substrates *Science* **311** 208.

[125]. Chen Y and Crosby A J 2014 High aspect ratio wrinkles via substrate prestretch *Adv. Mater.* **26** 5626.

[126]. Sun J-Y, Xia S, Moon M-W, Oh K H and Kim K-S 2012 Folding wrinkles of a thin stiff layer on a soft substrate *Proc. Royal Soc. London Ser. A* **468** 932–953.

[127]. Khelif A, Achaoui Y, Benchabane S, Laude V and Aoubiza B 2010 Locally resonant surface acoustic wave band gaps in a two-dimensional phononic crystal of pillars on a surface *Phys. Rev. B* **81** 214303.

[128]. Pennec Y, Vasseur J-O, Djafari-Rouhani B, Dobrzyński L and Deymier P A 2010 Two-dimensional phononic crystals: Examples and applications *Surf. Sci. Rep.* **65** 229-291.

[129]. Lucklum R 2014 Phononic crystals and metamaterials - Promising new sensor platforms *Procedia Eng.* **87** 40-45.

[130]. Olsson III R H, El-Kady I F, Su M F, Tuck M R and Fleming J G 2008 Microfabricated VHF acoustic crystals and waveguides *Sens. Actuators A* **145** 87-93.

[131]. Mohammadi S, Eftekhar A A, Pourabolghasem R and Adibi A 2011 Simultaneous high-Q confinement and selective direct piezoelectric excitation of flexural and extensional lateral vibrations in a silicon phononic crystal slab resonator *Sens. Actuators A* **167** 524-530.

[132]. Safavi-Naeini A H and Painter O 2010 Design of optomechanical cavities and waveguides on a simultaneous bandgap phononic-photonic crystal slab *Opt. Express* **18** 14926-14943.

[133]. Triantafyllidis N, Nestorovic M D and Schraad M W 2006 Failure surfaces for finitely strained two-phase periodic solids under general in-plane loading *J. Appl. Mech.***73** 505–516.

[134]. Ruzzene M, Scarpa F and Soranna F 2003 Wave beaming effects in two-dimensional cellular structures *J. Smart Mater. Struct.* **12** 363-372

[135]. Ruzzene M and Scarpa F 2005 Directional and band-gap behavior of periodic auxetic lattices *Phys. Status solidi B* **242** 665–680.

[136]. Introduction to Comsol Multiphysics. User Guide.

[137]. M. Attard 2003 Finite strain - isotropic hyperelasticity *Int. J. Solids Struct.* **40** 4353–4378.